\documentclass[reprint,groupedaddress,superscriptaddress,aps,prd,preprintnumbers,onecolumn,11pt,fleqn,nofootinbib,nobibnote,eqsecnum,floatfix,showpacs,preprintnumbers,a4paper]{revtex4-2}

\usepackage{xcolor,color}
\usepackage{bm,amsmath,amssymb,dsfont}
\usepackage{graphicx}
\usepackage{afterpage}
\usepackage{enumerate}

\long\def\comment#1{ }

\newcommand{\beq}{\begin{equation}}
\newcommand{\eeq}{\end{equation}}
\newcommand{\bal}{\begin{align}}
\newcommand{\eal}{\end{align}}

\newcommand{\del}{\partial}

\newcommand{\nn}{\nonumber\\}
\newcommand{\rmd}{{\rm d}}
\newcommand{\dif}{{\rm d}}

\newcommand{\bk}{\bm{k}}

\newcommand{\bx}{\bm{x}}
\newcommand{\by}{\bm{y}}

\newcommand{\br}{\bm{r}}
\newcommand{\bb}{\bm{b}}

\newcommand{\mcal}{\mathcal}

\newcommand{\bP}{\bm{P}}

\begin{document}

\title{Incoherent diffractive dijet production in electron DIS off nuclei at high energy}

\author{Benjamin Rodriguez-Aguilar}
\email{benjaroagui@gmail.com}
\affiliation{European Centre for Theoretical Studies in Nuclear Physics and Related Areas (ECT*)\\
and Fondazione Bruno Kessler, Strada delle Tabarelle 286, 38123 Villazzano (TN), Italy}
\affiliation{Physics Department, Trento University, Via Sommarive 14, 38123 Povo (TN), Italy}

\author{D.N.~Triantafyllopoulos}
\email{trianta@ectstar.eu}
\affiliation{European Centre for Theoretical Studies in Nuclear Physics and Related Areas (ECT*)\\
and Fondazione Bruno Kessler, Strada delle Tabarelle 286, 38123 Villazzano (TN), Italy}

\author{S.Y.~Wei}
\email{shuyi@sdu.edu.cn}
\affiliation{Key Laboratory of Particle Physics and Particle Irradiation (MOE), Institute of frontier and interdisciplinary science, Shandong University, Qingdao, Shandong 266237, China}

\date{\today}

\begin{abstract}
We study incoherent diffractive dijet production in electron-nucleus deep inelastic scattering at small $x_{\rm \scriptscriptstyle Bj}$ within the Color Glass Condensate. We follow the general approach of \cite{Mantysaari:2019hkq} but we focus on the correlation limit, that is, when the momentum transfer $\Delta_{\perp}$ and the gluon saturation momentum $Q_s$ of the nucleus are much smaller than the individual jet momentum $P_{\perp}$. We arrive at analytic expressions for the dijet cross section, which can be written as a sum of four terms which exhibit factorization: each such term is a product between a hard factor, which includes the decay of the virtual photon to the $q\bar{q}$ pair, and a semihard one which involves the dipole-nucleus scattering amplitude. We further calculate the azimuthal anisotropies $\langle \cos 2 \phi \rangle$ and $\langle \cos 4 \phi \rangle$. They are of the same order in the hard momentum $P_{\perp}$, but the $\langle \cos 4 \phi \rangle$ is logarithmically suppressed due to its dependence on the semihard factor. Finally, in order to extend the validity of our result towards the perturbative domain, we calculate the first higher kinematic twist, i.e.~the correction of relative order $\Delta_{\perp}^2/P_{\perp}^2$.

\end{abstract}

\maketitle

\section{Introduction}
\label{sec:intro}

It has been known for quite some time that the perturbative part of the light-cone wavefunction (LCWF) of a hadron or a nucleus contains gluon modes whose occupation numbers rise rapidly when $x$ becomes very small \cite{Lipatov:1976zz,Kuraev:1977fs,Balitsky:1978ic,Mueller:1993rr}, with $x$ being the longitudinal momentum fraction carried by the gluon under consideration. Such an increase can eventually lead to violation of unitarity limits: for example, at fixed impact parameter, the magnitude of the amplitude for a small projectile color dipole to scatter off the hadron/nucleus should never exceed unity. The problem finds its solution within perturbative QCD, since non-linear dynamics start to develop for the highly occupied modes, they tame their growth and lead to the phenomenon of gluon saturation \cite{Gribov:1984tu,Mueller:1985wy,Mueller:2001fv}. A dynamical semihard scale, the saturation momentum $Q_s$, is generated and serves as the boundary between a dilute regime, in which gluons have transverse momenta $k_{\perp} > Q_s$, and a dense regime, where gluons have transverse momenta $k_{\perp} < Q_s$. To a good accuracy, this scale grows according to $Q_s^2(A,x) \propto \Lambda^2 A^{1/3} (1/x)^\lambda$, where $\Lambda$ is the QCD scale, $A$ is the atomic number of the nucleus and $\lambda \simeq 0.25$, thus justifying the use of weak coupling methods for sufficiently large nuclei and/or at small-$x$. The Color Glass Condensate \cite{Iancu:2003xm,Gelis:2010nm,Kovchegov:2012mbw} has emerged as a modern QCD effective theory which incorporates gluon saturation. By calculating the emission of soft gluons in the presence of a strong background color field, there have been derived non-linear small-$x$ evolution equations \cite{Balitsky:1995ub,Kovchegov:1999yj,JalilianMarian:1997jx,JalilianMarian:1997gr,Kovner:2000pt,Weigert:2000gi,Iancu:2000hn,Iancu:2001ad,Ferreiro:2001qy} for gluon correlators pertinent to the determination of cross sections or other physical observables.  

The typical way to probe such a system is by collisions at high energy with a smaller projectile. Proton-nucleus ($pA$) collisions, Deep Inelastic Scattering (DIS) of electrons off hadrons or nuclei ($ep$ and $eA$), or even ultra-peripheral nucleus-nucleus collisions (UPCs) in which a quasi real photon emitted by the one nucleus scatters off the other nucleus, provide the main laboratories for exploring gluon saturation. Over the last few years there has been a great interest in correlations among the particles produced in the final state of these collisions. In this context, one of the simplest and most studied processes is the inclusive production of two jets in the forward projectile direction \cite{Marquet:2007vb,Dominguez:2011wm,Metz:2011wb,Dominguez:2011br,Mueller:2013wwa,Kotko:2015ura,Dumitru:2015gaa,vanHameren:2016ftb,Marquet:2016cgx,Kotko:2017oxg,Dumitru:2018kuw,Iancu:2018hwa,Klein:2019qfb,Mantysaari:2019hkq,Iancu:2020mos,Hatta:2021jcd,Boussarie:2021ybe,Caucal:2021ent,Taels:2022tza}. In $ep$ and $eA$ collisions the two jets are initiated by the $q\bar{q}$ pair to which the virtual photon fluctuates, while in $pA$ collisions they are mostly due to a quark-gluon pair emerging from the splitting of a collinear quark within the proton. Here one is mostly interested in the so-called ``correlation limit'' \cite{Dominguez:2011wm}: this refers to the regime in which the transverse momenta of the two jets are much harder than both the saturation momentum and the imbalance between them, with the second condition meaning that they propagate almost back-to-back in the transverse plane. To leading order in perturbation theory, the dijet imbalance is the same as the momentum transferred from the target to the projectile and therefore in the kinematics of interest it can be controlled by the physics of saturation. Similar arguments apply to the case of dihadron production \cite{Albacete:2010pg,Stasto:2011ru,Lappi:2012nh,Iancu:2013dta,Zheng:2014vka,Marquet:2017xwy,Albacete:2018ruq,Bergabo:2022tcu,Bergabo:2023wed}, since the perturbative part of the process remains the same.

Of equal value is the diffractive production of jets (or hadrons) in the same types of collisions \cite{Bartels:1999tn,Altinoluk:2015dpi,Hatta:2016dxp,Hagiwara:2017fye,Mantysaari:2019csc,Salazar:2019ncp,Mantysaari:2019hkq,Iancu:2021rup,Hatta:2022lzj,Beuf:2022kyp,Iancu:2022lcw}. In diffraction there is no net color exchange between the projectile partons and the hadronic/nuclear target and as a consequence one observes a rapidity gap, i.e.~an angular region between the two colliding objects which is void of particles. Focusing on $ep$ and $eA$ collisions, on the projectile side one must require the partonic fluctuation of the virtual photon to remain in a color singlet state after the scattering with the target. Regarding the target side, the hadron or nucleus may remain intact or not after the collision and we respectively refer to the process as coherent or incoherent diffraction. In order to calculate the total diffraction, we must average the cross section over all target configurations. If one averages at the amplitude level and then takes the square, one obtains only the coherent component. Subtracting the latter from the total diffractive cross section, we get the incoherent component and thus it becomes clear that incoherent diffraction is directly related to fluctuations in the target wave-function \cite{Good:1960ba}. For instance, in an appealing picture, exclusive vector-meson production in $ep$ DIS has been used to constrain geometrical fluctuations in the shape of the proton \cite{Mantysaari:2016ykx,Mantysaari:2016jaz}, while further related analytical work on the same ``hot spots'' model has been done in \cite{Demirci:2022wuy}. But with a more homogeneous target, like a large nucleus, the momentum transfer to the projectile is of the order of the inverse nucleus diameter, which is a very small non-perturbative scale and thus completely uninteresting to our purposes. Still, a different mechanism has been studied in \cite{Mantysaari:2019hkq} (see also the earlier and closely related work in \cite{Marquet:2010cf}): the sub-leading in $N_c$ piece (with $N_c$ the number of colors) of the gluon correlator which corresponds to diffraction, leads to incoherent scattering with a momentum transfer that can reach large perturbative values. As we shall discuss in Sect.~\ref{sec:corr}, this piece originates from the color exchange between different substructures inside the nucleus and as such it may be regarded as a color fluctuation. Notice that the relevant gluon correlator is to be calculated within the CGC framework and thus it involves minimal model dependence. 

The transverse momenta of the jets considered in \cite{Mantysaari:2019hkq} were rather generic and therefore the approach was necessarily numerical. Our scope here is to give an analytical insight in the correlation limit and the outline of this work is as follows. In Sect.~\ref{sec:incoherent} we present a short review of the derivation of the general formula which gives the $q\bar{q}$ contribution to dijet production. Then we isolate the part of the scattering which leads to incoherent diffraction. In Sect.~\ref{sec:corr} we evaluate the corresponding diffractive correlator in the correlation limit assuming the Gaussian approximation. In Sect.~\ref{sec:cross-section} we give the cross sections for both transversely and longitudinally polarized photons in a form which separates the hard from the semihard dynamics. In Sect.~\ref{sec:GDs} we analyze the momentum dependence of the semihard factor. Having all the elements at our disposal, in Sect.~\ref{sec:numerics} we present our results for the average (over the azimuthal angle) cross section and for the elliptic and quadrangular anisotropies. In the correlation limit, the latter quantify the distribution in the orientation of the imbalance w.r.t.~to the axis defined by the almost back-to-back jets. In Sect.~\ref{sec:kin} we evaluate the first higher kinematic twist and point out the changes that it brings to the elliptic anisotropy. Finally we conclude and discuss potential future developments in Sect.~\ref{sec:conc}. Some technical parts are presented in the four appendices at the end of the paper.

\section{Incoherent diffractive dijet production and kinematics}
\label{sec:incoherent}

First, we briefly review the derivation of the cross section for incoherent diffractive dijet production in high energy deep inelastic scattering of a virtual photon $\gamma^*$ off a generic nucleus with atomic number $A$.  Let us define the kinematics of the incoming particles. We shall work in a frame where they are both ultra-relativistic and carry zero transverse momentum. More precisely, in light-cone notation, the photon with virtuality $q^{\mu}q_{\mu}=-Q^2$ is a right mover which carries the 4-momentum $q^{\mu}=(q^+,-Q^2/2q^+,\bm{0})$, while the nucleus is a left mover with 4-momentum $P^{\mu}_N = (0,P_N^-,\bm{0})$ per nucleon, i.e.~the nucleon mass is set to zero. 

 At high energy it is convenient to adopt the dipole picture, in which the virtual photon first decays into a $q\bar{q}$ pair. The lifetime $\tau_{d}$ of such a color dipole is of the order of $ 2 q^+/Q^2$ and it is much larger than the longitudinal extent $L= R_A/\gamma$ of the boosted nucleus, where $R_A$ is its radius and $\gamma$ its Lorentz factor. Indeed, with $R_A \simeq A^{1/3} R_N$ and $\gamma \simeq P_N^- R_N$, where $R_N$ is the nucleon radius, the aforementioned condition $\tau_d \gg L$ is equivalent to $x_{\rm \scriptscriptstyle Bj} A^{1/3} \ll 1$. This inequality is satisfied by assumption, since  $x_{\rm \scriptscriptstyle Bj} \equiv Q^2/2 q \!\cdot\! P_N = Q^2/2 q^+P_N^-$ is typically of the order of $10^{-2}$ or smaller.  
 
 In this viewpoint, all the QCD dynamics is contained in the scattering of the color dipole off the nucleus. The latter can be at saturation, thus generating a strong color field and it becomes necessary to take into account multiple scattering. In the kinematics of interest the pair scatters eikonally and gives rise to a dijet system in the forward direction (of the virtual photon). We consider diffractive production in which the $q\bar{q}$ pair remains in its initial color state but acquires a transverse momentum $\bm{\Delta}$ from the nucleus. In the case of a homogeneous nucleus, this can occur only with the exchange of color between different nucleons. In turn, this means that the scattering is incoherent from the perspective of the nucleus and that the overall process is suppressed by $1/N_c^2$ in the multicolor limit. This process was numerically studied in \cite{Mantysaari:2019hkq} in the regime where all momenta are semihard, i.e.~when $k_{1\perp} \sim k_{2\perp} \sim \Delta_{\perp} \sim Q_s$, with $\bk_1$ and $\bk_2$ the final transverse momenta of the quark and the antiquark, respectively, and where $k_{1\perp} \equiv |\bk_1|$, etc. In the current work we will give an analytical calculation in the correlation limit $k_{1\perp} \sim k_{2\perp} \gg Q_s,\Delta_{\perp}$, where we point out that there is no need to make any assumption regarding the relation between the scales $\Delta_{\perp}$ and $Q_s$. Clearly the most interesting case is when $\Delta_{\perp}$ is of the order of (or smaller than) the semihard scale $Q_s$, so that unitarity corrections and gluon saturation are important, despite the fact that two hard jets are being produced. 
   
In the framework of light-cone perturbation theory (LCPT), the general expressions for dijet cross sections have been derived more than a decade ago \cite{Dominguez:2011wm,Dominguez:2011br}. In what follows, we shall follow the notation and conventions adopted in Refs.~\cite{Iancu:2018hwa,Iancu:2020mos,Iancu:2022gpw}, properly adjusted to the case of diffractive production. Let us start by considering the $q\bar{q}$ component in the LCWF of a virtual photon. In order to avoid a proliferation of equations, we shall deal explicitly only with the case of a transverse virtual photon, while for the case of longitudinal polarization we will just present the final result. In momentum space the direct amplitude (DA) reads
\begin{align}
	\label{qqmom}
	\big|\gamma_{\scriptscriptstyle T}^{i}(q)\big\rangle_{q\bar{q}} 
	=\!\sum_{\lambda_{1,2}= \pm 1/2}\,
	\sum_{\alpha,\beta=1}^{N_c}
	\delta_{\alpha\beta}
	\int_{0}^{1} & \rmd
	\vartheta_1 \rmd \vartheta_2 \,
	\delta(1 - \vartheta_1 - \vartheta_2)
	\int \rmd^{2}\bm{k}'_1  
	\rmd^{2}\bm{k}'_{2}\,
	\delta^{(2)}(\bm{k}'_1 +\bm{k}'_2)
	\nn 
	\times\,
	& \psi^i_{\lambda_{1}\lambda_{2}}(\vartheta_1,\bk'_1)\,
	\big|{q}_{\lambda_{1}}^{\alpha}(\vartheta_1, \bk'_1)\,
	\bar{q}_{\lambda_{2}}^{\beta}(\vartheta_2, \bk'_2)
	\big\rangle,
	\end{align}
where $i=1,2$ stands for the polarization index of the transverse photon, while $\alpha$ and $\beta$ are color indices in the fundamental representation. $\lambda_1$, $\bk'_1$ and $\vartheta_1$ are the helicity state, the transverse momentum and the longitudinal momentum fraction (with respect to the virtual photon) of the quark and similarly for the antiquark. Notice that we have used a prime to denote the transverse momenta before the scattering, since they are different from the final ones $\bk_1$ and $\bk_2$. The $q\bar{q}$ amplitude is given by
\begin{align}
	\label{psiqq}
	\psi^i_{\lambda_1 \lambda_{2}} (\vartheta, \bk) =
	\sqrt{\frac{q^+}{2}}\,\frac{ee_{f}}{(2\pi)^3}\,
	\frac{\varphi_{\lambda_{1}\lambda_{2}}^{il}
	(\vartheta)\,k^{l}}{\bm{k}^2 +
    \vartheta(1-\vartheta) Q^{2}},
\end{align}
with $e_f$ the fractional electric charge of the quark flavor under consideration. The function 
\begin{align}
	\label{phidef}
	\varphi_{\lambda_1\lambda_{2}}^{il}(\vartheta)
	=
	\delta_{\lambda_{1}\lambda_{2}}
	\left[(2\vartheta-1)\delta^{il}
	+ 2 i\varepsilon^{il} \lambda_{1}\right]
\end{align}		
appearing in the numerator in Eq.~\eqref{psiqq} represents the helicity structure of the photon splitting vertex, while the combination in the respective denominator arises, according to the LCPT rules, from the energy denominator
\begin{align}
	\label{EDqq}
	E_{q\bar{q}} - E_{\gamma} = 
	\frac{1}{2q^+}
	\left(
	\frac{\bk_1^{\prime\, 2}}{\vartheta_1} + 
	\frac{\bk_2^{\prime\, 2}}{\vartheta_2} + {Q^2} 
	\right)
	=\frac{1}{2 q^+\vartheta_1\vartheta_2}
	\left({\bk_1^{\prime 2}} +{\vartheta_1  \vartheta_2} Q^2
	\right).
\end{align} 

Next, we make a Fourier transform according to
\begin{align}
	\big|{q}_{\lambda_{1}}^{\alpha}(\vartheta_1, \bk'_1)\,
	\bar{q}_{\lambda_{2}}^{\beta}(\vartheta_2, \bk'_2)
	\big\rangle = \int \rmd^2\bx\, \rmd^2\by\,
	e^{-i \bk'_1\cdot \bx - i \bk'_2\cdot \by }\,
	\big|{q}_{\lambda_{1}}^{\alpha}(\vartheta_1, \bx)\,
	\bar{q}_{\lambda_{2}}^{\beta}(\vartheta_2, \by)
\big\rangle,	
\end{align}
where clearly $\bx$ and $\by$ are the corresponding transverse coordinates of the quark and the antiquark. A straightforward integration over the momenta $\bk'_1$ and $\bk'_2$ leads to
\begin{align}
	\label{qqcoord}
	\big|\gamma_{\scriptscriptstyle T}^{i}(q)\big\rangle_{q\bar{q}} 
	=\!\sum_{\lambda_{1,2}= \pm 1/2}\,
	\sum_{\alpha,\beta=1}^{N_c}
	\delta_{\alpha\beta}
	\int_{0}^{1} & \rmd
	\vartheta_1 \rmd \vartheta_2 \,
	\delta(1 - \vartheta_1 - \vartheta_2)
	\int \rmd^{2}\bx \, 
	\rmd^{2}\by
	\nn 
	\times\,
	& \widetilde{\psi}^i_{\lambda_{1}\lambda_{2}}(\vartheta_1,\br)\,
	\big|{q}_{\lambda_{1}}^{\alpha}(\vartheta_1, \bx)\,
	\bar{q}_{\lambda_{2}}^{\beta}(\vartheta_2, \by)
	\big\rangle,
	\end{align} 
with $\br = \bx - \by$ and where the coordinate space $q\bar{q}$ amplitude reads
\begin{align}
	\label{psiqqr}
	\widetilde{\psi}^i_{\lambda_{1}\lambda_{2}}(\vartheta,\br) = 
	-\sqrt{\frac{q^+}{2}}\frac{i e e_f}{(2\pi)^2}\,
	\varphi_{\lambda_1\lambda_{2}}^{il}(\vartheta)\,
	\frac{\bar{Q}r^l}{r} K_1(\bar{Q}r).
\end{align}
In the above, we have defined $\bar{Q}^2 = \vartheta (1 - \vartheta) Q^2$ and the dipole transverse size $r = |\br|$, while $K_1$ is the usual modified Bessel function of the second kind which suppresses exponentially the decay of the virtual photon to dipoles with size much larger than $1/\bar{Q}$.

The advantage of the above representation is that we can easily take into account the multiple scattering of the dipole off the target color field, since the transverse coordinate of a parton interacting with a soft gluon is a good quantum number. In the kinematics we are interested in, the nucleus can be viewed as a Lorentz contracted shockwave and the scattering of each parton in the virtual photon LCWF is given by a Wilson line: $V(\bx)$ for the quark and $V^{\dagger}(\by)$ for the antiquark, where
\begin{align}
	\label{wilson}
	V(\bm{x})={\rm T}
	\exp\left[i g \int \rmd x^{+} 
	t^{a} A^{-}_a(x^{+},\bm{x})\right].
\end{align}
Here T stands for time ordering in the light-cone time $x^+$, $g$ is the QCD coupling, $t^a$ is a SU($N_c$) color matrix in the fundamental representation and $A^{-}_a$, with $a=1, \dots, N_c^2-1$, is the only non-vanishing component of the target color field in the Lorentz gauge $\del_{\mu} A_a^{\mu}=0$. The inclusion of the multiple scattering simply amounts to the replacement
\begin{align}
	\label{scat}	
	\delta_{\alpha \beta}
	\rightarrow
	\big[V(\bx) V^\dagger(\by) - \mathds{1} \big]_{\alpha \beta}	
\end{align}
in Eq.~\eqref{qqcoord}. Still, the above corresponds to the inclusive production of a dijet which is not exactly what we want. In order to have a diffractive process to the order of accuracy, the projectile color dipole must scatter elastically and emerge as a color singlet in the final state. Such a projection is simply achieved by making the further replacement
\begin{align}
	\label{scatdiff}
	\big[V(\bx) V^\dagger(\by) - \mathds{1} \big]_{\alpha \beta}
	\rightarrow	
	\left\{ \frac{1}{N_c}\,
	{\rm tr} \big[V(\bx) V^\dagger(\by)\big] - 1 \right\} 
	\delta_{\alpha \beta}
	\equiv [S(\bx,\by) - 1]\, \delta_{\alpha \beta},
\end{align} 
where $S(\bx,\by)$ is the $S$-matrix for the elastic scattering of the color dipole $(\bx,\by)$. All in all, for evaluating the complete $q\bar{q}$ component $\big|\gamma_{\scriptscriptstyle T}^{i}(q)\big\rangle_{q\bar{q}}^{\rm \scriptscriptstyle D} $ of the virtual photon LCWF which determines the diffractive dijet production one must implement the above modification in Eq.~\eqref{qqcoord}, that is
\begin{align}
	\label{qqdiff}
	\big|\gamma_{\scriptscriptstyle T}^{i}(q)\big\rangle_{q\bar{q}}^{\rm \scriptscriptstyle D} 
	=\!\sum_{\lambda_{1,2}= \pm 1/2}\,
	\sum_{\alpha,\beta=1}^{N_c}
	\delta_{\alpha\beta}
	\int_{0}^{1} & \rmd
	\vartheta_1 \rmd \vartheta_2 \,
	\delta(1 - \vartheta_1 - \vartheta_2)
	\int \rmd^{2}\bx \, 
	\rmd^{2}\by
	\nn 
	\times\,
	& \widetilde{\psi}^i_{\lambda_{1}\lambda_{2}}(\vartheta_1,\br)\,
	[S(\bx,\by) - 1]\,
	\big|{q}_{\lambda_{1}}^{\alpha}(\vartheta_1, \bx)\,
	\bar{q}_{\lambda_{2}}^{\beta}(\vartheta_2, \by)
	\big\rangle.
	\end{align} 

Finally we come to calculate the corresponding cross section and we do so by calculating the number of quarks and antiquarks in the final state, namely 
\begin{align}
	\label{crossdef}
 	\frac{\rmd \sigma_{\rm \scriptscriptstyle D}^{\gamma_{\scriptscriptstyle T}^* A
  \rightarrow q\bar qX}}
  {\rmd k_1^+
  \rmd k_2^+
  \rmd^{2}\bm{k}_{1}
  \rmd^{2}\bm{k}_{2}}
  \,(2\pi)\,\delta(q^{+}\!-k_1^+\!-k_2^+)
	=\frac{1}{2}\!\!
	{\phantom{\big\langle}}_{q\bar{q}}^{\,\,\rm \scriptscriptstyle D}\big\langle \gamma_{\scriptscriptstyle T}^{i}(q)\big|\hat{N}_{q}(k_{1})\,\hat{N}_{\bar{q}}(k_{2})\big|\gamma_{\scriptscriptstyle T}^{i}(q)\big\rangle_{q\bar{q}}^{\rm \scriptscriptstyle D},
\end{align}
with $\hat{N}_{q}(k_{1})$, $\hat{N}_{\bar{q}}(k_{2})$ the respective quark and antiquark number operators and where we have averaged over the polarization of the transverse photon. From this point on, using the appropriate normalization of these number operators (cf.~Appendix C in \cite{Iancu:2018hwa}), straightforward algebra leads to (taking into account all relevant flavors and with $\alpha_{\rm em} = e^2/4\pi$)
\begin{align}
\label{transdiff}
\hspace*{-0.1cm}
	\frac{\rmd\sigma_{\scriptscriptstyle \rm D}^{\gamma_{\scriptscriptstyle T}^* A\rightarrow q\bar q X}}
	{\rmd \vartheta_1 \rmd \vartheta_2\, 
	\rmd^{2}\bm{k}_{1}
  \rmd^{2}\bm{k}_{2}}
	= \,&
	\frac{\alpha_{\rm em} \,N_{c}}{2\pi^2}
	\left(\sum e_{f}^{2}\right)\,
	\delta(1- \vartheta_1 -\vartheta_2)
	\left(\vartheta_1^{2}+ \vartheta_2^{2} \right)
	\int \frac{\rmd^2 \bx}{2\pi}\,
	\frac{\rmd^2 \by}{2\pi}\,
	\frac{\rmd^2 \bar{\bx}}{2\pi}\,
	\frac{\rmd^2 \bar{\by}}{2\pi}\,
	\nn & \hspace*{-0.9cm}
	\times e^{- i \bk_1 \cdot (\bx - \bar{\bx})
	- i \bk_2 \cdot (\by - \bar{\by})}
	\frac{\bm{r} \!\cdot\! \bar{\bm{r}}}{r \bar{r}}\,
	\bar{Q}^2
	K_{1}(\bar{Q}r)
	K_{1}(\bar{Q}{\bar{r}})
	\big\langle [S(\bx,\by) - 1] 
	[S(\bar{\by},\bar{\bx})  - 1]\big\rangle,
\end{align}
where the bar denotes coordinates in the complex conjugate amplitude (CCA).   The average in the last factor is to be taken over all possible target field configurations, with a weight function suitable to the longitudinal scale, or equivalently the rapidity, of interest. We shall work in the framework of the Color Glass Condensate (CGC): at moderate rapidities we can resort to the McLerran-Venugopalan (MV) model \cite{McLerran:1993ka,McLerran:1993ni}, whereas at higher rapidities we should better rely on the solution to the BK equation \cite{Balitsky:1995ub,Kovchegov:1999yj} or the JIMWLK equation \cite{JalilianMarian:1997jx,JalilianMarian:1997gr,Kovner:2000pt,Weigert:2000gi,Iancu:2000hn,Iancu:2001ad,Ferreiro:2001qy}. The precise rapidity is determined by the kinematics and we shall return to this matter at the end of the current section.

At this point we make a change of variables. In the DA and in coordinate space we go from $\bx$ and $\by$ to the separation $\br$ and the center of energy $\bb$ according to $\bx = \bb + \vartheta_2 \br$ and $\by = \bb - \vartheta_1 \br$, while in momentum space we replace $\bk_1$ and $\bk_2$ by the dijet imbalance $\bm{\Delta}$ and the relative momentum $\bP$ according to $\bk_1 = \bP + \vartheta_1 \bm{\Delta}$ and $\bk_2 = -\bP + \vartheta_2 \bm{\Delta}$. It is rather important to point out that, for our process, this imbalance is equal to the momentum transferred from the target, since in the final state there are no other particles between the dijet and the nucleus. With the analogous change of variables in the CCA we immediately find that the phase in the Fourier transform in Eq.~\eqref{transdiff} becomes
\begin{align}
	\label{phase}
	\bk_1 \!\cdot\! (\bx - \bar{\bx})
	+ \bk_2 \!\cdot\!  (\by - \bar{\by}) = 
	\bP \!\cdot\!  (\br - \bar{\br}) + 
	\bm{\Delta} \!\cdot\!   (\bb - \bar{\bb}).
\end{align}
When the nucleus is homogeneous, CGC correlators depend only on the differences between any two transverse coordinates. In particular, the one appearing in Eq.~\eqref{transdiff} can depend only on three vectors, which we choose to be $\br$, $\bar{\br}$ and the difference between the impact parameters in the DA and the CCA, i.e.~$\bm{B} \equiv \bm{b} - \bar{\bb}$. First, in view of Eq.~\eqref{phase}, this means that one of the two center of energy vector integrations can be trivially performed to give the total area $S_{\perp}$ of the nucleus. Second, if we keep only the factorized piece $\langle T(\br) \rangle \langle T^*(\bar{\br}) \rangle$, with $T = 1 -S$ the $T$-matrix, we find that the integration over $\bm{B}$ leads to a $\delta$-function in $\bm{\Delta}$ (more precisely this is smeared to momenta up to a small non-perturbative scale, but to our purposes such a scale can be taken equal to zero). Thus, in order to have a non-zero momentum transfer $\bm{\Delta}$, we must consider the ``connected'' part of the correlator in Eq.~\eqref{transdiff}. The resulting cross section will be suppressed by a color factor $1/N_c^2$ and the corresponding process is incoherent as we discuss in Sect.~\ref{sec:corr}. Putting everything together we arrive at the differential cross section in the transverse sector \cite{Mantysaari:2019hkq}
\begin{align}
\label{sigmab}
	\frac{\rmd\sigma_{\scriptscriptstyle \rm D}^{\gamma_{\scriptscriptstyle T}^* 
	A\rightarrow q\bar q X}}
	{\rmd \vartheta_1 \rmd \vartheta_2\, 
	\rmd^{2} \bm{P} \rmd^{2} \bm{\Delta}}
	= \,&
	\frac{S_{\perp} \alpha_{\rm em} \,N_{c}}{4\pi^3}
	\left(\sum e_{f}^{2}\right)\,
	\delta(1- \vartheta_1 -\vartheta_2)
	\left(\vartheta_1^{2}+ \vartheta_2^{2} \right)
	\int \frac{\rmd^2 \bm{B}}{2\pi}\,
	\frac{\rmd^2\bm{r}}{2\pi}\,
	\frac{\rmd^2\bar{\bm{r}}}{2\pi}
	\nn
	& \times
	e^{- i \bm{\Delta} \cdot \bm{B}
	- i \bm{P} \cdot (\bm{r}-\bar{\bm{r}})}\,
	\frac{\bm{r} \!\cdot\! \bar{\bm{r}}}{r \bar{r}}\,
	\bar{Q}^2\,
	K_{1}\big(\bar{Q}r\big)
	K_{1}\big(\bar{Q}{\bar{r}}\big)
	\mcal{W}_{\rm D}
	(\bm{r},\bar{\bm{r}},\bm{B}),
\end{align}
where the ``diffractive'' correlator $\mcal{W}_{\rm D}$ contains all the QCD dynamics in the CGC framework and is defined as 
\begin{align}
	\label{wdiff}
	\mcal{W}_{\rm D}
	(\bm{r},\bar{\bm{r}},\bm{B}) = 
	\left \langle 
	S(\bm{x},\bm{y}) S(\bar{\bm{y}},\bar{\bm{x}}) 
	\right \rangle
	-
	\left \langle 
	S(\bm{x},\bm{y}) 
	\right \rangle 
	\left \langle
	S(\bar{\bm{y}},\bar{\bm{x}}) 
	\right \rangle.
\end{align}
The corresponding expression for a longitudinally polarized virtual photon can be obtained from Eq.~\eqref{sigmab} via the replacements $ \vartheta_1^2 + \vartheta_2^2 \to 4 \vartheta_1 \vartheta_2$ and $(\br/r) K_1(\bar{Q} r) \to K_0(\bar{Q} r)$ and similarly for the respective factor in the CCA. We would also like to mention that Eq.~\eqref{sigmab} gives the cross section per unit longitudinal momentum fractions of the ``jets''. It is more standard to rewrite it in terms of rapidities, and for this to happen it is enough to multiply with a factor $\vartheta_1 \vartheta_2$. In this work we are interested in the scenario that the two jets share almost equally the longitudinal momentum of the virtual photon, i.e.~the two fractions are considered to be of the order of one-half. Then, such a multiplicative factor does not bring any significant parametric dependence and if needed it can be trivially restored at the end of the calculation.    

The correlators in Eq.~\eqref{wdiff} must be evaluated at the rapidity gap $Y_{\rm gap} = \ln 1/ x_{\rm gap}$ of the diffractive process. Here $x_{\rm gap}$ is defined as the fraction of the target nucleon longitudinal momentum $P_N^-$ which is transferred to the diffractive $q\bar{q}$ system. This is determined by the conservation law
\begin{align}
	\label{xgap}
	x_{\rm gap} P_N^- =
	\frac{1}{2 q^+}
	\left(
	\frac{k_{1\perp}^2}{\vartheta_1}
	+\frac{k_{2\perp}^2}{\vartheta_2}
	+Q^2
	\right)	
	\,\Longrightarrow\,
	 x_{\rm gap} = x_{\rm \scriptscriptstyle Bj}
	 \left(
	 1 + \frac{P_{\perp}^2}{\bar{Q}^2}
	 + \frac{\Delta_{\perp}^2}{Q^2}
	 \right).
\end{align}
We will be interested in the kinematic regime $P_{\perp} \sim \bar{Q} \gg \Delta_{\perp}$. As a consequence the term $\Delta_{\perp}^2/Q^2$ can be safely dropped, while the term $P_{\perp}^2/\bar{Q}^2$ is of the order of one, so that
\begin{align}
	\label{Ygap}
	Y_{\rm gap} \simeq\, \ln \frac{1}{x_{\rm \scriptscriptstyle Bj}}
	 - \ln
	 \left(
	 1 + \frac{P_{\perp}^2}{\bar{Q}^2}
	 \right).
\end{align}
This means that the production of the dijet ``consumes'' roughly one unit of the available longitudinal phase space, but since $x_{\rm \scriptscriptstyle Bj} \ll 1$ such a reduction of the gap is subdominant at the leading logarithmic level. Hence, although it is not hard to keep the second term in Eq.~\eqref{Ygap} when calculating the correlators, we shall neglect such a term and therefore assume that the gap is  $P_{\perp}$-independent. As typically done, one can start with the MV model initial condition at $Y_0 = \ln 1/x_0$ with $x_0 \simeq 10^{-2}$ and then evolve with the BK or the JIMWLK equation for a rapidity interval $\Delta Y = Y_{\rm gap} - Y_0 \,\simeq\, \ln x_0/x_{\rm \scriptscriptstyle Bj}$.

Eq.~\eqref{Ygap} also explains why we prefer to study only the symmetric case in which $\vartheta_1$ and $\vartheta_2$ are roughly equal to each other. Still assuming that $Q$ is a hard scale, let us see what happens in the unbalanced situation where $\vartheta_1$ (or $\vartheta_2$) gets small enough so that $\bar{Q}$ becomes much smaller than the hard momentum $P_{\perp}$. Then, although our calculation and results which follow in the next sections remain generally valid (since the correlation limit is not sensitive to the value of $\bar{Q}$), one of the two jets becomes very soft in its longitudinal momentum. As a consequence, it ``eats up'' a significant part of the rapidity space between the other jet and the target and therefore, as evident in Eq.~\eqref{Ygap}, the diffractive gap gets smaller leading to a less interesting situation. 

\section{Averaging in the CGC}
\label{sec:corr}

\begin{figure}
	\begin{center}
		\includegraphics[width=0.45\textwidth]{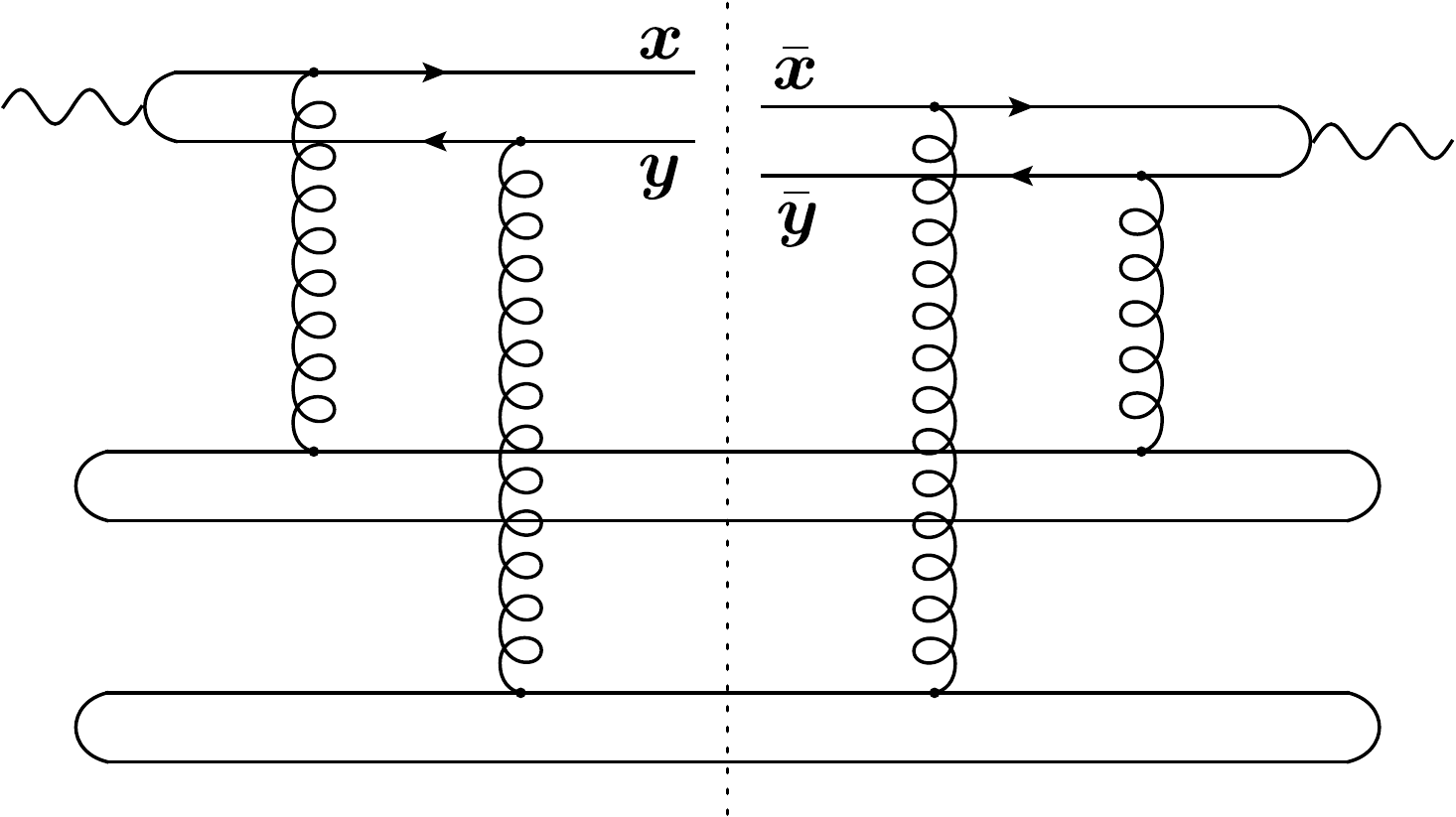}
		\hspace*{0.08\textwidth}
		\includegraphics[width=0.45\textwidth]{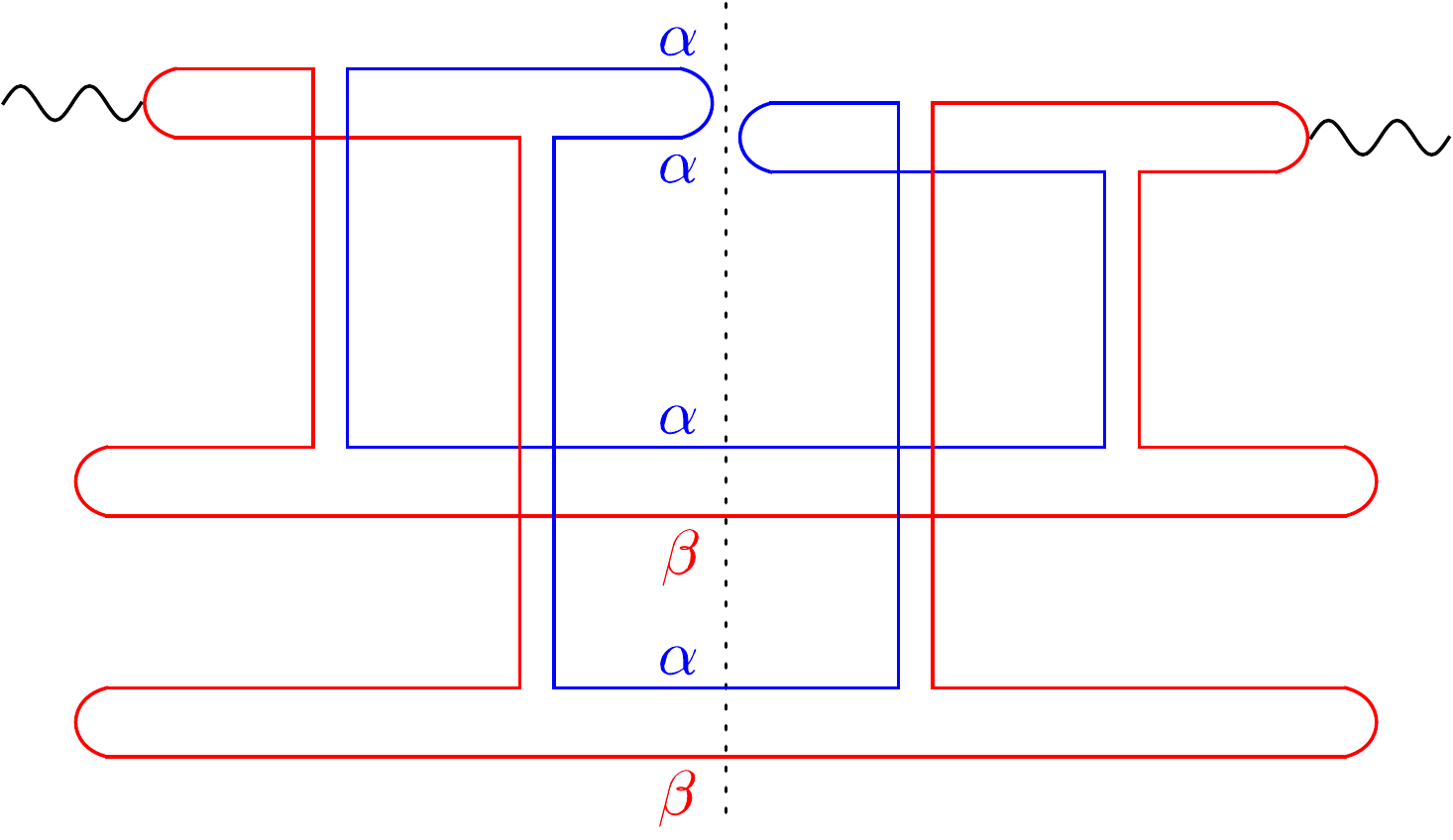}
	\end{center}
	\caption{\small Left panel: A diagram for incoherent diffractive dijet production at the level of four-gluon exchange. It corresponds to Eq.~\eqref{contract} which is part of the product $\langle T(\bm{x},\bar{\bm{y}})\rangle \langle T(\bm{y},\bar{\bm{x}})\rangle$. Each of the two lower double lines, which correspond to colorless substructures (nucleons) in the nucleus, exchanges only one gluon with the projectile $q\bar{q}$ pair in the DA (and one in the CCA), and thus it undergoes inelastic scattering. Right panel: The same in the large-$N_c$ limit, in which the gluons are represented by double lines. The substructures are not colorless in the final state, thus indicating that the nucleus does not remain intact. Two independent colors (in the fundamental representation), $\alpha$ and $\beta$, flow in the diagram which is of the order of $g^8 N_c^2$.}
\label{fig:inc}
\end{figure}

The incoherent diffractive cross section involves the correlator $\mcal{W}_{\rm D}(\bm{r},\bar{\bm{r}},\bm{B})$ defined in Eq.~\eqref{wdiff}, which we will calculate analytically in the correlation limit $r,\bar{r} \ll B,1/Q_s$, i.e.~when $P_{\perp} \gg \Delta_{\perp},Q_s$. For a warmup, and in order to get an insight to the dynamics, we shall first work at lowest order in the number of gluon exchanges. Within each projectile dipole, one in the DA and one in the CCA, there are two gluon hookings and therefore we can make use of the expansion
\begin{align}
	\label{sdip2g}
	S(\bm{x},\bm{y}) \simeq 1 - \frac{g^2}{4 N_c}
	\left(\alpha^a_{\bm{x}} - \alpha^a_{\bm{y}} \right)^2
	\quad 
	\mathrm{with}
	\quad
	\alpha^a_{\bm{x}} \equiv 
	\int_{-\infty}^{\infty} \!\dif x^+
	A^{-}_a(x^{+},\bm{x})
\end{align}
and similarly for $S(\bar{\bm{y}},\bar{\bm{x}})$. Obviously this expansion is valid only in the absence of unitarity corrections, that is when $r,\bar{r}, B \ll 1/Q_s$ or, equivalently, when $P_{\perp}, \Delta_{\perp} \gg Q_s$. We will assume that the CGC wavefunction is Gaussian:  this is valid by definition in the MV model, while it is very accurate even after JIMWLK evolution is included \cite{Dumitru:2011vk,Iancu:2011ns,Iancu:2011nj,Alvioli:2012ba}. When taking the expectation values in Eq.~\eqref{wdiff}, only the connected part survives, that is, only terms which involve the contraction of one field in the DA with a field in the CCA are relevant. For example, one such term is (cf.~Fig.~\ref{fig:inc})
\begin{align}
	\label{contract}
	\hspace*{-0.1cm}
	\frac{g^4}{4 N_c^2}
	\sum_{a,b}
	\big\langle \alpha^a_{\bm{x}}\, 
	\alpha^b_{\bar{\bm{y}}} \big\rangle 
	\big\langle \alpha^a_{\bm{y}}\, 
	\alpha^b_{\bar{\bm{x}}} \big\rangle =
	\frac{g^4}{4 N_c^2}\sum_{a}
	\big\langle \alpha^a_{\bm{x}}\, 
	\alpha^a_{\bar{\bm{y}}} \big\rangle 
	\big\langle \alpha^a_{\bm{y}}\, 
	\alpha^a_{\bar{\bm{x}}} \big\rangle =
	\sum_{a,b} 
	\frac{g^4}{4 N_c^2 (N_c^2-1)} 
	\big\langle \alpha^a_{\bm{x}}\, 
	\alpha^a_{\bar{\bm{y}}} \big\rangle 
	\big\langle \alpha^b_{\bm{y}}\, 
	\alpha^b_{\bar{\bm{x}}} \big\rangle,
\end{align}
where we have explicitly shown the summations over color indices (which refer to gluons) to avoid any confusion. For the first equality, we have just used the fact that the correlator of two fields is diagonal in color. In turn this means that only one gluon color flows in the l.h.s. as implied by the single summation over $a$. On the contrary, in the last expression two independent gluon colors are summed and therefore we have divided by the number of colors for the equation to be valid (assuming that the Gaussian approximation gives equal weight to all the color components of the gauge field). Thus, the connected piece which we are calculating is suppressed at large-$N_c$ as anticipated. Putting together all the possible terms, one can write $\mcal{W}_{\rm D}$ as
\begin{align}
	\label{wdiff4g1}
	\mcal{W}_{\rm D}
	(\bm{r},\bar{\bm{r}},\bm{B}) 
	\simeq
	\frac{g^4}{8 N_c^2 (N_c^2-1)} 
	\left[ 
	\big\langle 
	\alpha^a_{\bm{x}}\, 
	\alpha^a_{\bar{\bm{x}}} 
	\big\rangle
	+
	\big\langle 
	\alpha^a_{\bm{y}}\, 
	\alpha^a_{\bar{\bm{y}}} 
	\big\rangle
	-
	\big\langle 
	\alpha^a_{\bm{x}}\, 
	\alpha^a_{\bar{\bm{y}}} 
	\big\rangle
	-
	\big\langle 
	\alpha^a_{\bm{y}}\, 
	\alpha^a_{\bar{\bm{x}}} 
	\big\rangle 
	\right]^2.
\end{align}
We are free to add and subtract ``equal point correlators'' like 
$\big\langle \alpha^a_{\bm{x}}\, \alpha^a_{\bm{x}} \big\rangle$, so that eventually we can reconstruct average values of the scattering amplitudes (at the level of two-gluon exchange), i.e.
\begin{align}
	\label{wdiff4g2}
	\mcal{W}_{\rm D}
	(\bm{r},\bar{\bm{r}},\bm{B}) 
	\simeq
	\frac{1}{2 (N_c^2-1)} 
	\left[  
	\langle T(\bm{x},\bar{\bm{y}})\rangle 
	+
	\langle T(\bm{y},\bar{\bm{x}}) \rangle
	-
	\langle T(\bm{x},\bar{\bm{x}})\rangle
	-
	\langle T(\bm{y},\bar{\bm{y}})\rangle
	\right]^2.
\end{align}

\begin{figure}
	\begin{center}
		\includegraphics[width=0.48\textwidth]{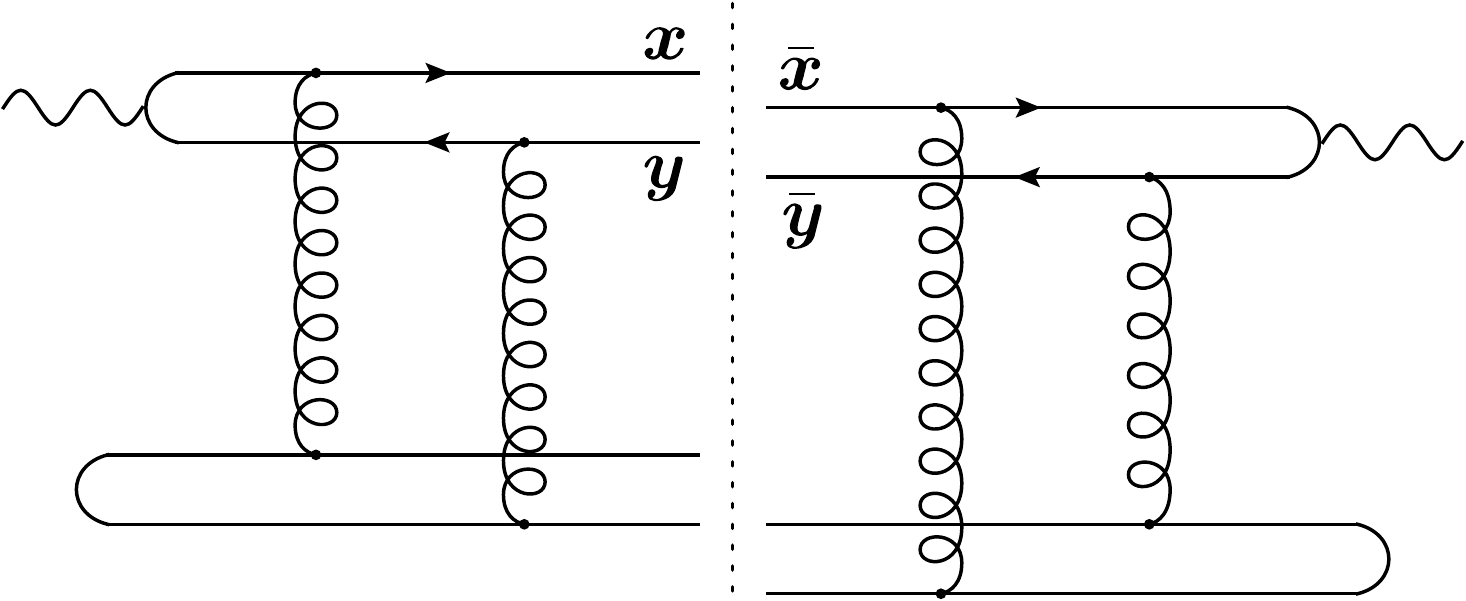}
		\hspace*{0.08\textwidth}
		\includegraphics[width=0.32\textwidth]{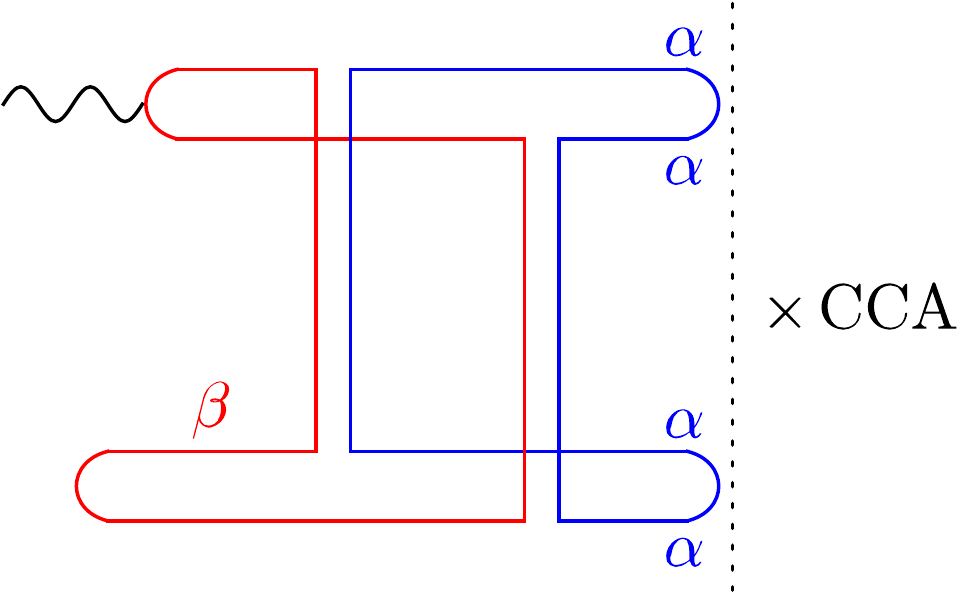}
	\end{center}
	\caption{\small Left panel: A diagram for coherent diffractive dijet production, in general proportional to $\langle T(\bm{x},\bm{y})\rangle \langle T(\bar{\bm{y}},\bar{\bm{x}})\rangle$, at the level of four-gluon exchange. A colorless substructure (nucleon) of the nucleus exchanges two gluons with the projectile $q\bar{q}$ pair in the DA (and the same, for a potentially different nucleon, in the CCA) and thus it undergoes elastic scattering. Right panel: The same in the large-$N_c$ limit, in which the gluons are represented by double lines. The nucleon remains intact. Two independent colors  flow in the DA and another two in the CCA, thus leading to a total diagram which is of the order of $g^8 N_c^4$.}
\label{fig:coh}
\end{figure}

In Fig.~\ref{fig:inc} we give a typical diagram corresponding to incoherent diffraction at the level of four-gluon exchange and we illustrate the color flow in the large-$N_c$ limit. Comparing to the respective coherent diffraction diagrams in Fig.~\ref{fig:coh}, we observe that the color exchange (necessary for incoherent scattering) between the colorless substructures (nucleons) of the target nucleus leads to a $1/N_c^2$ suppression. 

The separations $\bm{r}$ and $\bar{\bm{r}}$ are small, since they are constrained by their conjugate momentum $\bm{P}$ which is hard. Then in Eq.~\eqref{wdiff4g2} we can expand all amplitudes around $\bm{B}$, given that they all involve one leg in the DA and the other leg in the CCA. It is a straightforward exercise to show that the lowest order term in the linear combination in the square bracket is of quadratic order, which is due to the fact that we are still calculating an elastic process for the projectile dipole. More precisely we find 
\begin{align}
 \label{tttt}
 \mcal{T}(\bm{x},\bar{\bm{y}}) 
	+
	\mcal{T}(\bm{y},\bar{\bm{x}}) 
	-
	\mcal{T}(\bm{x},\bar{\bm{x}})
	-
	\mcal{T}(\bm{y},\bar{\bm{y}})
	\simeq r^i \bar{r}^j \partial^i \partial^j \mcal{T}(\bm{B}),	\end{align}
	where, for the purposes of a more compact notation, a calligraphic symbol is used for the average value of a quantity, for example here $\mcal{T} \equiv \langle T \rangle$. Thus, substitution of Eq.~\eqref{tttt} into Eq.~\eqref{wdiff4g2} leads to \begin{align}
	\label{wdiff4g3}
	\mcal{W}_{\rm D}
	(\bm{r},\bar{\bm{r}},\bm{B})
	\simeq
	\frac{1}{2 (N_c^2-1)}\, 
	r^i \bar{r}^j 
	r^k \bar{r}^l
	\partial^i \partial^j \mcal{T}(\bm{B})
	\partial^k \partial^l \mcal{T}(\bm{B}),
\end{align}
which is valid at the level of four-gluon exchange and in the Gaussian approximation.

\begin{figure}
	\begin{center}
		\includegraphics[width=0.6\textwidth]{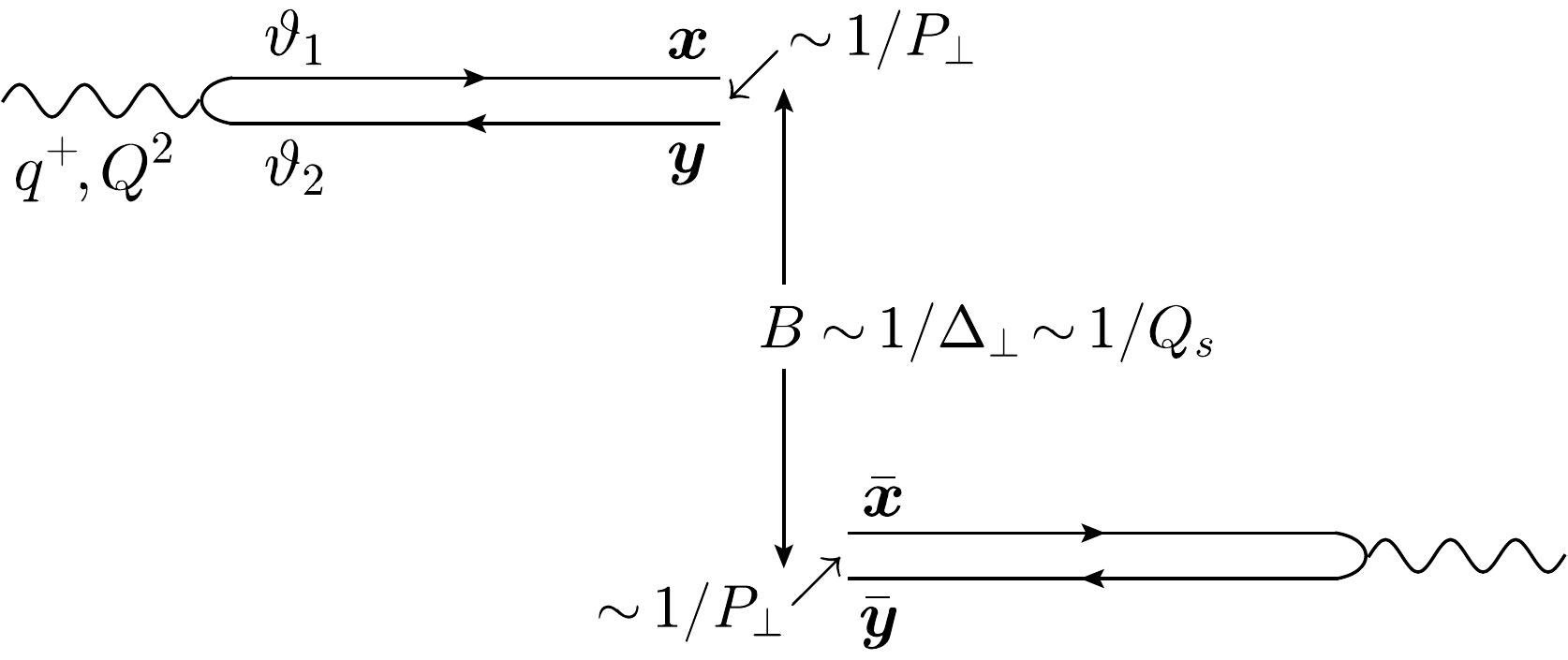}
	\end{center}
	\caption{\small The typical dipole configuration that determines the incoherent diffractive dijet cross section in Eq.~\eqref{sigmab} in the most interesting kinematic regime  $P_{\perp} \gg \Delta_{\perp} \sim Q_s$ (a subspace of the regime defined by the correlation limit $P_{\perp} \gg \Delta_{\perp}, Q_s$). The $q\bar{q}$ dipole in the DA is separated from the one in the CCA by a large distance $B \sim 1/\Delta_{\perp} \sim 1/Q_s$ when compared to the dipole sizes themselves that satisfy $r,\bar{r} \sim 1/P_{\perp} \ll 1/Q_s$.}
\label{fig:sizes}
\end{figure}

Now we would like to generalize the above to the case of larger values of $B$, since it is the coordinate space variable conjugate to the momentum $\Delta_{\perp}$ which we would like to be of the order of (or smaller than) $Q_s$. A typical configuration, as illustrated in  Fig.~\ref{fig:sizes}, is when $\bx$ and $\by$ in the DA remain close together, the same for $\bar{\bx}$ and $\bar{\by}$ in the CCA, but the pair in the DA is far away from the pair in the CCA. Such an extension (always in the Gaussian approximation) for $\mcal{W}_{\rm D}
	(\bm{r},\bar{\bm{r}},\bm{B})$ which includes
	unitarity corrections for the scattering amplitude $\mcal{T}(\bm{B})$ is
	not that complicated and is given in Appendix \ref{app:dipdip}. We find
\begin{align}
	\label{wdiffsat}
	\mcal{W}_{\rm D} (\bm{r},\bar{\bm{r}},\bm{B})
	\simeq
	\frac{C_F}{2 N_c^3}\,
	r^i \bar{r}^j 
	r^k \bar{r}^l\,
	\Phi(B) \,
	[\partial^i \partial^j \ln \mcal{S}_g(B)]\,
	[\partial^k \partial^l \ln \mcal{S}_g(B)],
\end{align}
with $C_F=(N_c^2-1)/2N_c$ and where, in order to have more compact expressions in the following sections, we have conveniently defined the dimensionless scalar quantity
\begin{align}
	\label{F0}
	\Phi(B) = \frac{\mcal{S}_g(B) - 1 - 
	\ln \mcal{S}_g(B) }
	{\ln^2\mcal{S}_g(B)}.
\end{align}
In the above equations 
\begin{align}
	\label{sgg}
	\mcal{S}_g(B) = [\mcal{S}(B)]^{N_c/C_F}, 
\end{align}
valid in the Gaussian approximation, is the average dipole $S$-matrix for the scattering of a gluon-gluon (or adjoint) dipole, which is further assumed to depend only on the magnitude of $\bm{B}$.  Still, one notices that there is angular correlation between $\br$ and $\bm{B}$, and thus there will also be one between $\bP$ and $\bm{\Delta}$.  Clearly, Eq.~\eqref{wdiffsat} reduces to Eq.~\eqref{wdiff4g3} as it should when $B \ll 1/Q_s$ and the two equations share a similar form. Notice that Eq.~\eqref{wdiffsat} exhibits a ``tensorial'' factorization: the power-law dependence on the small fundamental dipoles $\br$ and $\bar{\br}$  has been factored out from the (more complicated) dependence on the large adjoint dipole $\bm{B}$. We shall shortly see that such a property remains valid for the respective conjugate momenta $\bP$ and $\bm{\Delta}$. Finally, we would like to point out that, although the appearance of the adjoint dipole should not come as a surprise (cf.~Fig.~\ref{fig:sizes}), the particular dependence on $\mcal{S}_g(B)$ is non-trivial.

\section{The cross section for incoherent diffractive dijet production}
\label{sec:cross-section}

Due to the form of the correlator $\mcal{W}_{\rm D}$ in Eq.~\eqref{wdiffsat}, the integrals over $\bm{r}$ and $\bar{\bm{r}}$ in Eq.~\eqref{sigmab} factorize from the one over $\bm{B}$. Thus, we are allowed to write 
\begin{align}
\label{transdiff1}
	\frac{\rmd\sigma_{\scriptscriptstyle \rm D}^{\gamma_{\scriptscriptstyle T}^* 
	A\rightarrow q\bar q X}}
	{\rmd \vartheta_1 \rmd \vartheta_2\, 
	\rmd^{2} \bm{P} \rmd^{2} \bm{\Delta}}
	= \,&
	\frac{S_{\perp} \alpha_{\rm em} \,N_{c}}{4\pi^3}
	\left(\sum e_{f}^{2}\right)\,
	\delta(1- \vartheta_1 -\vartheta_2)
	\left(\vartheta_1^{2}+ \vartheta_2^{2} \right)
	\frac{C_F}{2 N_c^3}
	\left| \mcal{A}_{\rm D}^T \right|^2,
\end{align}
where, in order to get rid of simple multiplicative factors, we have defined the ``reduced'' cross section $\left| \mcal{A}_{\rm D}^T \right|^2$ as
\begin{align}
	\label{ad2}
	\left| \mcal{A}_{\rm D}^T \right|^2 = 
	H_T^{iks}(\bP,\bar{Q})\, 
	H_T^{jls*}(\bP,\bar{Q})\, 
	\mcal{G}_{\rm D}^{ij,kl}(\bm{\Delta}). 
\end{align}
This is one of the central results of the current work: at leading twist\footnote{It should be clear that the twist refers to inverse powers of the hard momentum $P_{\perp}$. Inverse powers of the imbalance $\Delta_{\perp}$ are resummed to all orders.} the incoherent diffractive dijet cross section can be decomposed into the calculation of a factor which depends only on the hard momentum $\bP$ (and $\bar{Q}$) and a factor which depends only on the momentum transfer $\bm{\Delta}$, similarly to what happens in the inclusive dijet cross section \cite{Dominguez:2011wm,Dominguez:2011br}. Notice that such a factorization does not hold at the level of the amplitude.\footnote{On the contrary, factorization at the amplitude level occurs at the case of coherent diffractive dijet production \cite{Iancu:2021rup,Iancu:2022lcw,Hatta:2022lzj}, where no momentum is transferred from the target to the projectile, but the imbalance in the dijet transverse momentum is generated by the emission (from the quark or the antiquark) of a gluon.}
 
The hard factor $H_T^{iks}H_T^{jls*}$ in Eq.~\eqref{ad2} involves
\begin{align}
	\label{hiks}
	H_T^{iks}(\bP,\bar{Q}) = \int 
	\frac{\rmd^2 \bm{r}}{2\pi}\,
	e^{- i \bm{P} \cdot \bm{r}}\,
	\frac{r^i r^k r^s}{r}\,
	\bar{Q}
	K_{1}\big(\bar{Q}{r}\big),
\end{align}
which is symmetric in all indices and therefore the result can be expressed in terms of $P^i P^k P^s$, $\delta^{ik}P^s$ and the permutations of the latter with equal weight. It can be calculated in terms of two independent scalar integrals and eventually we obtain
\begin{align}
	\label{hiks1}
	H_T^{iks}(\bP,\bar{Q}) = -\frac{2 i}
	{(P_{\perp}^2 + \bar{Q}^2)^2}
	\big(\delta^{ik} P^s + 
	\delta^{is} P^k + 
	\delta^{ks} P^i  \big)
	+
	\frac{8 i P^i P^k P^s}{(P_{\perp}^2 + \bar{Q}^2)^3}.
\end{align}

The semihard factor $\mcal{G}_{\rm D}^{ij,kl}(\bm{\Delta})$ in Eq.~\eqref{ad2} which incorporates the QCD dynamics via the non-linear scattering of the adjoint dipole off the nuclear target reads
\begin{align}
	\label{Gd}
	\mcal{G}_{\rm D}^{ij,kl}(\bm{\Delta}) = 
	\int \frac{\rmd^2 \bm{B}}{2\pi}\,
	e^{- i \bm{\Delta} \cdot \bm{B}}\,
	\Phi(B)\,
	[\partial^i \partial^j \ln \mcal{S}_g(B)]\,
	[\partial^k \partial^l \ln \mcal{S}_g(B)],
\end{align}
where we recall that $\Phi(B)$ has been defined in Eq.~\eqref{F0}. Given that $\mcal{S}_g$ depends only on the magnitude of $\bm{B}$, one can write the decomposition into a diagonal and a traceless part
\begin{align}
	\label{ddlogS}
	\partial^i \partial^j \ln \mcal{S}_g(B) = 
	\frac{\delta^{ij}}{2}\,
	F_+(B)
	+ \left(\frac{B^i B^j}{B^2} - 
	\frac{\delta^{ij}}{2} \right)
	 F_-(B), 
\end{align} 
where two more scalar function $F_+(B)$ and $F_-(B)$ have been introduced according to
\begin{align}
	\label{Fpm}
	F_{\pm}(B) = \frac{\del^2 \ln \mcal{S}_g(B) }{\del B^2} 
	\pm \frac{1}{B}\,\frac{\del \ln \mcal{S}_g(B) }{\del B}
\end{align}
and we also notice that $F_+(B) = \nabla^2 \ln \mcal{S}_g(B)$. It is rather useful to remember that these two functions have dimension of mass squared. The structure in Eq.~\eqref{ddlogS} appears also in the Weizs\"{a}cker-Williams gluon Transverse Momentum Distribution (TMD), cf.~Eq.~\eqref{tensortmd} in Appendix \ref{app:inclusive}. The difference here is that it is already present at the amplitude level, hence it has to be squared when evaluating the diffractive cross section as evident in Eq.~\eqref{Gd}. It is clear that the 4-rank tensor $\mcal{G}_{\rm D}^{ij,kl}(\bm{\Delta})$ can be decomposed into a few terms, each of which is proportional to $\delta^{ij} \delta^{kl}$ or $\delta^{ij} \Delta^k \Delta^l$ or $\Delta^i \Delta^j\Delta^k \Delta^l$ or the permutations of the first two structures. The coefficients in this decomposition, whose precise form is given in Appendix \ref{app:contract}, are given by five different scalar quantities which depend on the momentum transfer $\Delta_{\perp}$. Eventually the process under consideration probes only four of them [cf.~Eqs.~\eqref{ad2final} and \eqref{ad2lfinal} below], which are expressed in terms of the Fourier transforms 
\begin{align}
	\label{Gdplus}
	\mcal{G}_{\rm D}^{(+)}(\Delta_{\perp}) = 
	\int \frac{\rmd^2 \bm{B}}{2\pi}\,
	e^{- i \bm{\Delta} \cdot \bm{B}}\,
	\Phi(B)
	[F_+(B)]^2,
\end{align}
\begin{align}
	\label{Gdminus}
	\mcal{G}_{\rm D}^{(-)}(\Delta_{\perp}) = 
	\int \frac{\rmd^2 \bm{B}}{2\pi}\,
	e^{- i \bm{\Delta} \cdot \bm{B}}\,
	\Phi(B)
	[F_-(B)]^2,
\end{align}
\begin{align}
	\label{Gd1}
	\mcal{G}_{\rm D}^{(1)}(\Delta_{\perp}) = 
	\int \frac{\rmd^2 \bm{B}}{2\pi}\,
	e^{- i \bm{\Delta} \cdot \bm{B}}\,
	\Phi(B)
	F_+(B)
	F_-(B)
	\cos2\phi_{\scriptscriptstyle \Delta B}
\end{align}
and
\begin{align}
	\label{Gd2}
	\mcal{G}_{\rm D}^{(2)}(\Delta_{\perp}) = 
	\int \frac{\rmd^2 \bm{B}}{2\pi}\,
	e^{- i \bm{\Delta} \cdot \bm{B}}\,
	\Phi(B)
	[F_-(B)]^2
	\cos4\phi_{\scriptscriptstyle \Delta B},
\end{align}
with $\phi_{\scriptscriptstyle \Delta B}$ the angle between the vectors $\bm{\Delta}$ and $\bm{B}$. Following our discussion below Eq.~\eqref{Fpm}, we see that in coordinate space ($\bm{B}$-space) these functions are bilinears in the corresponding unpolarized and linearly polarized gluon distributions \cite{Dominguez:2011wm,Metz:2011wb,Dominguez:2011br} (which determine the inclusive dijet production), given in Eqs.~\eqref{alphaxG} and \eqref{alphaxh} for our convenience. Although we shall not attempt to provide for a precise identification in terms of number operators in the nucleus, we shall (perhaps abusively) refer to them as distributions for the ease of presentation. Notice that it is straightforward to further integrate over the angle between $\bm{B}$ and $\bm{\Delta}$ by making use of
\begin{align}
	\label{angular}
	\int \frac{\rmd^2 \bm{B}}{2\pi}\,
	e^{- i \bm{\Delta} \cdot \bm{B}}\,
	\mcal{F}(B)
	\cos 2n\phi_{\scriptscriptstyle \Delta B}
	= (-1)^n
	\int_0^{\infty} {\rmd B}\, B J_n(\Delta_{\perp} B) \mcal{F}(B), 
\end{align}
with $\mcal{F}(B)$ an arbitrary function of $B$, and thus express all the distributions in terms of one-dimensional integrals involving the first three even Bessel functions of the first kind, i.e~$J_n(\Delta_{\perp} B)$ with $n=0,2,4$.

Having both the hard and semihard tensors at our disposal, we can perform the contractions of the indices in Eq.~\eqref{ad2} (as detailed in Appendix \ref{app:contract}), to arrive at the transverse reduced cross section
\begin{align}
	\label{ad2final}
	\left| \mcal{A}_{\rm D}^T \right|^2 =\,& 
	\frac{4 P_{\perp}^2 (3 \bar{Q}^4 + P_{\perp}^4)}
	{(P_{\perp}^2 + \bar{Q}^2)^6}\, 
	\mcal{G}_{\rm D}^{(+)}(\Delta_{\perp}) 
	+
	\frac{8 \bar{Q}^4 P_{\perp}^2}
	{(P_{\perp}^2 + \bar{Q}^2)^6}\,
	\mcal{G}_{\rm D}^{(-)}(\Delta_{\perp})
	\nn & +
	\frac{ 16 \bar{Q}^2 P_{\perp}^2 
	(\bar{Q}^2 - P_{\perp}^2) \cos 2\phi}
	{(P_{\perp}^2 + \bar{Q}^2)^6}\,
	\mcal{G}_{\rm D}^{(1)}(\Delta_{\perp})
	-\frac{ 8 \bar{Q}^2 P_{\perp}^4 \cos 4\phi}
	{(P_{\perp}^2 + \bar{Q}^2)^6}\,
	\mcal{G}_{\rm D}^{(2)}(\Delta_{\perp}),
\end{align}
where now $\phi$ is the angle between the hard jet momentum $\bm{P}$ and the momentum transfer $\bm{\Delta}$.

Concerning the longitudinal cross section, as we already mentioned one merely makes the replacements $\vartheta_1^2 + \vartheta_2^2 \to 4 \vartheta_1 \vartheta_2$ and $(\br/r) K_1(\bar{Q} r) \to K_0(\bar{Q} r)$ and similarly for the respective factor in the CCA in Eq.~\eqref{sigmab}. This leads just to a modification of the hard tensorial factor, while the semihard one remains the same as it should, more precisely we have
\begin{align}
	\label{ad2l}
	\left| \mcal{A}_{\rm D}^L \right|^2 = 
	H_L^{ik}(\bP,\bar{Q})\, 
	H_L^{jl}(\bP,\bar{Q})\, 
	\mcal{G}_{\rm D}^{ij,kl}(\bm{\Delta}),
\end{align}
with
\begin{align}
	\label{hik}
	H_L^{ik}(\bP,\bar{Q}) & = 
	\int 
	\frac{\rmd^2 \bm{r}}{2\pi}\,
	e^{- i \bm{P} \cdot \bm{r}}\,
	r^i r^k\,\bar{Q}
	K_{0}\big(\bar{Q}{r}\big)
	\nn
	& =
	\frac{4 \bar{Q} (\bar{Q}^2 - P_{\perp}^2)}
	{(P_{\perp}^2 + \bar{Q}^2)^3}\frac{\delta^{ik}}{2} -
	\frac{8 \bar{Q} P_{\perp}^2}{(P_{\perp}^2 + \bar{Q}^2)^3}
	\left(\frac{P^i P^k}{P_{\perp}^2} - 
	\frac{\delta^{ik}}{2} \right).
\end{align}
Performing the contractions (again cf.~Appendix \ref{app:contract}) we get
\begin{align}
	\label{ad2lfinal}
	\left| \mcal{A}_{\rm D}^L \right|^2 =\,& 
	\frac{2 \bar{Q}^2 (\bar{Q}^4  - 2 \bar{Q}^2 P_{\perp}^2 + 5P_{\perp}^4 )}
	{(P_{\perp}^2 + \bar{Q}^2)^6}\, 
	\mcal{G}_{\rm D}^{(+)}(\Delta_{\perp}) 
	+
	\frac{2 \bar{Q}^2 (\bar{Q}^2 - P_{\perp}^2)^2}
	{(P_{\perp}^2 + \bar{Q}^2)^6}\,
	\mcal{G}_{\rm D}^{(-)}(\Delta_{\perp})
	\nn & 
	-
	\frac{ 16 \bar{Q}^2 P_{\perp}^2 
	(\bar{Q}^2 - P_{\perp}^2) \cos 2\phi}
	{(P_{\perp}^2 + \bar{Q}^2)^6}\,
	\mcal{G}_{\rm D}^{(1)}(\Delta_{\perp})
	+\frac{ 8 \bar{Q}^2 P_{\perp}^4 \cos 4\phi}
	{(P_{\perp}^2 + \bar{Q}^2)^6}\,
	\mcal{G}_{\rm D}^{(2)}(\Delta_{\perp}).
\end{align}
 
Thus, we have been able to express both the transverse and longitudinal components of the incoherent diffractive dijet production as a linear combination of four distributions $\mcal{G}_{\rm D}^{(n)}(\Delta_{\perp})$. All the corresponding coefficients are determined analytically and depend on the two hard momenta $P_{\perp}$ and $\bar{Q}$, but also on the angle between $\bP$ and $\bm{\Delta}$. If we assume that $P_{\perp} \sim {\bar{Q}}$, which is indeed the typical value of the magnitude of the jet momenta as dictated by the photon decay vertex, we immediately see that the all hard coefficients fall like $1/P_{\perp}^6$. This was to be expected since $H_T^{iks}(\bP,\bar{Q})$ in Eq.~\eqref{hiks1} and $H_L^{ik}(\bP,\bar{Q})$ in Eq.~\eqref{hik} fall like $1/P_{\perp}^3$. 

Regarding the angular dependence, there is a one-to-one correspondence between the angle $\phi_{\scriptscriptstyle \Delta B}$ and the angle $\phi$: for example,  $\mcal{G}_{\rm D}^{(1)}(\Delta_{\perp})$ which involves $\cos2\phi_{\scriptscriptstyle \Delta B}$ leads to the $\cos2\phi$ dependence in the cross section. Finally we notice that the coefficients of $\cos2\phi$ in the transverse and longitudinal sectors differ only in their sign. We recall that this is also a property of the inclusive dijet cross section \cite{Dominguez:2011br}. The same feature holds for the $\cos4\phi$ coefficients.

\section{Momentum dependence of the diffractive distributions}
\label{sec:GDs}

In this section we will try to get an insight to the four diffractive distributions $\mcal{G}_{\rm D}^{(n)}(\Delta_{\perp})$. First, we shall do simple analytic calculations and considerations in the regimes where the momentum transfer $\Delta_{\perp}$ is much larger than, of the order of, or much smaller than the saturation momentum $Q_s$. Then we will proceed to a direct numerical evaluation for any value of $\Delta_{\perp}$.

For definiteness we shall first assume that the dipole-nucleus scattering is described by the MV model \cite{McLerran:1993ka,McLerran:1993ni}, i.e.
\begin{align}
        \label{SgMVd}
        \mcal{S}_g(B) =  1 - \mcal{T}_g(B) =
        \exp \left(  
        -\frac{B^2 Q_{A}^2}{4} 
        \ln \frac{4}{B^2\Lambda^2} \right),
\end{align}
where the scale $Q_A^2$ is expressed in terms of the saturation momentum via the relation
\begin{align}
	\label{QAQs}
	Q_A^2 \ln \frac{Q_s^2}{\Lambda^2} = Q_s^2.
\end{align}
Notice that the above refers to the gluon saturation momentum which is enhanced by a factor of $N_c/C_F$ w.r.t.~the quark one. Then for the dimensionful  $B$-dependent factors in Eq.~\eqref{Fpm} one finds 
\begin{align}
	\label{FpmMV}
	F_+(B) = -Q_A^2 \ln (4/B^2 \Lambda^2) + 2 Q_A^2
	\,,\quad\,\, 
	F_-(B)= Q_A^2,
\end{align}
while for the dimensionless $\Phi(B)$ defined in Eq.~\eqref{F0} we find the piecewise expression 
\begin{align}
	\label{Phicases}
	\Phi(B) \simeq 
	\begin{cases}
	 1/2 \quad &\mathrm{for} \quad B\ll 1/Q_s,
	 \\*[0.2cm]
	 \mcal{O}(1) \quad &\mathrm{for} \quad B \sim 1/Q_s,
	 \\*[0.2cm]
	 \displaystyle
	 -\frac{1}{\ln \mcal{S}_g(B)} = 
	 \frac{4}{B^2 Q_A^2 \ln(4/B^2 \Lambda^2)} \quad &\mathrm{for} \quad B \gg 1/Q_s.
	\end{cases}
\end{align}
Even though $\Phi(B)$ is of the same order in the first two cases, we have distinguished between them in order to point out that the normalization is under control when $B\ll 1/Q_s$. The last line shows that the unitarity corrections lead naturally to a suppression at large $B$. 

When $\Delta_{\perp} \gg Q_s$, small dipoles such that $B \ll 1/Q_s$ naturally give the dominant contribution to all the distributions $\mcal{G}_{\rm D}^{(n)}(\Delta_{\perp})$. After rather simple integrations we obtain to logarithmic accuracy
\begin{align}
\hspace*{-0.2cm}
	\label{GDhigh}
	\mcal{G}_{\rm D}^{(+)}\!
	\simeq
	\frac{2 Q_A^4}{\Delta_{\perp}^2}
	\ln \frac{\Delta_{\perp}^2}{\Lambda^2},
	\quad\,\,
	\mcal{G}_{\rm D}^{(-)}\!
	\simeq
	\frac{Q_A^6}{3\Delta_{\perp}^4},
	\quad\,\,
	\mcal{G}_{\rm D}^{(1)}\!	
	\simeq
	\frac{Q_A^4}{\Delta_{\perp}^2}
	\ln \frac{\Delta_{\perp}^2}{\Lambda^2},
	\quad\,\,
	\mcal{G}_{\rm D}^{(2)}\!
	\simeq
	\frac{2Q_A^4}{\Delta_{\perp}^2}
	\quad\,\,
	\mathrm{for} \quad \Delta_{\perp} \gg Q_s.
\end{align}
Notice that $\mcal{G}_{\rm D}^{(+)}(\Delta_{\perp}) \simeq 2 \mcal{G}_{\rm D}^{(1)}(\Delta_{\perp})$ are the dominant distributions, while $\mcal{G}_{\rm D}^{(-)}(\Delta_{\perp})$ is power suppressed.\footnote{One must keep the first subleading term $-\mcal{T}_g/6$ in the expansion of $\Phi$, where the amplitude $\mcal{T}_g$ contains exchanges of only two gluons.}

In the regime that we are mostly interested in, that is when $\Delta_{\perp} \sim Q_s$, dipoles of size $B \sim 1/Q_s$ determine the bulk of the four distributions and we easily find that
\begin{align}
	\hspace*{0cm}
	\label{Gscale}
	\mcal{G}_{\rm D}^{(+)} 
	\sim Q_s^2\,,
	\qquad
	\mcal{G}_{\rm D}^{(-)}
	\sim \frac{Q_s^2}{\rho_A^2}\,,
	\qquad
	\mcal{G}_{\rm D}^{(1)}
	\sim \frac{Q_s^2}{\rho_A}\,,
	\qquad
	\mcal{G}_{\rm D}^{(2)}
	\sim \frac{Q_s^2}{\rho_A^2}
	\qquad
	\mathrm{for} \quad \Delta_{\perp} \sim Q_s,
\end{align} 
where recall that one can rewrite Eq.~\eqref{QAQs} as $Q_A^2 = Q_s^2/\rho_A$, where $\rho_A \equiv \ln Q_s^2/\Lambda^2$ is a sizable number for typical numbers of the gluon saturation momentum. As a consequence, $\mcal{G}_{\rm D}^{(+)}(Q_s)$ is still the largest of the four distributions to logarithmically accuracy, at least parametrically. One can further verify that such a property remains valid also for small momenta such that $\Lambda \ll \Delta_{\perp} \ll Q_s$, therefore $\mcal{G}_{\rm D}^{(+)}(\Delta_{\perp})$ is eventually the largest in the whole kinematic regime. In fact, the integration leading to $\mcal{G}_{\rm D}^{(+)}(\Delta_{\perp})$ is logarithmic for large $B$ and we obtain 
\begin{align}
	\label{GDsmall}
	\mcal{G}_{\rm D}^{(+)}(\Delta_{\perp})
	\simeq Q_A^2 \left( \ln^2 \frac{Q_s^2}{\Lambda^2} - \ln^2 \frac{\Delta_{\perp}^2}{\Lambda^2} \right)
	\quad\,\,
	\mathrm{for} \quad \Delta_{\perp} \ll Q_s,
\end{align}
which exhibits saturation.

Mostly for the completeness of presentation, we would like to comment on whether BK evolution \cite{Balitsky:1995ub,Kovchegov:1999yj} brings any kind of modifications to these estimates and in particular to the hierarchy among the distributions. This evolution leads to a softening in the perturbative tail of the amplitude and at asymptotically high energy the solution to the BK equation exhibits scaling, i.e.~the amplitude becomes a function of the single variable $B Q_s(Y)$, in a wide kinematic regime which extends to momenta well above $Q_s$ \cite{Iancu:2002tr,Mueller:2002zm,Munier:2003vc}. When $B \ll 1/Q_s$ such a solution (up to logarithmic factors) reads $\mcal{T}_g(B) \propto (B^2 Q_s^2)^\gamma$, where at leading order $\gamma \simeq 0.63$. With NLO corrections \cite{Balitsky:2006wa,Kovchegov:2006vj,Balitsky:2008zza} and the necessary collinear resummations \cite{Beuf:2014uia,Iancu:2015vea,Ducloue:2019ezk} included, the above remains a good solution with $\gamma$ a number satisfying $1/2 < \gamma <1$. It is possible to give analytic estimates of the      
distributions in the presence of BK evolution at very high energy. We will avoid doing so for the economy of the presentation, since the rapidity scale, at which the dipole-nucleus scattering amplitude should be evaluated, is not very large for phenomenological purposes, and thus we expect the $\mcal{G}_{\rm D}^{(n)}(\Delta_{\perp})$ to obey the hierarchy of the MV model expressions in Eqs.~\eqref{GDhigh} and \eqref{Gscale}.

\begin{figure}
	\begin{center}
		\includegraphics[width=0.45\textwidth]{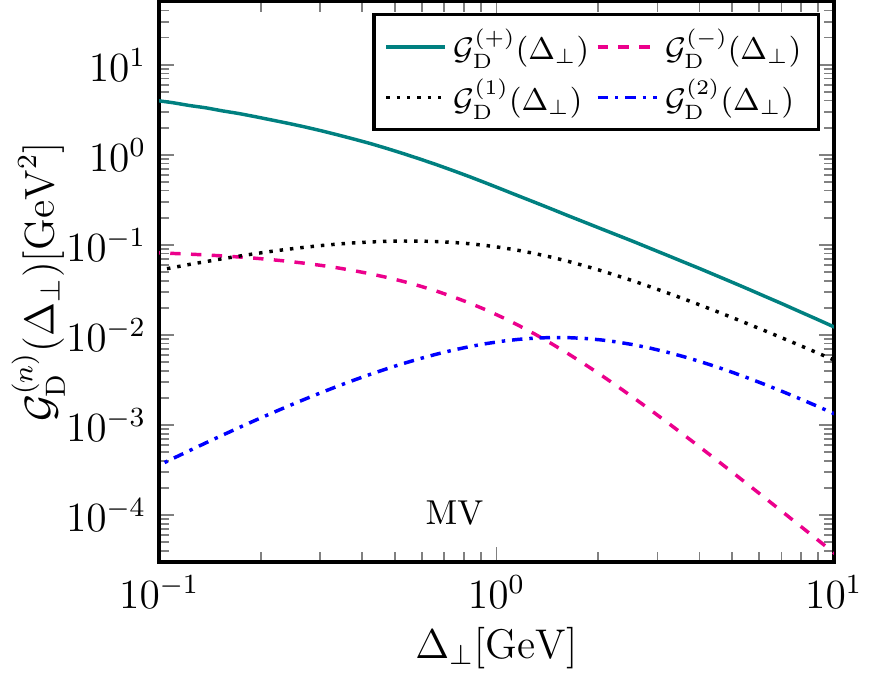}
	\hspace*{0.05\textwidth}
	\includegraphics[width=0.45\textwidth]{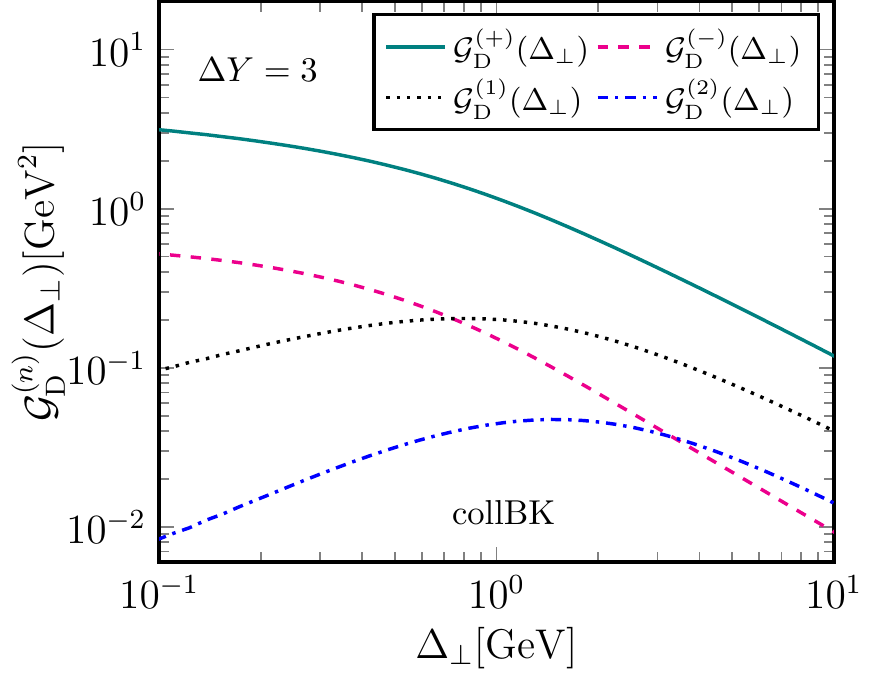}
	\end{center}
	\caption{\small The four diffractive distributions as functions of the transverse momentum $\Delta_{\perp}$. Left panel: the input amplitude $\mcal{T}_g(B)$ is given by the MV model with $Q_s^2=2$ GeV$^2$. Right panel: the input amplitude $\mcal{T}_g(B)$ is obtained by evolving the MV model initial condition to $\Delta Y=3$ with the collinearly improved BK equation \cite{Ducloue:2019ezk}.}
\label{fig:GD}
\end{figure}

In Fig.~\ref{fig:GD} we show the numerical evaluation of the four distributions as functions of $\Delta_{\perp}$. When $\mcal{T}_{g}(B)$ is determined by the MV model, we do see in the left panel that the two hierarchies for $\Delta_{\perp} \gg Q_s$ and $\Delta_{\perp} \sim Q_s$  in Eqs.~\eqref{GDhigh} and \eqref{Gscale} are confirmed. We further notice that $\mcal{G}_{\rm D}^{(1)}(\Delta_{\perp})$ and $\mcal{G}_{\rm D}^{(2)}(\Delta_{\perp})$ bend down at small $\Delta_{\perp}$. This is due to the particular dependence on the angle between $\bm{\Delta}$ and $\bm{B}$ which upon integration leads to a $J_2(\Delta_{\perp} B)$ and $J_4(\Delta_{\perp} B)$ for the corresponding distributions [cf.~Eqs.~\eqref{Gd1}, \eqref{Gd2} and \eqref{angular}]. The two Bessel functions vanish quadratically and quartically, respectively, at small argument and lead to the suppression of the distributions in the region where $\Delta$ is smaller than $Q_s$. Regarding the other two distributions, $\mcal{G}_{\rm D}^{(+)}(\Delta_{\perp})$ keeps growing logarithmically with decreasing $\Delta_{\perp}$, more precisely as the square of the logarithm of $\Delta_{\perp}^2$ as evident in Eq.~\eqref{GDsmall}, while $\mcal{G}_{\rm D}^{(-)}(\Delta_{\perp})$ grows more slowly. As expected, BK evolution\footnote{Here we solve the collinearly improved BK evolution with running coupling as developed in \cite{Ducloue:2019ezk}. The precise details and conventions are given in Appendix D of \cite{Iancu:2020jch}.} in a short interval does not bring strong changes, except in $\mcal{G}_{\rm D}^{(-)}(\Delta_{\perp})$ for $\Delta_{\perp} \gg Q_s$, since it is not any more power suppressed as in the MV model. However, it is worthwhile to mention that $\mcal{G}_{\rm D}^{(-)}(\Delta_{\perp})$ can be neglected for all practical purposes. Indeed, it is always much smaller than $\mcal{G}_{\rm D}^{(+)}(\Delta_{\perp})$, while they both give the same type of contribution: the one that is independent of the angle $\phi$ (for both transverse and longitudinal sectors). 

\begin{figure}
	\begin{center}
		\includegraphics[width=0.45\textwidth]{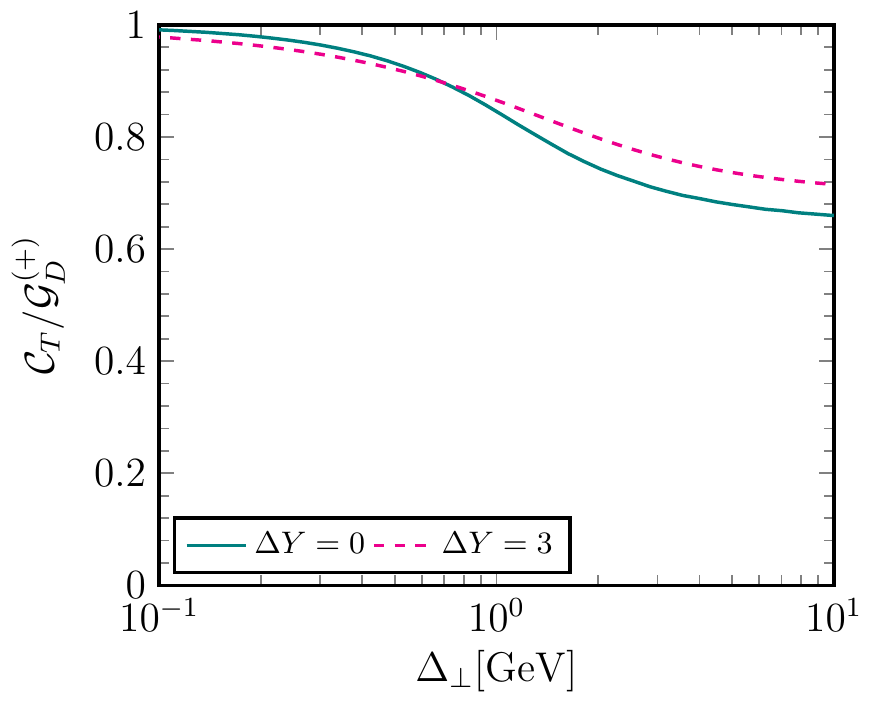}
	\hspace*{0.05\textwidth}
	\includegraphics[width=0.45\textwidth]{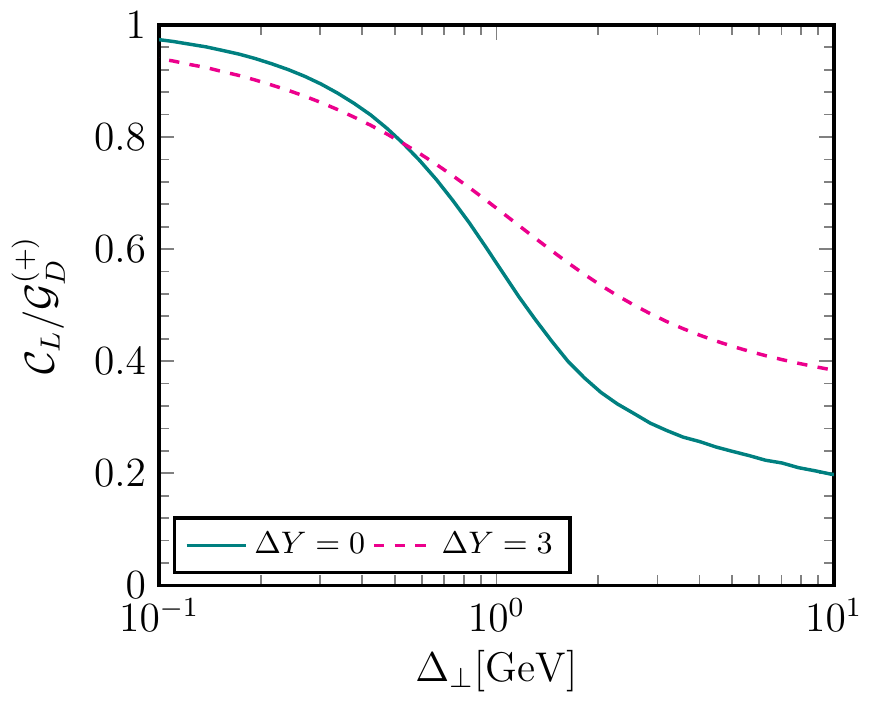}
	\end{center}
	\caption{\small The l.h.s.~of the two positivity constraints normalized to $\mcal{G}_{\rm D}^{(+)}(\Delta_{\perp})$. Left panel: transverse sector [cf.~Eq.~\eqref{constraintT}]. Right panel: longitudinal sector [cf.~Eq.~\eqref{constraintL}].}
\label{fig:pos}
\end{figure}

The fact that the transverse and longitudinal cross sections are positive, but at the same time they contain contributions proportional to $\cos n\phi$, induces certain constraints for the distributions. Since the averaged over $\phi$ reduced cross sections $\mcal{N}_0^{\lambda}$ [anticipating the notation introduced in the next Section, cf.~Eq.~\eqref{Ns}] are explicitly positive, we can equivalently consider the ratios $|\mcal{A}_{\rm D}^{\lambda}|^2/\mcal{N}_0^{\lambda}$. Then, at each sector, we minimize the ratio as a function of $P_{\perp}^2$ and $\phi$ and require that its value at the minimum is non-negative. Let us first look at the longitudinal case which will turn out to be more interesting. A full analysis is rather involved, hence, to get a feeling, we first keep only the $\mcal{G}_{\rm D}^{(+)}(\Delta_{\perp})$ and $\mcal{G}_{\rm D}^{(1)}(\Delta_{\perp})$ terms in Eq.~\eqref{ad2lfinal}; since the latter is the only distribution which can be comparable to the former one in the MV model and in a certain regime of $\Delta_{\perp}$, this looks a reasonable assumption to start with in general. We find that $|\mcal{A}_{\rm D}^{L}|^2/\mcal{N}_0^{L}$ has a global minimum at $\phi=0$ and $P_{\perp} = \bar{Q}/\sqrt{3}$ and one arrives at the positivity constraint $\mcal{G}_{\rm D}^{(+)}(\Delta_{\perp}) - 2 \mcal{G}_{\rm D}^{(1)}(\Delta_{\perp}) \geq 0$. It is rather amusing that such a constraint is saturated in the MV model for $\Delta_{\perp} \gg Q_s$, if we keep only the leading logarithmically enhanced pieces of the two distributions as given in Eq.~\eqref{GDhigh}. This is very similar to the property satisfied by the unpolarized and linearly polarized gluon distributions \cite{Metz:2011wb}. A more detailed calculation that includes also $\mcal{G}_{\rm D}^{(2)}(\Delta_{\perp}) \ll \mcal{G}_{\rm D}^{(+),(1)}(\Delta_{\perp})$ [while it still neglects $\mcal{G}_{\rm D}^{(-)}(\Delta_{\perp})$ assuming that it is small compared to $\mcal{G}_{\rm D}^{(+)}(\Delta_{\perp})$ like in the MV model] leads to
\begin{align}
	\label{constraintL}
	\mcal{C}_L(\Delta_{\perp})
	\equiv
	\mcal{G}_{\rm D}^{(+)}(\Delta_{\perp}) - 2 \mcal{G}_{\rm D}^{(1)}(\Delta_{\perp}) + \frac{1}{2}\, \mcal{G}_{\rm D}^{(2)}(\Delta_{\perp}) \geq 0.
\end{align}
Evaluating more precisely the distributions by keeping all the $1/\Delta_{\perp}^2$ terms (and not only those enhanced by a logarithm) in the MV model, we can show analytically that the l.h.s.~in the above is strictly positive for large $\Delta_{\perp} \gg Q_s$. Similarly, the corresponding positivity constraint in the transverse sector reads
\begin{align}
	\label{constraintT}
	\mcal{C}_T(\Delta_{\perp})
	\equiv
	\mcal{G}_{\rm D}^{(+)}(\Delta_{\perp}) - \frac{2}{3}\, \mcal{G}_{\rm D}^{(1)}(\Delta_{\perp}) - \frac{1}{2}\, \mcal{G}_{\rm D}^{(2)}(\Delta_{\perp}) \geq 0.
\end{align}
In Fig.~\ref{fig:pos} we numerically confirm the validity of Eqs.~\eqref{constraintL} and \eqref{constraintT} in a wide region of $\Delta_{\perp}$, where the distributions are calculated either by using the MV model or the solution to the collinearly improved BK equation. 

\section{Analytical estimates and numerical results for the cross sections and the azimuthal anisotropies}
\label{sec:numerics}

Using the general expressions in Sect.~\ref{sec:cross-section} and the behavior of the distributions in Sect.~\ref{sec:GDs} we shall now study the incoherent diffractive dijet cross section. To this end, let us be more specific about the quantities to be calculated. First, we shall consider the cross section obtained after averaging over the relative angle $\phi$ between the momenta $\bm{\Delta}$ and $\bm{P}$, namely
\begin{align}
	\label{sigmaave}
	\frac{\rmd\sigma_{\scriptscriptstyle \rm D}^{\gamma_\lambda^* 
	A\rightarrow q\bar q X}}
	{\dif \Pi}
	=
	\int_0^{2\pi} 
	\frac{\dif \phi}{2\pi}\,
	\frac{\rmd\sigma_{\scriptscriptstyle \rm D}^{\gamma_\lambda^*  
	A\rightarrow q\bar q X}}
	{\rmd \vartheta_1 \rmd \vartheta_2\, 
	\rmd^{2} \bm{P} \rmd^{2} \bm{\Delta}},
\end{align}
with $\dif \Pi = (2\pi)^2 P_{\perp} \dif P_{\perp} \Delta_{\perp} \dif \Delta_{\perp} \dif \vartheta_1 \dif \vartheta_2$. It is clear from Eqs.~\eqref{ad2final} and \eqref{ad2lfinal}, that the coefficients of $\mcal{G}_{\rm D}^{(1)}(\Delta_{\perp})$ and $\mcal{G}_{\rm D}^{(2)}(\Delta_{\perp})$ vanish due to the angle averaging and therefore the above quantity probes a combination of $\mcal{G}_{\rm D}^{(+)}(\Delta_{\perp})$ and $\mcal{G}_{\rm D}^{(-)}(\Delta_{\perp})$. As already pointed out at the end of Sect.~\ref{sec:GDs}, in practice only $\mcal{G}_{\rm D}^{(+)}(\Delta_{\perp})$ contributes to the cross section. Thus, for simplicity, $\mcal{G}_{\rm D}^{(-)}(\Delta_{\perp})$ will not be shown in the following analytical expressions, although it will be kept in all numerical calculations. In order to facilitate our notation, let us note that the dependence of the reduced cross sections in Eqs.~\eqref{ad2final} and  \eqref{ad2lfinal} on the momenta $\bm{P}$ and $\bm{\Delta}$ may be written as
\begin{align}
	\label{Ns}
	\big| 
	\mcal{A}_{\rm D}^{\lambda} \big|^2 = 
	\mcal{N}_0^{\lambda}(P_{\perp},\Delta_{\perp}) 
	+ 2 \sum_{n=1}^{\infty} \mcal{N}_{n}^{\lambda}(P_{\perp},\Delta_{\perp}) \cos n \phi,
\end{align}
where it is clear that the only non vanishing terms at the leading twist level in which we have been working so far, are $\mcal{N}_n^{\lambda}$ with $n=0,2,4$. Then the dependence on $P_{\perp}$ and $\Delta_{\perp}$ (and $\bar{Q}$) of the average cross section is encoded in 
\begin{align}
	\label{ADave}
	\mcal{N}_0^{T}
	\simeq
	\frac{4 P_{\perp}^2 (3 \bar{Q}^4 + P_{\perp}^4)}
	{(P_{\perp}^2 + \bar{Q}^2)^6}\, 
	\mcal{G}_{\rm D}^{(+)}(\Delta_{\perp}) 
	\, ,\quad\,
	\mcal{N}_0^{L}
	\simeq 
	\frac{2 \bar{Q}^2 (\bar{Q}^4  - 2 \bar{Q}^2 P_{\perp}^2 + 5P_{\perp}^4 )}
	{(P_{\perp}^2 + \bar{Q}^2)^6}\, 
	\mcal{G}_{\rm D}^{(+)}(\Delta_{\perp}), 
\end{align}
while all the remaining prefactors can be readily restored making use of Eq.~\eqref{transdiff1}. In order to get a better feeling for the magnitude of the incoherent diffractive cross section, let us make a direct comparison with the respective inclusive cross section defined analogously to Eq.~\eqref{sigmaave}. The latter is analytically known in the correlation limit \cite{Dominguez:2011wm,Dominguez:2011br} and is reviewed in Appendix \ref{app:inclusive}. It is a trivial exercise to obtain the ratio
\begin{align}
	\label{ratioditT}
	\frac{\rmd\sigma_{\scriptscriptstyle \rm D}^{\gamma_T^* 
	A\rightarrow q\bar q X}/\dif \Pi}
	{\rmd\sigma_{\scriptscriptstyle \rm inc}^{\gamma_T^* 
	A\rightarrow q\bar q X}/\dif \Pi} = \frac{1}{2 N_c^2}\,
	\frac{4 P_{\perp}^2 (3 \bar{Q}^4 + P_{\perp}^4)}
	{(P_{\perp}^4 + \bar{Q}^4) (P_{\perp}^2 + \bar{Q}^2)^2}\,
	\frac{\mcal{G}_{\rm D}^{(+)}(\Delta_{\perp})}
	{(4 \pi^3 \alpha_s/C_F)\, xG(\Delta_{\perp})}
\end{align}
for the transverse polarizations and the ratio
\begin{align}
	\label{ratioditL}
	\frac{\rmd\sigma_{\scriptscriptstyle \rm D}^{\gamma_L^* 
	A\rightarrow q\bar q X}/\dif \Pi}
	{\rmd\sigma_{\scriptscriptstyle \rm inc}^{\gamma_L^* 
	A\rightarrow q\bar q X}/\dif \Pi} = \frac{1}{2 N_c^2}\,
	\frac{\bar{Q}^4 -2 \bar{Q}^2 P_{\perp}^2 + 5 P_{\perp}^4}
	{P_{\perp}^2 (P_{\perp}^2 + \bar{Q}^2)^2}\,
	\frac{\mcal{G}_{\rm D}^{(+)}(\Delta_{\perp})}
	{(4 \pi^3 \alpha_s/C_F)\, xG(\Delta_{\perp})},
\end{align}
for the longitudinal one. Here $xG(\Delta_{\perp})$ is the unpolarized unintegrated gluon distribution in the target, cf.~Eq.~\eqref{alphaxG}, and it is dimensionless. The last fraction, which encodes the semihard dynamics, is of the order of $Q_s^2$ when  $\Delta_{\perp} \sim Q_s$, and generally it is a slowly varying function of $\Delta_\perp$. Hence, when $P_{\perp} \sim \bar{Q}$, one sees that both ratios scale as
\begin{align}
	\label{Rscale}
	\frac{\rmd\sigma_{\scriptscriptstyle \rm D}^{\gamma_\lambda^* 
	A\rightarrow q\bar q X}/\dif \Pi}
	{\rmd\sigma_{\scriptscriptstyle \rm inc}^{\gamma_\lambda^* 
	A\rightarrow q\bar q X}/\dif \Pi} \sim \frac{1}{N_c^2}\, \frac{Q_s^2}{P_{\perp}^2}, 
\end{align}
and exhibit a double suppression: in addition to the $1/N_c^2$ factor discussed earlier in Sect.~\ref{sec:corr}, the $1/P_{\perp}^6$ fall-off in the diffractive case is much stronger than the $1/P_{\perp}^4$ one in the inclusive case. Still, the precise $P_{\perp}$ dependence of the ratio is quite different for the two polarizations and it has to do with the value of $P_{\perp}$ w.r.t.~$\bar{Q}$. In the transverse case the ratio vanishes quadratically at small $P_{\perp}$ while it falls like $1/P_{\perp}^2$ at large  $P_{\perp}$, thus showing a clear maximum at a value $P_{\perp}^*$ which is proportional to $\bar{Q}$ (the numerical solution to a simple algebraic equation gives $P_{\perp}^*\simeq 0.75 \bar{Q}$). On the contrary, in the longitudinal case the ratio scales like $1/P_{\perp}^2$ at both small and large $P_{\perp}$ (but with a different coefficient in each regime). Such detailed features have already been observed in the numerical solution in \cite{Mantysaari:2019hkq} (cf.~the right panel in Fig.~3 there) and what we point out here is that they can be analytically understood in the correlation limit.

Next, we define the azimuthal anisotropies, which are sensitive to the relative angle between  $\bm{\Delta}$ and $\bm{P}$, as the dimensionless ratios
\begin{align}
	\label{cosnphi}
	\langle \cos n \phi \rangle_{\lambda} = 
	\int_0^{2\pi}
	\frac{\dif \phi}{2\pi}\,
	\cos n \phi \,
	\frac{\rmd\sigma_{\scriptscriptstyle \rm D}^{\gamma_\lambda^* 
	A\rightarrow q\bar q X}}
	{\rmd \vartheta_1 \rmd \vartheta_2\, 
	\rmd^{2} \bm{P} \rmd^{2} \bm{\Delta}}
	\bigg/
	\int_0^{2\pi} 
	\frac{\dif \phi}{2\pi}\,
	\frac{\rmd\sigma_{\scriptscriptstyle \rm D}^{\gamma_\lambda^* 
	A\rightarrow q\bar q X}}
	{\rmd \vartheta_1 \rmd \vartheta_2\, 
	\rmd^{2} \bm{P} \rmd^{2} \bm{\Delta}}
	= \frac{\mcal{N}_n^{\lambda}}{\mcal{N}_0^{\lambda}},
\end{align}
with $n$ an (even) integer. By straightforward inspection of Eqs.~\eqref{ad2final} and \eqref{ad2lfinal} we immediately see that for both transverse and longitudinal virtual photons, the non-vanishing anisotropies are the elliptic $\langle \cos 2\phi \rangle_{\lambda}$ and the quadrangular $\langle \cos 4\phi \rangle_{\lambda}$ ones, for which we obtain
\begin{align}
	\label{cos2phi}
	\langle \cos 2 \phi \rangle_{\scriptscriptstyle T}
	\simeq
	\frac{2 \bar{Q}^2 (\bar{Q}^2 - P_{\perp}^2)}
	{3 \bar{Q}^4+ P_{\perp}^4}\, 
	\frac{\mcal{G}_{\rm D}^{(1)}(\Delta_{\perp})}
	{\mcal{G}_{\rm D}^{(+)}(\Delta_{\perp})} 
	\, ,\quad\,
	\langle \cos 2 \phi \rangle_{\scriptscriptstyle L}
	\simeq 
	\frac{4 P_{\perp}^2 (P_{\perp}^2 - \bar{Q}^2)}
	{\bar{Q}^4 - 2 \bar{Q}^2 P_{\perp}^2 + 5 P_{\perp}^4}\,
	\frac{\mcal{G}_{\rm D}^{(1)}(\Delta_{\perp})}
	{\mcal{G}_{\rm D}^{(+)}(\Delta_{\perp})} 
\end{align}
and
\begin{align}
	\label{cos4phi}
	\langle \cos 4 \phi \rangle_{\scriptscriptstyle T}
	\simeq
	- \frac{\bar{Q}^2 P_{\perp}^2}
	{3 \bar{Q}^4+ P_{\perp}^4}\, 
	\frac{\mcal{G}_{\rm D}^{(2)}(\Delta_{\perp})}
	{\mcal{G}_{\rm D}^{(+)}(\Delta_{\perp})} 
	\, ,\quad\,
	\langle \cos 4 \phi \rangle_{\scriptscriptstyle L}
	\simeq 
	\frac{2 P_{\perp}^4}
	{\bar{Q}^4 - 2 \bar{Q}^2 P_{\perp}^2 + 5 P_{\perp}^4}\,
	\frac{\mcal{G}_{\rm D}^{(2)}(\Delta_{\perp})}
	{\mcal{G}_{\rm D}^{(+)}(\Delta_{\perp})}. 
\end{align}

\begin{figure}
\begin{center}
	\includegraphics[width=0.45\textwidth]{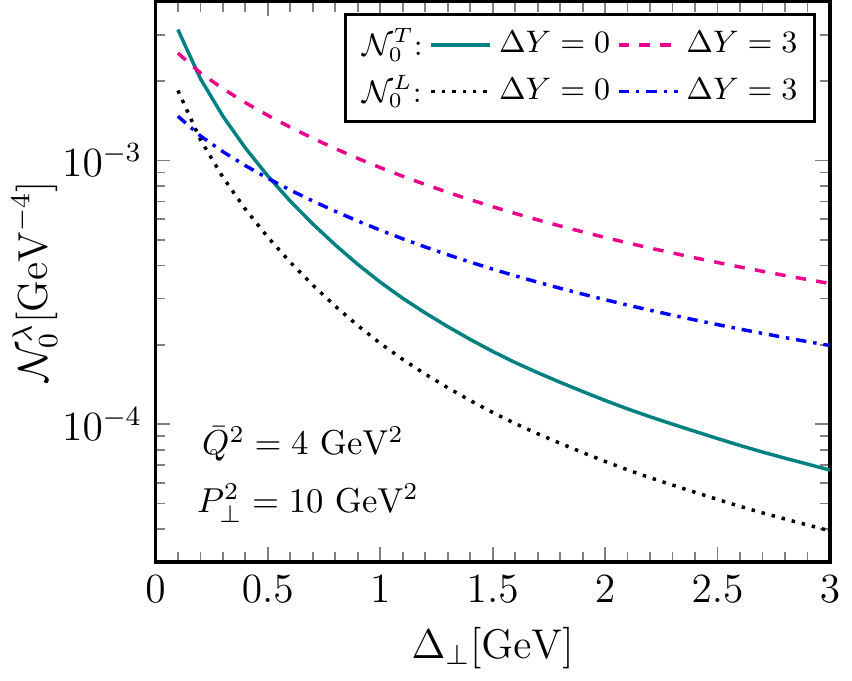}
	\hspace*{0.05\textwidth}
	\includegraphics[width=0.45\textwidth]{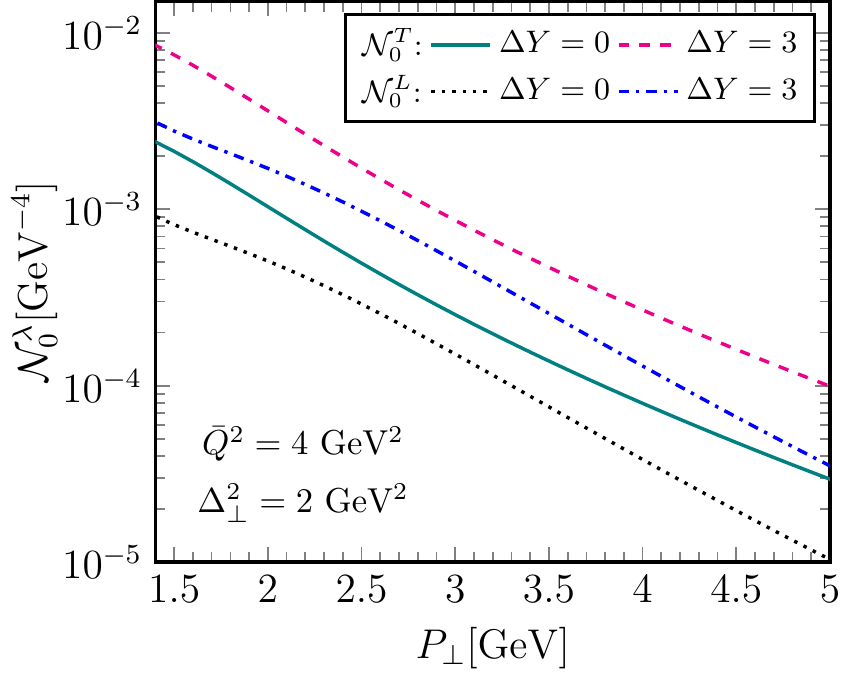}
\end{center}
	\caption{\small The averaged over angle reduced incoherent diffractive dijet cross section at leading twist. Left panel: as a function of the momentum transfer $\Delta_{\perp}$ for fixed relative momentum $P_{\perp}$. Right panel: as a function of $P_{\perp}$ for fixed $\Delta_{\perp}$.}
\label{fig:cs}
\end{figure}
A few useful observations are in order: 
\begin{enumerate}[(i)]
	\item As already implied in the discussion at the end of Sect.~\ref{sec:cross-section}, $\langle \cos 2\phi \rangle_{\lambda}$ is proportional to $\mcal{G}_{\rm D}^{(1)}(\Delta_{\perp})$ while $\langle \cos 4\phi \rangle_{\lambda}$ is proportional to $\mcal{G}_{\rm D}^{(2)}(\Delta_{\perp})$.\vspace*{-0.2cm}
	\item The sign of $\langle \cos 2\phi \rangle_{\scriptscriptstyle T}$ is opposite to the one of $\langle \cos 2\phi \rangle_{\scriptscriptstyle L}$ and similarly for the $\langle \cos 4\phi \rangle_{\lambda}$ anisotropies, that is 
\begin{align}
	\label{negative}
	\langle \cos 2\phi \rangle_{\scriptscriptstyle T}
	\langle \cos 2\phi \rangle_{\scriptscriptstyle L} \leq 0
	\qquad
	\mathrm{and}
	\qquad
	\langle \cos 4\phi \rangle_{\scriptscriptstyle T}
	\langle \cos 4\phi \rangle_{\scriptscriptstyle L} < 0.
\end{align}
\vspace*{-0.7cm} 
	\item As we have indicated above, $\langle \cos 2\phi \rangle_{\lambda}$ can vanish. Indeed, as readily observed in Eqs.~\eqref{cos2phi} and \eqref{cos4phi}, we have 
\begin{align}
	\label{zero}
	\langle \cos 2\phi \rangle_{\scriptscriptstyle T} = 
	\langle \cos 2\phi \rangle_{\scriptscriptstyle L} = 0
	\quad
	\mathrm{when}
	\quad 
	P_{\perp} = \bar{Q}
	\quad
	\mathrm{for\,\,any}
	\quad
	\Delta_{\perp}.
	\end{align}	 
	Similar zeros have been observed in coherent diffractive dijet production \cite{Salazar:2019ncp}. We shall elaborate on this in Sect.~\ref{sec:kin}: we will see that such a property is still valid when we include the first higher twist correction in the longitudinal sector, but it is violated in the transverse one.
	\vspace*{-0.2cm} 
	\item The $\langle \cos 4\phi \rangle_{\lambda}$ anisotropies seem to be parametrically of the same order of the $\langle \cos 2\phi \rangle_{\lambda}$ ones, since all the hard coefficients are of the same order in $P_{\perp}$ (they fall like $1/P_{\perp}^6$ assuming that $P_{\perp}$ and $\bar{Q}$ are of the same order). This is in sharp contrast to the case of inclusive dijet production, where in the correlation limit only $\langle \cos 2\phi \rangle_{\lambda}$ are non-zero at leading twist. In that process, $\langle \cos 4\phi \rangle_{\lambda}$ anisotropies can be generated when higher twists are taken into account \cite{Dumitru:2016jku}, but naturally they are suppressed by powers of either $Q_s^2/P_{\perp}^2$ or $\Delta_{\perp}^2/P_{\perp}^2$.
\end{enumerate}

In Fig.~\ref{fig:cs} we show the results for the averaged over angle reduced cross sections. As expected, both the transverse and longitudinal cross sections exhibit a much harder tail in $P_{\perp}$ than in $\Delta_{\perp}$ at high momenta. With decreasing $\Delta_{\perp}$, when $\Delta_{\perp} < Q_s$, the cross section shows a logarithmic increase, as it inherits the properties of the dominant distribution $\mcal{G}_{\rm D}^{(+)}(\Delta_{\perp})$ [cf.~Eq.~\eqref{GDsmall}].

\begin{figure}
\begin{center}
	\includegraphics[width=0.45\textwidth]{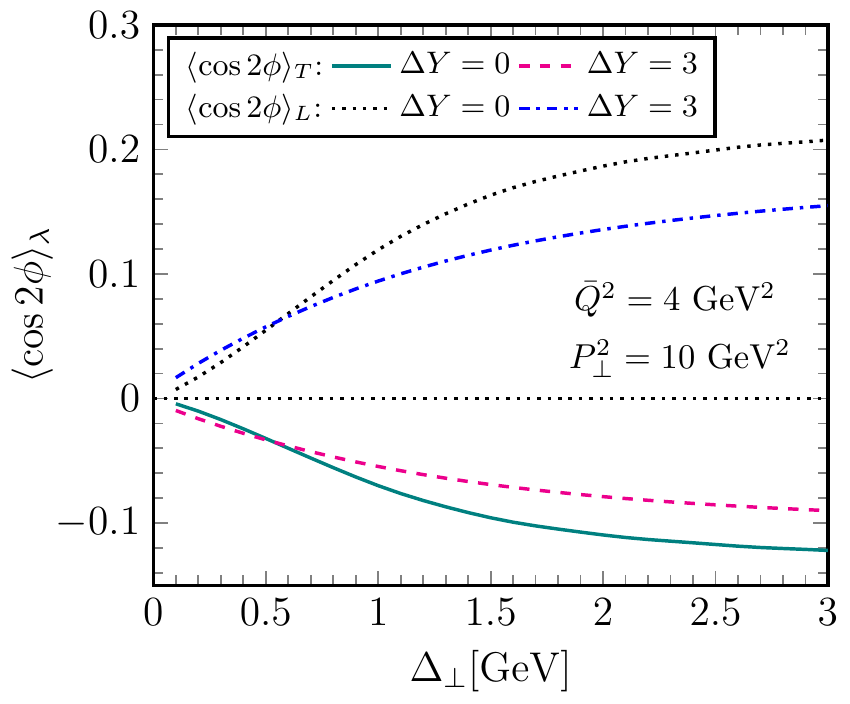}
	\hspace*{0.05\textwidth}
	\includegraphics[width=0.45\textwidth]{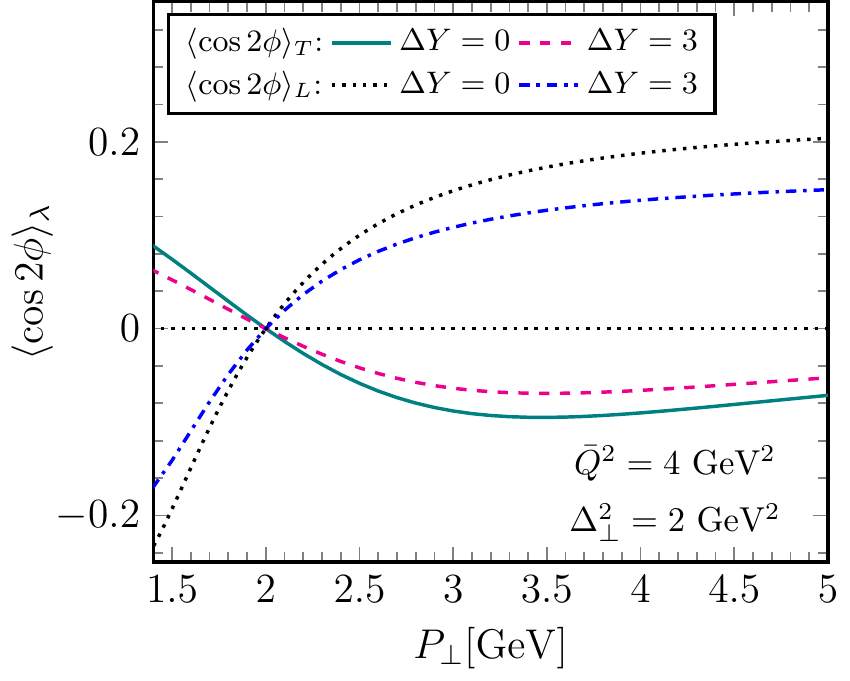}
\end{center}
	\caption{\small Diffractive $\langle \cos 2 \phi \rangle_{\lambda}$ azimuthal anisotropies at leading twist. Left panel: as a function of the momentum transfer $\Delta_{\perp}$ for fixed relative momentum $P_{\perp}$. Right panel: as a function of $P_{\perp}$ for fixed $\Delta_{\perp}$.}
\label{fig:cos2phi}
\end{figure}

\begin{figure}
\begin{center}
	\includegraphics[width=0.45\textwidth]{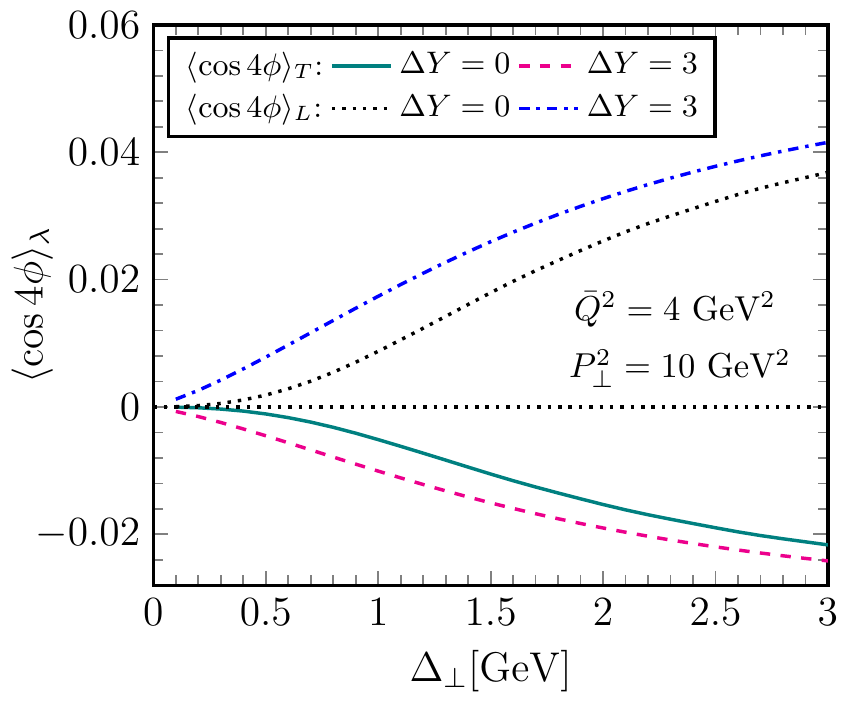}
	\hspace*{0.05\textwidth}
	\includegraphics[width=0.45\textwidth]{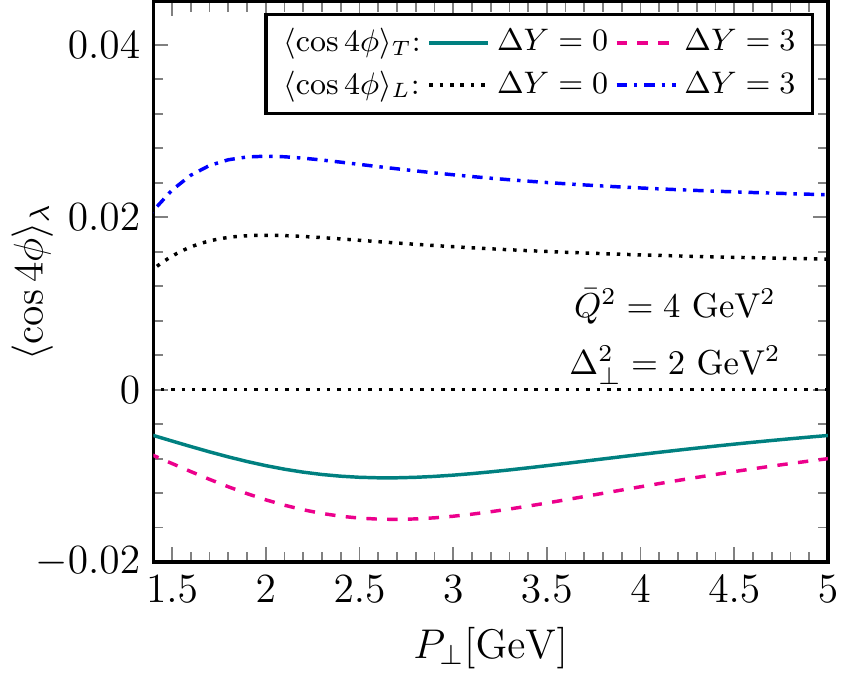}
\end{center}
	\caption{\small Diffractive $\langle \cos 4 \phi \rangle_{\lambda}$ azimuthal anisotropies at leading twist. Left panel: as a function of the momentum transfer $\Delta_{\perp}$ for fixed relative momentum $P_{\perp}$. Right panel: as a function of $P_{\perp}$ for fixed $\Delta_{\perp}$.}
\label{fig:cos4phi}
\end{figure}

The $\langle \cos 2 \phi \rangle_{\lambda}$ azimuthal anisotropies are illustrated in Fig.~\ref{fig:cos2phi}. One sees that the onset of saturation, i.e.~a decreasing momentum transfer $\Delta$, brings about a smaller anisotropy. As a function of the relative momentum $P_{\perp}$ the two anisotropies trivially vanish at $P_{\perp} =\bar{Q}$ due to the form of the hard factor, while the transverse one displays a maximum in its absolute value and then vanishes for very large $P_{\perp}$. These features, except the zero at $P_{\perp} =\bar{Q}$, are also shared by the $\langle \cos 4 \phi \rangle_{\lambda}$ anisotropies shown in Fig.~\ref{fig:cos4phi}. In the MV model, and to some extent in the BK solution, the semihard factors satisfy $\mcal{G}_{\rm D}^{(2)}(\Delta_{\perp}) \ll \mcal{G}_{\rm D}^{(1)}(\Delta_{\perp})$ for all the momenta of interest due to the presence of large logarithmic factors, cf.~Eqs.~\eqref{GDhigh} and \eqref{Gscale}. Thus, we eventually see that $\langle \cos 4 \phi \rangle_{\lambda}$ is small (although it is far from being zero) when compared to $\langle \cos 2 \phi \rangle_{\lambda}$, even though they are both of the same twist.

\section{Higher twist kinematic corrections}
\label{sec:kin}

The problem we are dealing with involves three scales, therefore we can form the two dimensionless ratios $Q_s/P_{\perp}$ and $\Delta_\perp/P_{\perp}$ which in our approach are taken to be small. Indeed, what we have developed so far is valid only in the limit where the jet momentum $P_{\perp}$ is much larger than both the saturation scale $Q_s$ and the momentum transfer $\Delta$. One way to improve the description is to calculate higher twists, either of the form $(Q_s^2/P_{\perp}^2)^n$, with $n$ a positive integer, which are ``genuine saturation'' effects, or of the form $(\Delta_{\perp}^2/P_{\perp}^2)^n$ which are ``kinematic corrections'' \cite{Altinoluk:2019fui,Altinoluk:2019wyu,Fujii:2020bkl} (see also \cite{Boussarie:2020vzf}). In order to fully address the saturation twists one must resort to a numerical approach \cite{Mantysaari:2019hkq}, but since we are interested in hard jets, we shall not be concerned with them in what follows. Regarding the kinematic twists, one can rely on the improved TMD framework and in principle one could resum them to all orders by following the procedure developed in \cite{Altinoluk:2019wyu}. For example, such a task has already been performed for the case of inclusive dijet production \cite{Boussarie:2021ybe}. The implementation of this resummation to the diffractive problem under consideration is not straightforward, thus our goal will be more modest. We will calculate only the first higher order kinematic twist, but we will see that this is enough to capture the main features of the anisotropies in the relevant kinematic regime.\footnote{We would like to comment that we have used this approach to calculate the cross section for inclusive dijets as a test. It turns out that the first kinematic correction describes correctly the main features also for this process in the relevant kinematic regime, like capturing in the $\langle \cos 2\phi \rangle_{\scriptscriptstyle T}$ anisotropy a global minimum with a value $\sim -0.2$ around $P_{\perp} \sim \Delta_{\perp} \sim 2$GeV \cite{Mantysaari:2019hkq}.}

Our starting point is again the general expression in Eq.~\eqref{sigmab} for the transverse cross section and similarly for the longitudinal one. At next to leading twist one discovers that there is explicit dependence of the reduced cross section on the fractions $\vartheta_1$ and $\vartheta_2$, and for simplicity we shall take $\vartheta_1=\vartheta_2=1/2$ in what follows.\footnote{At leading twist the only dependence turned out to be implicit, i.e.~it came only via $\bar{Q}$. In general, a simple inspection of Eq.~\eqref{sigmab} shows that any explicit dependence may arise only due to the changes of variables from ($\bx,\by$) and ($\bk_1,\bk_2$) to ($\bb,\br$) and ($\bm{\Delta},\bP$) which involve the two fractions.}  Now one expands the correlator $\mathcal{W}_{\rm D}(\bm{r},\bar{\bm{r}},\bm{B})$ defined in Eq.~\eqref{wdiff} to the 6th order in the sizes $r$ and/or $\bar{r}$ and this is to be compared with the 4th order expansion in Eq.~\eqref{wdiffsat}. These two extra powers in the small dipoles lead to an additional $1/P_{\perp}^2$ factor after taking the Fourier transform. At the same time, it is clear that there will be two more derivatives which eventually lead to a factor proportional to either $\Delta_{\perp}^2$ or $Q_s^2$, and only the former classifies as kinematic correction. Since $\ln \mcal{S}_g (B)$ scales like $Q_s^2$, we want to keep only the terms which have the smallest number of such factors, that is two, like in the leading twist term in Eq.~\eqref{wdiffsat}. In fact straightforward algebra shows there is only one such term at next to leading twist, more precisely one finds that the first order kinematic corrections to be added to the reduced cross sections in Eqs.~\eqref{ad2} and \eqref{ad2l} read
\begin{align}
	\label{ad2cort}
	&\delta\!\left|\mcal{A}_{\rm D}^T \right|^2 =
	\frac{1}{6}\, 
	H_T^{ikmns}(\bP,\bar{Q})\, 
	H_T^{jls*}(\bP,\bar{Q})\, 
	\mcal{G}_{\rm D}^{ij,klmn}(\bm{\Delta}),
	\\
	\label{ad2corl}
	&\delta\!\left|\mcal{A}_{\rm D}^L \right|^2 =
	\frac{1}{6}\, 
	H_L^{ikmn}(\bP,\bar{Q})\, 
	H_L^{jl*}(\bP,\bar{Q})\, 
	\mcal{G}_{\rm D}^{ij,klmn}(\bm{\Delta}),
\end{align}
for the transverse and longitudinal sectors respectively. The hard factors $H_T^{ikmns}(\bP,\bar{Q})$ and $H_L^{ikmn}(\bP,\bar{Q})$ are the obvious extensions of Eqs.~\eqref{hiks} and \eqref{hik} to tensors with five and four indices and they can be calculated analytically, while the semihard tensor reads
\begin{align}
	\label{Gdcor}
	\mcal{G}_{\rm D}^{ij,klmn}(\bm{\Delta}) = 
	\int \frac{\rmd^2 \bm{B}}{2\pi}\,
	e^{- i \bm{\Delta} \cdot \bm{B}}\,
	\Phi(B)\,
	[\partial^i \partial^j \ln \mcal{S}_g(B)]\,
	[\partial^k \partial^l \partial^m \partial^n \ln \mcal{S}_g(B)].
\end{align}
One can perform the tensor decomposition of the above and subsequently the contraction with the hard factors. We shall not present any more details on this, rather we will only give the final expressions for the MV model in Appendix \ref{app:nltwist}. Here it suffices to observe that the tensor in Eq.~\eqref{Gdcor} has mass dimensions four, contrary to $\mcal{G}_{\rm D}^{ij,kl}(\bm{\Delta})$ defined in Eq.~\eqref{Gd} which has mass dimension two. The two extra powers come from the two extra derivatives w.r.t.~the large dipole $B$ in the last square bracket. This leads to an additional factor $\sim 1/B^2$ in the integrand and eventually to a factor $\Delta_{\perp}^2$ after the Fourier transform is taken. For example, in the MV model and up to logarithms, the integrand scales likes $Q_A^4/B^2$ and hence the integral like $Q_A^4$, to be compared with the $Q_A^4$ and  $Q_A^4/\Delta_{\perp}^2$ scaling of the respective leading twist quantities. Thus, as anticipated, we confirm that the corrections Eqs.~\eqref{ad2cort} and \eqref{ad2corl} are suppressed by a $\Delta_{\perp}^2/P_{\perp}^2$ when compared to the leading twist result.

\begin{figure}
\begin{center}
	\includegraphics[width=0.45\textwidth]{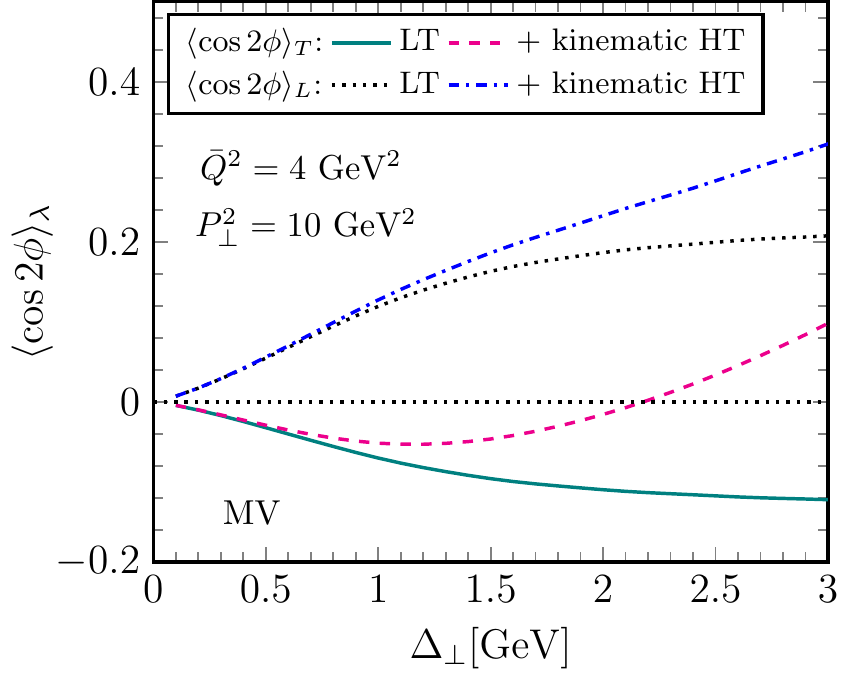}
	\hspace*{0.05\textwidth}
	\includegraphics[width=0.45\textwidth]{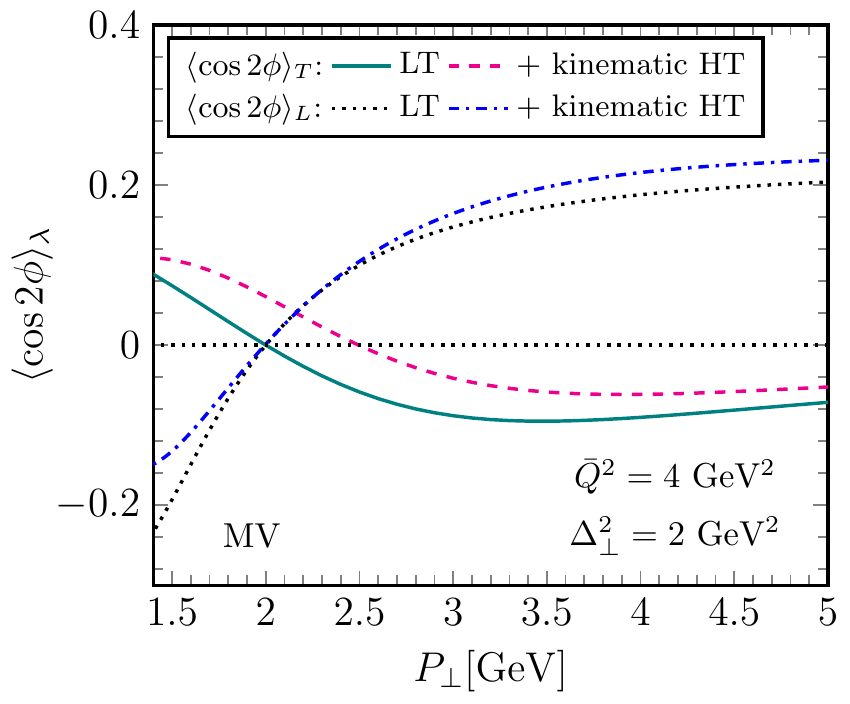}
\end{center}
	\caption{\small Diffractive $\langle \cos 2 \phi \rangle_{\lambda}$ azimuthal anisotropies including the first higher kinematic twist. Left panel: as a function of the momentum transfer $\Delta_{\perp}$ for fixed relative momentum $P_{\perp}$. Right panel: as a function of  $P_{\perp}$ for fixed $\Delta_{\perp}$.}
\label{fig:cos2phinext}
\end{figure}

\vskip 4pt \noindent In Fig.~\ref{fig:cos2phinext} we show $\langle \cos 2 \phi \rangle_{\lambda}$ in the MV model when the first higher kinematic twist is included.
\begin{enumerate}[(i)]
\item Regarding the longitudinal elliptic anisotropy, the higher twist induces a reduction when $\Delta_{\perp}$ gets larger than $Q_s$, but there are no significant qualitative changes. It still exhibits a sign change as a function of $P_{\perp}$ at the point $P_{\perp}=\bar{Q}$ [cf.~Eqs.~\eqref{cos2phi} and \eqref{dL2}], which is independent of $\Delta_{\perp}$. \vspace*{-0.2cm}
\item However for the transverse sector the situation is now somewhat different, since the kinematic correction [cf.~Eq.~\eqref{dT2}], contrary to the leading twist piece, does not vanish when $P_{\perp} = \bar{Q}$. By inspection of the analytical expressions in Appendix \ref{app:nltwist}, one can show that the value of $P_{\perp}$ at which the sign change of $\langle \cos 2 \phi \rangle_{\scriptscriptstyle T}$ happens, increases with $\Delta_{\perp}$. \vspace*{-0.2cm}
\item Furthermore, we observe that  $\langle \cos 2 \phi \rangle_{\scriptscriptstyle T}$ develops a minimum as a function of $\Delta_{\perp}$ when $P_{\perp}>\bar{Q}$. Indeed, while under this condition the leading twist contribution is negative [cf.~Eq.~\eqref{cos2phi}], one can verify that the kinematic correction (cf.~Appendix \ref{app:nltwist}) is positive and increasing with $\Delta_{\perp}$. \vspace*{-0.2cm}
\end{enumerate}
We would like to emphasize that all these features, obtained when $P_{\perp}^2$ is (much) larger than $Q_s^2$, are in very good qualitative and quantitative agreement with the ``all twists'' numerical solution and the related discussion presented in \cite{Mantysaari:2019hkq} (cf.~Fig.~3 there).  

\section{Conclusion and perspectives}
\label{sec:conc} 

In this work we have studied the process of incoherent diffractive dijet production introduced in \cite{Mantysaari:2019hkq} in $eA$ collisions, for which we have provided an analytical insight in the correlation limit, i.e.~our approach is valid in the regime where the momenta $k_{1\perp}$ and $k_{2\perp}$ of the two jets are much larger than both the saturation scale $Q_s$ and the dijet imbalance $\Delta_{\perp}$, the latter being equal to the transverse momentum transferred from the nucleus. No assumption has been made regarding the magnitude of $\Delta_{\perp}$ when compared to $Q_s$, hence inverse powers of $\Delta_{\perp}^2$ are resummed to all orders. Although the cross section is suppressed by $1/N_c^2$ (when, for example, compared to the one for inclusive dijets \cite{Dominguez:2011wm} or for exclusive dijets with the nucleus remaining intact \cite{Salazar:2019ncp}), as realized in \cite{Mantysaari:2019hkq} it provides for a mechanism which leads to a significant dijet imbalance: it can be much larger than the inverse inhomogeneity of the nucleus, for example of the order of $Q_s$. 

At leading twist, we have managed to separate the hard dynamics, i.e.~the dependence on the relative momentum $P_{\perp}$ and $\bar{Q}$, from the semihard one, i.e.~from the dependence on the imbalance $\Delta_{\perp}$. Then, we have computed the first higher kinematic twist, that is the correction of relative order $\Delta_{\perp}^2/P_{\perp}^2$. This led to a rather good analytical description of the azimuthal anisotropies in the extended regime where $\Delta_{\perp}$ may start to be comparable to $P_{\perp}$. Needless to say, it would be desirable to perform a complete resummation of all the $(\Delta_{\perp}^2/P_{\perp}^2)^n$ terms by using more elegant techniques, like the improved TMD approach \cite{Altinoluk:2019wyu}.

We have also seen that at the leading twist level the differential diffractive dijet cross section follows a $1/P_{\perp}^6$ power law (assuming that $P_{\perp}\sim \bar{Q}$). This falloff is much faster than the $1/P_{\perp}^4$ of the inclusive dijet cross section. While here we have calculated only the $q\bar{q}$ contribution, when going to the next to leading order in the strong coupling, we should consider an additional parton, more precisely a gluon, in the wavefunction of the virtual photon. When this gluon is emitted far from the $q\bar{q}$ pair a large gluon-gluon dipole emerges (with one leg being the small $q\bar{q}$ pair which remains in a color-octet state after the gluon emission and the other leg being the emitted gluon itself). Such gluon-gluon dipole scatters strongly with the target nucleus and the resulting $q\bar{q}g$ contribution to the cross section falls like $1/P_{\perp}^4$. This has been thoroughly studied and computed in coherent diffractive dijet production \cite{Iancu:2021rup,Iancu:2022lcw,Hatta:2022lzj} and the same idea should be applicable also to the incoherent process we considered in the present work. 

Finally, we are aware of the fact that there are mechanisms, other than those studied here (and in \cite{Mantysaari:2019hkq}), which have an impact to the azimuthal anisotropies. For example soft gluon final state radiation induces a sizable elliptic anisotropy in dijet events \cite{Hatta:2020bgy,Hatta:2021jcd} and thus should be calculated and combined with the one we have presented here.

\begin{acknowledgements}
	D.N.T.~would like to thank Edmond Iancu and Farid Salazar for valuable discussions. S.Y.W.~is supported by the Taishan fellowship of the Shandong Province for junior scientists.
\end{acknowledgements}

\appendix

\section{The double dipole correlator in the correlation limit}
\label{app:dipdip}

In this Appendix we will compute Eq.~\eqref{wdiff}, i.e., the ``connected'' part of the average of a product of two dipole $S$-matrices in the fundamental representation, to lowest order in $r$ and $\bar{r}$ and for an arbitrary number of colors $N_c$. That is, we assume that $r,\bar{r} \ll B, 1/Q_s$, while there isn't any special constraint between the sizes $B$ and $1/Q_s$. We will work in the Gaussian approximation, which is valid to excellent accuracy even after including JIMWLK evolution \cite{Dumitru:2011vk,Iancu:2011ns,Iancu:2011nj,Alvioli:2012ba}, and we shall further assume that the average dipole $S$-matrix depends only on the magnitude of the dipole separation, that is, $\mcal{S}(\bx,\by) = \mcal{S}(|\bx-\by|)$. Then, in a convenient to our purposes notation, we have \cite{Dominguez:2008aa}
\begin{align}
	\label{wdiffgen}
	\left \langle 
	S(\bm{x},\bm{y}) S(\bar{\bm{y}},\bar{\bm{x}}) 
	\right \rangle
	= \,& 
	\mcal{S}(\bx,\by) \, \mcal{S}(\bar{\by},\bar{\bx})\,
	e^{-\frac{F}{2} + \frac{f}{N_c^2}}
	\nn
	& \times 
	\left[
	\left(
	\frac{\sqrt{D} + F}{2 \sqrt{D}}
	-\frac{f}{N_c^2 \sqrt{D}}
	\right)
	e^{\frac{\sqrt{D}}{2}}
	+\left(
	\frac{\sqrt{D} - F}{2 \sqrt{D}}
	+\frac{f}{N_c^2 \sqrt{D}}
	\right)
	e^{-\frac{\sqrt{D}}{2}}
	\right],
\end{align}
where we have defined
\begin{align}
\hspace*{-0.2cm}
	\label{deltaff}
	\hspace*{-0.1cm}
	D = F^2 - \frac{4}{N_c^2} f (F-f),
	\quad
	F= \frac{N_c}{2 C_F} 
	\ln \frac{\mcal{S}(\bx,\by)\, 
	\mcal{S}(\bar{\by},\bar{\bx})}
	{\mcal{S}(\bx,\bar{\bx})\,
	\mcal{S}(\by,\bar{\by})}
	\quad \mathrm{and}
	\quad
	f= 
	\frac{N_c}{2 C_F}
	\ln \frac{\mcal{S}(\bx,\bar{\by})\, 
	\mcal{S}(\by,\bar{\bx})}
	{\mcal{S}(\bx,\bar{\bx})\,
	\mcal{S}(\by,\bar{\by})}.
\end{align}
Let us start by studying the limit of the various above quantities to lowest order in $r$ and $\bar{r}$. First, recalling that $\bx - \bar{\by} = \bm{B} + \vartheta_2 \br +\vartheta_1 \bar{\br}$, we have
\begin{align}
	\label{lnsexpand}
	\ln \mcal{S}(\bx,\bar{\by}) 
	\simeq
	\ln \mcal{S}(B) 
	+(\vartheta_2 r + \vartheta_1 \bar{r})^i 
	\del^i \ln \mcal{S}(B)
	+\frac{1}{2}\,
	(\vartheta_2 r + \vartheta_1 \bar{r})^i 
	(\vartheta_2 r + \vartheta_1 \bar{r})^j 
	\del^i
	\del^j \ln \mcal{S}(B),
\end{align}
and similarly for the other three dipole correlators appearing in the argument of the logarithm in $f$. Since they all involve one position in the DA and one in the CCA, it is obvious that the leading terms cancel. One easily checks that the terms linear in the dipole size cancel too, and combining the second order terms we get
\begin{align}
	\label{f2nd}
	f \simeq
	\frac{N_c}{2 C_F}\,
	r^i \bar{r}^j\, \del^i \del^j \ln \mcal{S}(B)
	=\frac{1}{2}\,
	r^i \bar{r}^j\, \del^i \del^j \ln \mcal{S}_g(B),
\end{align}
which can be viewed as the generalization of Eq.~\eqref{tttt}. We have expressed the above in terms of the gluon-gluon dipole correlator $\mcal{S}_g(B)$ already introduced in Eq.~\eqref{sgg} in the Gaussian approximation. On the contrary, in the same regime $F$ becomes independent of $r$ and $\bar{r}$, namely
\begin{align}
	\label{Fzeroth}
	F \simeq - 
	\frac{N_c}{2 C_F}
	\ln \mcal{S}^2(B)
	= - \ln \mcal{S}_g(B), 
\end{align}
and we shall see in a moment that it is not necessary to take into account the higher order terms that vanish when $r,\bar{r}\to 0$. Given the limiting behavior in Eqs.~\eqref{f2nd} and \eqref{Fzeroth}, we expand $\sqrt{D}$ to second order in $f$ to obtain
\begin{align}
	\label{sqrtdelta}
	\sqrt{D} \simeq
	F - \frac{2 f}{N_c^2}
	+\frac{4 C_F f^2 }{N_c^3 F}.
\end{align}
Then we return to Eq.~\eqref{wdiffgen} and expand for small $f$ to find that the linear terms cancel, while the quadratic ones give
\begin{align}
	\label{ssmss}
	\left \langle
	S(\bm{x},\bm{y}) S(\bar{\bm{y}},\bar{\bm{x}}) 
	\right \rangle
	\simeq
	\mcal{S}(\bx,\by) \, \mcal{S}(\bar{\by},\bar{\bx})
	\left[
	1 + 
	\frac{2 C_F}{N_c^3}\,
	\frac{f^2}{F^2}
	\left(
	e^{-F} -1 + F
	\right)
	\right].
\end{align}
We see that the correction is already quartic in the small dipole sizes $r$ and $\bar{r}$ as expected and therefore we can safely let $\br=\bar{\br}=0$ when calculating $F$, that is, we can use Eq.~\eqref{Fzeroth}. We also notice that the expansion is safe for any $B$; even when $F$ is small (i.e.~for $B \ll 1/Q_s$), the term in the parenthesis upon expansion becomes of the order of $F^2$, thus cancelling the denominator in the prefactor there. Furthermore, to the order of accuracy, $\mcal{S}(\bx,\by) \, \mcal{S}(\bar{\by},\bar{\bx})$ can be set equal to unity when multiplying the second term in the square bracket. Using Eqs.~\eqref{f2nd} and \eqref{Fzeroth} (in terms of $\mcal{S}_g$) we arrive at the final expression for the correlator defined in Eq.~\eqref{wdiff}
\begin{align}
	\label{wdiffapp}
	\mcal{W}_{\rm D} (\bm{r},\bar{\bm{r}},\bm{B})
	\simeq
	\frac{C_F}{2 N_c^3} 
	\frac{\mcal{S}_g(B) - 1 - 
	\ln \mcal{S}_g(B)}
	{\ln^2\mcal{S}_g(B)}\,
	r^i \bar{r}^j 
	r^k \bar{r}^l\, 
	[\partial^i \partial^j \ln \mcal{S}_g(B)]\,
	[\partial^k \partial^l \ln \mcal{S}_g(B)].
\end{align}
We further point out that, after the trivial replacement $2 C_F/N_c^3 \simeq 1/N_c^2$, Eq.~\eqref{ssmss} is the exact large-$N_c$ limit of Eq.~\eqref{wdiffgen}, i.e.~it is valid for generic dipoles sizes. 

\section{Inclusive dijet production}
\label{app:inclusive}

To assist the discussion in Sect.~\ref{sec:numerics} we wish to review here the cross section for inclusive dijet production \cite{Dominguez:2011wm,Dominguez:2011br}. Our starting expression is Eq.~\eqref{sigmab}, but we must replace $\mcal{W}_{\rm D}$ with [cf.~the discussion after Eq.~\eqref{scat}] 
\begin{align}
	\label{wincl}
	\mcal{W}_{\rm inc} (\br,\bar{\br},\bm{B}) =
	\frac{1}{N_c}\,
	\big \langle
	{\rm tr} \big[V(\bx) V^{\dagger}(\by) 
	V(\bar{\by}) V^{\dagger}(\bar{\bx}) \big]
	-{\rm tr} 
	\big[V(\bx) V^{\dagger}(\by) \big]
	- {\rm tr} 
	\big[ V(\bar{\by}) V^{\dagger}(\bar{\bx}) \big]
	+1
	\big\rangle. 
\end{align}
As usual we calculate the above using the Gaussian approximation. In the correlation limit we single out the semihard dynamics into the gluon TMD which is defined as (using the notation of the present work)
\begin{align}
\label{tensortmd}
	xG^{ij}(\bm{\Delta}) = \frac{C_F}{4\pi^3 \alpha_s}
	\int \frac{\dif^2 \bm{B}}{2\pi}\,e^{- i \bm{\Delta}\cdot \bm{B}}\,
	\frac{1 - \mcal{S}_g(B)}{\ln \mcal{S}_g(B)}\,\del^i \del^j 
	\ln\mcal{S}_g(B)
	 = 
	 \frac{\delta^{ij}}{2}\,xG(\Delta_{\perp}) + 
	 \hat{\Delta}^{ij}\,xh(\Delta_{\perp}).  
\end{align}
For our convenience we have defined the traceless tensor 
\begin{align}
	\label{Deltaij}
	\hat{\Delta}^{ij} 
	\equiv 
	\frac{\Delta^i \Delta^j}{\Delta_{\perp}^2} -
	\frac{\delta^{ij}}{2},
\end{align}
while the unpolarized and polarized unintegrated gluon distribution of the nucleus are given by (see also Eq.~\eqref{ddlogS})	
\begin{align}
	\label{alphaxG}
	xG(\Delta_{\perp}) = 
	\frac{C_F}{4\pi^3 \alpha_s}
	\int_0^\infty \dif BB\, J_0(\Delta_{\perp}B)\,
	\frac{1 - \mcal{S}_g(B)}{\ln \mcal{S}_g(B)}\,
	F_+(B)
\end{align}
and
\begin{align}
	\label{alphaxh}
	xh(\Delta_{\perp}) = 
	-\frac{C_F}{4\pi^3 \alpha_s}
	\int_0^\infty \dif BB\, J_2(\Delta_{\perp}B)\,
	\frac{1 - \mcal{S}_g(B)}{\ln \mcal{S}_g(B)}\,
	F_-(B).
\end{align}
Contracting with the corresponding hard factors one arrives at
\begin{align}
	\label{sigmaincT}
	\frac{\rmd\sigma_{\scriptscriptstyle \rm inc}^{\gamma_{T}^{*} 
	A\rightarrow q\bar q X}}
	{\dif \vartheta_1 \dif 
	\vartheta_2 \rmd^{2} \bm{P} \rmd^{2} \bm{\Delta}} = \,& 
	S_{\perp} \alpha_{\rm em}
	\left(\sum e_{f}^{2}\right)\,
	\delta(1- \vartheta_1 -\vartheta_2)
	\left(\vartheta_1^{2}+ \vartheta_2^{2} \right)
	\nn
	&\times \left[\frac{P_{\perp}^4 + \bar{Q}^4}{(P_{\perp}^2 + \bar{Q}^2)^4}\,
	\alpha_s xG (\Delta_{\perp})
	- \frac{2 \bar{Q}^2 P_{\perp}^2}{(P_{\perp}^2 + \bar{Q}^2)^4}\,
	\alpha_s xh (\Delta_{\perp}) \cos 2 \phi
	\right]
\end{align}
and 
\begin{align}
	\label{sigmaincL}
	\frac{\rmd\sigma_{\scriptscriptstyle \rm inc}^{\gamma_{L}^{*} 
	A\rightarrow q\bar q X}}
	{\dif \vartheta_1 \dif 
	\vartheta_2 \rmd^{2} \bm{P} \rmd^{2} \bm{\Delta}} = \,& 
	S_{\perp} \alpha_{\rm em}
	\left(\sum e_{f}^{2}\right)\,
	\delta(1- \vartheta_1 -\vartheta_2)
	(4 \vartheta_1 \vartheta_2)
	\nn
	&\times \frac{2 \bar{Q}^2 P_{\perp}^2}{(P_{\perp}^2 + \bar{Q}^2)^4}
	\big[
	\alpha_s xG (\Delta_{\perp})
	+
	\alpha_s xh (\Delta_{\perp}) \cos 2 \phi
	\big].
\end{align}

We would like to add that in the standard approach \cite{Mulders:2000sh} the gluon TMD is defined through gauge invariant field correlators, which in general are non-perturbative quantities. In the small-$x$ limit they reduce to correlators of Wilson lines as in here and the two approaches agree with each other \cite{Dominguez:2011wm,Metz:2011wb,Dominguez:2011br}. In the presence of gluon saturation, a dynamical scale $Q_s \gg \Lambda$ is generated and therefore the problem can be addressed with weak coupling methods. Thus the CGC approach adopted in this work provides for a way to calculate such correlators from ``first principles''. This is what has been done in writing the above expressions under the additional assumption of the Gaussian approximation.

\section{The semihard factor and contractions}
\label{app:contract}

Starting from Eq.~\eqref{Gd} (together with Eq.~\eqref{ddlogS}), we find that the semihard factor admits a tensor decomposition, in which the coefficients are given by five different scalar quantities which depend only on the magnitude $\Delta_{\perp}$, namely
\begin{align}
	\label{Gddec}
	\hspace*{-0.4cm}
	\mcal{G}_{\rm D}^{ij,kl}(\bm{\Delta}) = \,& 
	\frac{\delta^{ij}}{2}\,
	\frac{\delta^{kl}}{2}
	\left[
	\mcal{G}_{\rm D}^{(+)}(\Delta_{\perp}) - 
	\frac{1}{2}\,
	\mcal{G}_{\rm D}^{(-)}(\Delta_{\perp})
	\right]
	+ 
	\left(\frac{\delta^{ik}}{2}\,
	\frac{\delta^{jl}}{2}+
	\frac{\delta^{il}}{2}\,
	\frac{\delta^{jk}}{2}
	\right)
	\frac{1}{2}\,
	\mcal{G}_{\rm D}^{(-)}(\Delta_{\perp})
	\nn
	 & + 
	\left(
	\frac{\delta^{ij}}{2} \hat{\Delta}^{kl}
	+\frac{\delta^{kl}}{2} \hat{\Delta}^{ij}
	\right)
	\left[
	\mcal{G}_{\rm D}^{(1)}(\Delta_{\perp}) - 
	\frac{2}{3}\,
	\mcal{G}_{\rm D}^{(1')}(\Delta_{\perp})
	\right]
	\nn
	&+
	\left(
	\frac{\delta^{ik}}{2} \hat{\Delta}^{jl}
	+\frac{\delta^{jl}}{2} \hat{\Delta}^{ik}
	+\frac{\delta^{il}}{2} \hat{\Delta}^{jk}
	+\frac{\delta^{jk}}{2} \hat{\Delta}^{il}
	\right) 
	\frac{1}{3}\,
	\mcal{G}_{\rm D}^{(1')}(\Delta_{\perp})
	\nn
	&+
	\left(
	\hat{\Delta}^{ij} \hat{\Delta}^{kl}
	-\frac{1}{2}\,
	\frac{\delta^{ij}}{2}\,
	\frac{\delta^{kl}}{2}
	+\hat{\Delta}^{ik} \hat{\Delta}^{jl}
	-\frac{1}{2}\,
	\frac{\delta^{ik}}{2}\,
	\frac{\delta^{jl}}{2}
	+\hat{\Delta}^{il} \hat{\Delta}^{jk}
	-\frac{1}{2}\,
	\frac{\delta^{il}}{2}\,
	\frac{\delta^{jk}}{2}
	\right)
	\frac{1}{3}\,
	\mcal{G}_{\rm D}^{(2)}(\Delta_{\perp}).
\end{align}
The traceless tensor $\hat{\Delta}^{ij}$ has already been defined in Eq.~\eqref{Deltaij}, while the four functions $\mcal{G}_{\rm D}^{(+)}(\Delta_{\perp})$, $\mcal{G}_{\rm D}^{(-)}(\Delta_{\perp})$, $\mcal{G}_{\rm D}^{(1)}(\Delta_{\perp})$ and $\mcal{G}_{\rm D}^{(2)}(\Delta_{\perp})$ have been written in the Sect.~\ref{sec:cross-section}. For completeness we also give $\mcal{G}_{\rm D}^{(1')}$ (which cancels when computing the cross section) which reads
\begin{align}
	\label{Gd1prime}
	\mcal{G}_{\rm D}^{(1')}(\Delta_{\perp}) = 
	\int \frac{\rmd^2 \bm{B}}{2\pi}\,
	e^{- i \bm{\Delta} \cdot \bm{B}}\,
	\Phi(B)
	[F_-(B)]^2
	\cos2\phi_{\scriptscriptstyle \Delta B},
\end{align}
with $\phi_{\scriptscriptstyle \Delta B}$ the angle between the vectors $\bm{\Delta}$ and $\bm{B}$. Now we wish to bring the hard factors in a similar form. Defining again the traceless tensor
\begin{align}
	\label{Pij}
	\hat{P}^{ij} 
	\equiv 
	\frac{P^i P^j}{P_{\perp}^2} -
	\frac{\delta^{ij}}{2},
\end{align}
and ``squaring'' the transverse hard factor in Eq.~\eqref{hiks1} we arrive after some algebra at
\begin{align}
\label{HTsquared}
	H_T^{iks}(\bP,\bar{Q})\, 
	H_T^{jls*}(\bP,\bar{Q}) = \,& 
	\frac{16 P^2 (3 \bar{Q}^4 + 2 \bar{Q}^2 P_{\perp}^2 - P_{\perp}^4)}
	{(P_{\perp}^2 + \bar{Q}^2)^6}\, 
	\frac{\delta^{ik}}{2}\, \frac{\delta^{jl}}{2}
	\nn
	& +
	\frac{16 P^2 (\bar{Q}^4 - \bar{Q}^2 P_{\perp}^2 + P_{\perp}^4)}
	{(P_{\perp}^2 + \bar{Q}^2)^6}\, 
	\left(
	\frac{\delta^{ij}}{2}\, \frac{\delta^{kl}}{2}+
	\frac{\delta^{il}}{2}\, \frac{\delta^{kj}}{2}
	\right)
	\nn 
	& +
	\frac{16 P_{\perp}^2 (\bar{Q}^4 - P_{\perp}^4) }
	{(P_{\perp}^2 + \bar{Q}^2)^6}
	\left(\frac{\delta^{ik}}{2} \hat{P}^{jl} 
	+ \frac{\delta^{jl}}{2} \hat{P}^{ik}\right) 
	\nn
	& +
	\frac{8 P_{\perp}^2 (\bar{Q}^2 - P_{\perp}^2)^2 }
	{(P_{\perp}^2 + \bar{Q}^2)^6}
	\left
	(\frac{\delta^{ij}}{2} \hat{P}^{kl} 
	+ \frac{\delta^{kl}}{2} \hat{P}^{ij}
	+\frac{\delta^{il}}{2} \hat{P}^{kj} 
	+ \frac{\delta^{kj}}{2} \hat{P}^{il}
	\right) 
	\nn  
	& -
	\frac{32 \bar{Q}^2 P_{\perp}^4}
	{(P_{\perp}^2 + \bar{Q}^2)^6}
	\left( 
	\hat{P}^{ij} \hat{P}^{kl} 
	-\frac{1}{2}\, \frac{\delta^{ij}}{2}\, \frac{\delta^{kl}}{2}
	+\hat{P}^{il} \hat{P}^{kj} 
	-\frac{1}{2}\, \frac{\delta^{il}}{2}\, \frac{\delta^{kj}}{2}
	\right).  
\end{align}
The longitudinal hard factor is much more straightforward to obtain since there is no index to contract. Starting from Eq.~\eqref{hik} we find
\begin{align}
	\label{HLsquared}
	H_L^{ik}(\bP,\bar{Q})\, 
	H_L^{jl}(\bP,\bar{Q}) = \,& 
	\frac{16 \bar{Q}^2 (\bar{Q}^4 - 2 \bar{Q}^2 P_{\perp}^2 + 3 P_{\perp}^4)}
	{(P_{\perp}^2 + \bar{Q}^2)^6}\, 
	\frac{\delta^{ik}}{2}\, \frac{\delta^{jl}}{2}
	\nn 
	& -
	\frac{32 \bar{Q}^2 P_{\perp}^2 (\bar{Q}^2 - P_{\perp}^2) }
	{(P_{\perp}^2 + \bar{Q}^2)^6}
	\left(\frac{\delta^{ik}}{2} \hat{P}^{jl} 
	+ \frac{\delta^{jl}}{2} \hat{P}^{ik}\right) 
	\nn 
	& +
	\frac{64 \bar{Q}^2 P_{\perp}^4}
	{(P_{\perp}^2 + \bar{Q}^2)^6}
	\left( \hat{P}^{ik} \hat{P}^{jl} 
	-\frac{1}{2}\, \frac{\delta^{ik}}{2}\, \frac{\delta^{jl}}{2}
	\right).  
\end{align} 
Using the formulae 
\begin{align}
	\label{form1}
	{\delta}^{ij} \hat{P}^{ij} = 0
	\,, \qquad
	\hat{P}^{ij} \hat{P}^{ij} = \frac{1}{2}
	\qquad \mathrm{and} \qquad
	\hat{P}^{ij} \hat{P}^{jk} = \frac{1}{4}\,\delta^{ik}
\end{align}
(and similarly for $\hat{\Delta}^{ij}$) and 
\begin{align}
	\label{form2}
	2 \hat{P}^{ij} \hat{\Delta}^{ji} & = 
	2\cos^2\phi - 1
	= \cos2\phi
\\
	8
	\hat{P}^{ij} \hat{\Delta}^{jk} 
	\hat{P}^{kl} \hat{\Delta}^{li} & = 
	8\cos^4\phi - 8\cos^2\phi + 1
	= \cos 4\phi
\end{align}
where $\phi$ is the angle between $\bm{P}$ and $\bm{\Delta}$, we can contract $\mcal{G}_{\rm D}^{ij,kl}$ in Eq.~\eqref{Gddec} with either $H_T^{iks} 
	H_T^{jls*}$ as expressed in Eq.~\eqref{HTsquared} or $H_L^{ik}H_L^{jl}$ as expressed in Eq.~\eqref{HLsquared} to get Eqs.~\eqref{ad2final} and \eqref{ad2lfinal}.

\section{Next to leading twist}
\label{app:nltwist}

Here we would like to refer to a few technical details regarding the calculation of the next to leading kinematic twist. Taking $\vartheta_1=\vartheta_2$ for simplicity, we expand $f$, which is defined in Eq.~\eqref{deltaff}, to 4th order in the small dipole size. Inserting it into Eq.~\eqref{ssmss} we arrive at Eqs.~\eqref{ad2cort} and \eqref{ad2corl}. The hard tensors $H_T^{ikmns}(\bP,\bar{Q})$ and $H_L^{ikmn}(\bP,\bar{Q})$ are the obvious extensions of $H_T^{iks}(\bP,\bar{Q})$ and $H_L^{ik}(\bP,\bar{Q})$ respectively: they just contain an additional $r^m r^n$ factor in the integrand. These rank-5 and rank-3 tensors are easily calculated, for example by simply taking the two partial derivatives w.r.t.~$P^m$ and $P^n$ (and multiplying with $i^2$) of the corresponding rank-3 and rank-2 tensors which are already given in Eqs.~\eqref{hiks} and \eqref{hik}. The tensor decomposition of the semihard factor in Eq.~\eqref{Gdcor} can also be achieved, although it is cumbersome and requires careful bookkeeping since the $S$-matrix $\mcal{S}_{g}(B)$ appearing in the integrand is generally not specified. Finally one performs the contractions of all the indices in Eqs.~\eqref{ad2cort} and \eqref{ad2corl}. 

In the following we shall give the results of the calculation for the situation that the scattering is described by the MV model. In such a case the integration in the semihard sector reduces to elementary integrations over the adjoint dipole size $B$. Having in mind the expression in Eq.~\eqref{Ns}, we first present the corrections to the averaged over angle reduced cross sections. For the transverse sector we have
\begin{align}
	\label{dTave}
	\delta \mcal{N}_{0}^T
	\simeq
	-\frac{16 P_{\perp}^2 \bar{Q}^2 (7 \bar{Q}^4 -7 \bar{Q}^2 P_{\perp}^2 + 2 P_{\perp}^4)}
	{(P_{\perp}^2 + \bar{Q}^2)^8}\, 	
	Q_A^4 
	\int \frac{\dif B} {B}\, 
	J_0(\Delta_{\perp} B) \Phi(B)
\end{align}
while for the longitudinal one
\begin{align}
	\label{dLave}
	\delta \mcal{N}_{0}^L
	\simeq
	-\frac{16 \bar{Q}^2 (\bar{Q}^6 -5 \bar{Q}^4 P_{\perp}^2 + 8 \bar{Q}^2 P_{\perp}^4 - 2 P_{\perp}^6)}
	{(P_{\perp}^2 + \bar{Q}^2)^8}\, 	
	Q_A^4 
	\int \frac{\dif B} {B}\, 
	J_0(\Delta_{\perp} B) \Phi(B).
	\end{align}
We immediately see that the above integrations diverge logarithmically at small $B$. However, we recall that our expansion scheme requires $r \ll B$, and therefore a UV cutoff equal to $c_0/P_{\perp}$ is assumed in Eqs.~\eqref{dTave} and \eqref{dLave}. The constant $c_0$ is of the order of one and we have checked that our results are not sensitive to its precise choice.

Regarding the corrections to the numerator of the $\langle \cos 2\phi \rangle_{\lambda}$ anisotropies we find for the transverse sector
\begin{align}
	\label{dT2}
	\hspace*{-0.5cm}
	\delta \mcal{N}_{2}^T = 
	\frac{8 P_{\perp}^2}{(P_{\perp}^2 + \bar{Q}^2)^8}\, 	
	Q_A^4
	\bigg[ 
	&(12 \bar{Q}^6 - 34 \bar{Q}^4 P_{\perp}^2 + 6 \bar{Q}^2 P_{\perp}^4 - 4 P_{\perp}^6)
	\int \frac{\dif B} {B}\, 
	J_2(\Delta_{\perp} B) \Phi(B)
	\nn
	& -(7 \bar{Q}^6 - 20 \bar{Q}^4 P_{\perp}^2 + 3 \bar{Q}^2 P_{\perp}^4 - 2 P_{\perp}^6)
	\int \frac{\dif B} {B}\, 
	J_2(\Delta_{\perp} B) \Phi(B) \ln \frac{4}{B^2 \Lambda^2}
	\bigg]
\end{align}
and for the longitudinal one
\begin{align}
	\label{dL2}
	\delta \mcal{N}_{2}^L
	= \frac{8 P_{\perp}^2 \bar{Q}^2 (\bar{Q}^2 -  P_{\perp}^2)}{(P_{\perp}^2 + \bar{Q}^2)^8}\, 	
	Q_A^4
	\bigg[ 
	&(15 P_{\perp}^2 - 13 \bar{Q}^2)
	\int \frac{\dif B} {B}\, 
	J_2(\Delta_{\perp} B) \Phi(B)
	\nn
	& +8(\bar{Q}^2 -  P_{\perp}^2)
	\int \frac{\dif B} {B}\, 
	J_2(\Delta_{\perp} B) \Phi(B) \ln \frac{4}{B^2 \Lambda^2}
	\bigg].
\end{align}
The above integrations are UV finite since the Bessel function $J_2(\Delta_{\perp} B)$ vanishes quadratically for small argument. We notice that now only the longitudinal anisotropy vanishes when  $P_{\perp} = \bar{Q}$. Similar expressions can be given for the remaining non-vanishing quantities $\delta \mcal{N}_{4}^\lambda$ and $\delta \mcal{N}_{6}^\lambda$ which contribute to the corresponding  $\langle \cos 4\phi \rangle_{\lambda}$ and $\langle \cos 6\phi \rangle_{\lambda}$ anisotropies.


\begin{thebibliography}{10}

\bibitem{Mantysaari:2019hkq}
H.~M\"antysaari, N.~Mueller, F.~Salazar, and B.~Schenke, ``{Multigluon
  Correlations and Evidence of Saturation from Dijet Measurements at an
  Electron-Ion Collider},''
  \href{http://dx.doi.org/10.1103/PhysRevLett.124.112301}{{\em Phys. Rev.
  Lett.} {\bf 124} (2020) no.~11, 112301},
  \href{http://arxiv.org/abs/1912.05586}{{\tt arXiv:1912.05586 [nucl-th]}}.

\bibitem{Lipatov:1976zz}
L.~Lipatov, ``{Reggeization of the Vector Meson and the Vacuum Singularity in
  Nonabelian Gauge Theories},''
{\em Sov.J.Nucl.Phys.} {\bf 23} (1976)  338--345.

\bibitem{Kuraev:1977fs}
E.~Kuraev, L.~Lipatov, and V.~S. Fadin, ``{The Pomeranchuk Singularity in
  Nonabelian Gauge Theories},''
{\em Sov.Phys.JETP} {\bf 45} (1977)  199--204.

\bibitem{Balitsky:1978ic}
I.~Balitsky and L.~Lipatov, ``{The Pomeranchuk Singularity in Quantum
  Chromodynamics},''
{\em Sov.J.Nucl.Phys.} {\bf 28} (1978)  822--829.

\bibitem{Mueller:1993rr}
A.~H. Mueller, ``{Soft gluons in the infinite momentum wave function and the
  BFKL pomeron},''
\href{http://dx.doi.org/10.1016/0550-3213(94)90116-3}{{\em Nucl.Phys.} {\bf
  B415} (1994)  373--385}.

\bibitem{Gribov:1984tu}
L.~Gribov, E.~Levin, and M.~Ryskin, ``{Semihard Processes in QCD},''
\href{http://dx.doi.org/10.1016/0370-1573(83)90022-4}{{\em Phys.Rept.} {\bf
  100} (1983)  1--150}.

\bibitem{Mueller:1985wy}
A.~H. Mueller and J.-w. Qiu, ``{Gluon Recombination and Shadowing at Small
  Values of x},'' \href{http://dx.doi.org/10.1016/0550-3213(86)90164-1}{{\em
  Nucl. Phys. B} {\bf 268} (1986)  427--452}.

\bibitem{Mueller:2001fv}
A.~H. Mueller, ``{Parton saturation: An Overview},''
\href{http://arxiv.org/abs/hep-ph/0111244}{{\tt arXiv:hep-ph/0111244
  [hep-ph]}}.

\bibitem{Iancu:2003xm}
E.~Iancu and R.~Venugopalan, ``{The color glass condensate and high energy
  scattering in QCD},''
\href{http://arxiv.org/abs/hep-ph/0303204}{{\tt arXiv:hep-ph/0303204}}.

\bibitem{Gelis:2010nm}
F.~Gelis, E.~Iancu, J.~Jalilian-Marian, and R.~Venugopalan, ``{The Color Glass
  Condensate},''
  \href{http://dx.doi.org/10.1146/annurev.nucl.010909.083629}{{\em
  Ann.Rev.Nucl.Part.Sci.} {\bf 60} (2010)  463--489},
\href{http://arxiv.org/abs/1002.0333}{{\tt arXiv:1002.0333 [hep-ph]}}.

\bibitem{Kovchegov:2012mbw}
Y.~V. Kovchegov and E.~Levin, {\em {Quantum chromodynamics at high energy}}.
\newblock {Cambridge University Press},
2012.
\newblock

\bibitem{Balitsky:1995ub}
I.~Balitsky, ``{Operator expansion for high-energy scattering},''
  \href{http://dx.doi.org/10.1016/0550-3213(95)00638-9}{{\em Nucl. Phys.} {\bf
  B463} (1996)  99--160},
\href{http://arxiv.org/abs/hep-ph/9509348}{{\tt arXiv:hep-ph/9509348}}.

\bibitem{Kovchegov:1999yj}
Y.~V. Kovchegov, ``{Small-x F2 structure function of a nucleus including
  multiple pomeron exchanges},''
  \href{http://dx.doi.org/10.1103/PhysRevD.60.034008}{{\em Phys. Rev.} {\bf
  D60} (1999)  034008},
\href{http://arxiv.org/abs/hep-ph/9901281}{{\tt arXiv:hep-ph/9901281}}.

\bibitem{JalilianMarian:1997jx}
J.~Jalilian-Marian, A.~Kovner, A.~Leonidov, and H.~Weigert, ``{The BFKL
  equation from the Wilson renormalization group},''
  \href{http://dx.doi.org/10.1016/S0550-3213(97)00440-9}{{\em Nucl. Phys.} {\bf
  B504} (1997)  415--431},
\href{http://arxiv.org/abs/hep-ph/9701284}{{\tt arXiv:hep-ph/9701284}}.

\bibitem{JalilianMarian:1997gr}
J.~Jalilian-Marian, A.~Kovner, A.~Leonidov, and H.~Weigert, ``{The Wilson
  renormalization group for low x physics: Towards the high density regime},''
  \href{http://dx.doi.org/10.1103/PhysRevD.59.014014}{{\em Phys.Rev.} {\bf D59}
  (1998)  014014},
\href{http://arxiv.org/abs/hep-ph/9706377}{{\tt arXiv:hep-ph/9706377
  [hep-ph]}}.

\bibitem{Kovner:2000pt}
A.~Kovner, J.~G. Milhano, and H.~Weigert, ``{Relating different approaches to
  nonlinear QCD evolution at finite gluon density},''
  \href{http://dx.doi.org/10.1103/PhysRevD.62.114005}{{\em Phys. Rev.} {\bf
  D62} (2000)  114005},
\href{http://arxiv.org/abs/hep-ph/0004014}{{\tt arXiv:hep-ph/0004014}}.

\bibitem{Weigert:2000gi}
H.~Weigert, ``{Unitarity at small Bjorken x},''
  \href{http://dx.doi.org/10.1016/S0375-9474(01)01668-2}{{\em Nucl. Phys.} {\bf
  A703} (2002)  823--860},
\href{http://arxiv.org/abs/hep-ph/0004044}{{\tt arXiv:hep-ph/0004044}}.

\bibitem{Iancu:2000hn}
E.~Iancu, A.~Leonidov, and L.~D. McLerran, ``{Nonlinear gluon evolution in the
  color glass condensate. I},''
  \href{http://dx.doi.org/10.1016/S0375-9474(01)00642-X}{{\em Nucl. Phys.} {\bf
  A692} (2001)  583--645},
\href{http://arxiv.org/abs/hep-ph/0011241}{{\tt arXiv:hep-ph/0011241}}.

\bibitem{Iancu:2001ad}
E.~Iancu, A.~Leonidov, and L.~D. McLerran, ``{The renormalization group
  equation for the color glass condensate},''
  \href{http://dx.doi.org/10.1016/S0370-2693(01)00524-X}{{\em Phys. Lett.} {\bf
  B510} (2001)  133--144},
\href{http://arxiv.org/abs/hep-ph/0102009}{{\tt arXiv:hep-ph/0102009}}.

\bibitem{Ferreiro:2001qy}
E.~Ferreiro, E.~Iancu, A.~Leonidov, and L.~McLerran, ``{Nonlinear gluon
  evolution in the color glass condensate. II},''
  \href{http://dx.doi.org/10.1016/S0375-9474(01)01329-X}{{\em Nucl. Phys.} {\bf
  A703} (2002)  489--538},
\href{http://arxiv.org/abs/hep-ph/0109115}{{\tt arXiv:hep-ph/0109115}}.

\bibitem{Marquet:2007vb}
C.~Marquet, ``{Forward inclusive dijet production and azimuthal correlations in
  pA collisions},''
  \href{http://dx.doi.org/10.1016/j.nuclphysa.2007.09.001}{{\em Nucl. Phys.}
  {\bf A796} (2007)  41--60},
\href{http://arxiv.org/abs/0708.0231}{{\tt arXiv:0708.0231 [hep-ph]}}.

\bibitem{Dominguez:2011wm}
F.~Dominguez, C.~Marquet, B.-W. Xiao, and F.~Yuan, ``{Universality of
  Unintegrated Gluon Distributions at small x},''
  \href{http://dx.doi.org/10.1103/PhysRevD.83.105005}{{\em Phys. Rev.} {\bf
  D83} (2011)  105005},
\href{http://arxiv.org/abs/1101.0715}{{\tt arXiv:1101.0715 [hep-ph]}}.

\bibitem{Metz:2011wb}
A.~Metz and J.~Zhou, ``{Distribution of linearly polarized gluons inside a
  large nucleus},'' \href{http://dx.doi.org/10.1103/PhysRevD.84.051503}{{\em
  Phys. Rev. D} {\bf 84} (2011)  051503},
  \href{http://arxiv.org/abs/1105.1991}{{\tt arXiv:1105.1991 [hep-ph]}}.

\bibitem{Dominguez:2011br}
F.~Dominguez, J.-W. Qiu, B.-W. Xiao, and F.~Yuan, ``{On the linearly polarized
  gluon distributions in the color dipole model},''
  \href{http://dx.doi.org/10.1103/PhysRevD.85.045003}{{\em Phys. Rev. D} {\bf
  85} (2012)  045003}, \href{http://arxiv.org/abs/1109.6293}{{\tt
  arXiv:1109.6293 [hep-ph]}}.

\bibitem{Mueller:2013wwa}
A.~Mueller, B.-W. Xiao, and F.~Yuan, ``{Sudakov double logarithms resummation
  in hard processes in the small-x saturation formalism},''
  \href{http://dx.doi.org/10.1103/PhysRevD.88.114010}{{\em Phys. Rev. D} {\bf
  88} (2013) no.~11, 114010}, \href{http://arxiv.org/abs/1308.2993}{{\tt
  arXiv:1308.2993 [hep-ph]}}.

\bibitem{Kotko:2015ura}
P.~Kotko, K.~Kutak, C.~Marquet, E.~Petreska, S.~Sapeta, and A.~van Hameren,
  ``{Improved TMD factorization for forward dijet production in dilute-dense
  hadronic collisions},'' \href{http://dx.doi.org/10.1007/JHEP09(2015)106}{{\em
  JHEP} {\bf 09} (2015)  106},
\href{http://arxiv.org/abs/1503.03421}{{\tt arXiv:1503.03421 [hep-ph]}}.

\bibitem{Dumitru:2015gaa}
A.~Dumitru, T.~Lappi, and V.~Skokov, ``{Distribution of Linearly Polarized
  Gluons and Elliptic Azimuthal Anisotropy in Deep Inelastic Scattering Dijet
  Production at High Energy},''
  \href{http://dx.doi.org/10.1103/PhysRevLett.115.252301}{{\em Phys. Rev.
  Lett.} {\bf 115} (2015) no.~25, 252301},
  \href{http://arxiv.org/abs/1508.04438}{{\tt arXiv:1508.04438 [hep-ph]}}.

\bibitem{vanHameren:2016ftb}
A.~van Hameren, P.~Kotko, K.~Kutak, C.~Marquet, E.~Petreska, and S.~Sapeta,
  ``{Forward di-jet production in p+Pb collisions in the small-x improved TMD
  factorization framework},''
  \href{http://dx.doi.org/10.1007/JHEP12(2016)034}{{\em JHEP} {\bf 12} (2016)
  034},
\href{http://arxiv.org/abs/1607.03121}{{\tt arXiv:1607.03121 [hep-ph]}}.

\bibitem{Marquet:2016cgx}
C.~Marquet, E.~Petreska, and C.~Roiesnel, ``{Transverse-momentum-dependent
  gluon distributions from JIMWLK evolution},''
  \href{http://dx.doi.org/10.1007/JHEP10(2016)065}{{\em JHEP} {\bf 10} (2016)
  065},
\href{http://arxiv.org/abs/1608.02577}{{\tt arXiv:1608.02577 [hep-ph]}}.

\bibitem{Kotko:2017oxg}
P.~Kotko, K.~Kutak, S.~Sapeta, A.~M. Stasto, and M.~Strikman, ``{Estimating
  nonlinear effects in forward dijet production in ultra-peripheral heavy ion
  collisions at the LHC},''
  \href{http://dx.doi.org/10.1140/epjc/s10052-017-4906-6}{{\em Eur. Phys. J. C}
  {\bf 77} (2017) no.~5, 353}, \href{http://arxiv.org/abs/1702.03063}{{\tt
  arXiv:1702.03063 [hep-ph]}}.

\bibitem{Dumitru:2018kuw}
A.~Dumitru, V.~Skokov, and T.~Ullrich, ``{Measuring the Weizs\"acker-Williams
  distribution of linearly polarized gluons at an electron-ion collider through
  dijet azimuthal asymmetries},''
  \href{http://dx.doi.org/10.1103/PhysRevC.99.015204}{{\em Phys. Rev. C} {\bf
  99} (2019) no.~1, 015204}, \href{http://arxiv.org/abs/1809.02615}{{\tt
  arXiv:1809.02615 [hep-ph]}}.

\bibitem{Iancu:2018hwa}
E.~Iancu and Y.~Mulian, ``{Forward trijet production in
  proton\textendash{}nucleus collisions},''
  \href{http://dx.doi.org/10.1016/j.nuclphysa.2019.02.003}{{\em Nucl. Phys. A}
  {\bf 985} (2019)  66--127}, \href{http://arxiv.org/abs/1809.05526}{{\tt
  arXiv:1809.05526 [hep-ph]}}.

\bibitem{Klein:2019qfb}
S.~R. Klein and H.~M\"antysaari, ``{Imaging the nucleus with high-energy
  photons},'' \href{http://dx.doi.org/10.1038/s42254-019-0107-6}{{\em Nature
  Rev. Phys.} {\bf 1} (2019) no.~11, 662--674},
  \href{http://arxiv.org/abs/1910.10858}{{\tt arXiv:1910.10858 [hep-ex]}}.

\bibitem{Iancu:2020mos}
E.~Iancu and Y.~Mulian, ``{Forward dijets in proton-nucleus collisions at
  next-to-leading order: the real corrections},''
  \href{http://dx.doi.org/10.1007/JHEP03(2021)005}{{\em JHEP} {\bf 03} (2021)
  005}, \href{http://arxiv.org/abs/2009.11930}{{\tt arXiv:2009.11930
  [hep-ph]}}.

\bibitem{Hatta:2021jcd}
Y.~Hatta, B.-W. Xiao, F.~Yuan, and J.~Zhou, ``{Azimuthal angular asymmetry of
  soft gluon radiation in jet production},''
  \href{http://dx.doi.org/10.1103/PhysRevD.104.054037}{{\em Phys. Rev. D} {\bf
  104} (2021) no.~5, 054037}, \href{http://arxiv.org/abs/2106.05307}{{\tt
  arXiv:2106.05307 [hep-ph]}}.

\bibitem{Boussarie:2021ybe}
R.~Boussarie, H.~M\"antysaari, F.~Salazar, and B.~Schenke, ``{The importance of
  kinematic twists and genuine saturation effects in dijet production at the
  Electron-Ion Collider},''
  \href{http://dx.doi.org/10.1007/JHEP09(2021)178}{{\em JHEP} {\bf 09} (2021)
  178}, \href{http://arxiv.org/abs/2106.11301}{{\tt arXiv:2106.11301
  [hep-ph]}}.

\bibitem{Caucal:2021ent}
P.~Caucal, F.~Salazar, and R.~Venugopalan, ``{Dijet impact factor in DIS at
  next-to-leading order in the Color Glass Condensate},''
  \href{http://dx.doi.org/10.1007/JHEP11(2021)222}{{\em JHEP} {\bf 11} (2021)
  222}, \href{http://arxiv.org/abs/2108.06347}{{\tt arXiv:2108.06347
  [hep-ph]}}.

\bibitem{Taels:2022tza}
P.~Taels, T.~Altinoluk, G.~Beuf, and C.~Marquet, ``{Dijet photoproduction at
  low x at next-to-leading order and its back-to-back limit},''
  \href{http://dx.doi.org/10.1007/JHEP10(2022)184}{{\em JHEP} {\bf 10} (2022)
  184}, \href{http://arxiv.org/abs/2204.11650}{{\tt arXiv:2204.11650
  [hep-ph]}}.

\bibitem{Albacete:2010pg}
J.~L. Albacete and C.~Marquet, ``{Azimuthal correlations of forward di-hadrons
  in d+Au collisions at RHIC in the Color Glass Condensate},''
  \href{http://dx.doi.org/10.1103/PhysRevLett.105.162301}{{\em Phys. Rev.
  Lett.} {\bf 105} (2010)  162301},
\href{http://arxiv.org/abs/1005.4065}{{\tt arXiv:1005.4065 [hep-ph]}}.

\bibitem{Stasto:2011ru}
A.~Stasto, B.-W. Xiao, and F.~Yuan, ``{Back-to-Back Correlations of Di-hadrons
  in dAu Collisions at RHIC},''
  \href{http://dx.doi.org/10.1016/j.physletb.2012.08.044}{{\em Phys.Lett.} {\bf
  B716} (2012)  430--434},
\href{http://arxiv.org/abs/1109.1817}{{\tt arXiv:1109.1817 [hep-ph]}}.

\bibitem{Lappi:2012nh}
T.~Lappi and H.~M{\"a}ntysaari, ``{Forward dihadron correlations in
  deuteron-gold collisions with the Gaussian approximation of JIMWLK},''
  \href{http://dx.doi.org/10.1016/j.nuclphysa.2013.03.017}{{\em Nucl.Phys.}
  {\bf A908} (2013)  51--72},
\href{http://arxiv.org/abs/1209.2853}{{\tt arXiv:1209.2853 [hep-ph]}}.

\bibitem{Iancu:2013dta}
E.~Iancu and J.~Laidet, ``{Gluon splitting in a shockwave},''
  \href{http://dx.doi.org/10.1016/j.nuclphysa.2013.07.012}{{\em Nucl.Phys.}
  {\bf A916} (2013)  48--78},
\href{http://arxiv.org/abs/1305.5926}{{\tt arXiv:1305.5926 [hep-ph]}}.

\bibitem{Zheng:2014vka}
L.~Zheng, E.~Aschenauer, J.~Lee, and B.-W. Xiao, ``{Probing Gluon Saturation
  through Dihadron Correlations at an Electron-Ion Collider},''
  \href{http://dx.doi.org/10.1103/PhysRevD.89.074037}{{\em Phys. Rev. D} {\bf
  89} (2014) no.~7, 074037}, \href{http://arxiv.org/abs/1403.2413}{{\tt
  arXiv:1403.2413 [hep-ph]}}.

\bibitem{Marquet:2017xwy}
C.~Marquet, C.~Roiesnel, and P.~Taels, ``{Linearly polarized small-$x$ gluons
  in forward heavy-quark pair production},''
  \href{http://dx.doi.org/10.1103/PhysRevD.97.014004}{{\em Phys. Rev. D} {\bf
  97} (2018) no.~1, 014004}, \href{http://arxiv.org/abs/1710.05698}{{\tt
  arXiv:1710.05698 [hep-ph]}}.

\bibitem{Albacete:2018ruq}
J.~L. Albacete, G.~Giacalone, C.~Marquet, and M.~Matas, ``{Forward dihadron
  back-to-back correlations in $pA$ collisions},''
  \href{http://dx.doi.org/10.1103/PhysRevD.99.014002}{{\em Phys. Rev. D} {\bf
  99} (2019) no.~1, 014002}, \href{http://arxiv.org/abs/1805.05711}{{\tt
  arXiv:1805.05711 [hep-ph]}}.

\bibitem{Bergabo:2022tcu}
F.~Bergabo and J.~Jalilian-Marian, ``{One-loop corrections to dihadron
  production in DIS at small x},''
  \href{http://dx.doi.org/10.1103/PhysRevD.106.054035}{{\em Phys. Rev. D} {\bf
  106} (2022) no.~5, 054035}, \href{http://arxiv.org/abs/2207.03606}{{\tt
  arXiv:2207.03606 [hep-ph]}}.

\bibitem{Bergabo:2023wed}
F.~Bergabo and J.~Jalilian-Marian, ``{Dihadron Production in DIS at Small $x$
  at Next to Leading Order: Transverse Photons},''
  \href{http://arxiv.org/abs/2301.03117}{{\tt arXiv:2301.03117 [hep-ph]}}.

\bibitem{Bartels:1999tn}
J.~Bartels, H.~Jung, and M.~Wusthoff, ``{Quark - anti-quark gluon jets in DIS
  diffractive dissociation},''
  \href{http://dx.doi.org/10.1007/s100520050618}{{\em Eur. Phys. J. C} {\bf 11}
  (1999)  111--125}, \href{http://arxiv.org/abs/hep-ph/9903265}{{\tt
  arXiv:hep-ph/9903265}}.

\bibitem{Altinoluk:2015dpi}
T.~Altinoluk, N.~Armesto, G.~Beuf, and A.~H. Rezaeian, ``{Diffractive Dijet
  Production in Deep Inelastic Scattering and Photon-Hadron Collisions in the
  Color Glass Condensate},''
  \href{http://dx.doi.org/10.1016/j.physletb.2016.05.032}{{\em Phys. Lett. B}
  {\bf 758} (2016)  373--383}, \href{http://arxiv.org/abs/1511.07452}{{\tt
  arXiv:1511.07452 [hep-ph]}}.

\bibitem{Hatta:2016dxp}
Y.~Hatta, B.-W. Xiao, and F.~Yuan, ``{Probing the Small- x Gluon Tomography in
  Correlated Hard Diffractive Dijet Production in Deep Inelastic Scattering},''
  \href{http://dx.doi.org/10.1103/PhysRevLett.116.202301}{{\em Phys. Rev.
  Lett.} {\bf 116} (2016) no.~20, 202301},
  \href{http://arxiv.org/abs/1601.01585}{{\tt arXiv:1601.01585 [hep-ph]}}.

\bibitem{Hagiwara:2017fye}
Y.~Hagiwara, Y.~Hatta, R.~Pasechnik, M.~Tasevsky, and O.~Teryaev, ``{Accessing
  the gluon Wigner distribution in ultraperipheral $pA$ collisions},''
  \href{http://dx.doi.org/10.1103/PhysRevD.96.034009}{{\em Phys. Rev. D} {\bf
  96} (2017) no.~3, 034009}, \href{http://arxiv.org/abs/1706.01765}{{\tt
  arXiv:1706.01765 [hep-ph]}}.

\bibitem{Mantysaari:2019csc}
H.~M\"antysaari, N.~Mueller, and B.~Schenke, ``{Diffractive Dijet Production
  and Wigner Distributions from the Color Glass Condensate},''
  \href{http://dx.doi.org/10.1103/PhysRevD.99.074004}{{\em Phys. Rev. D} {\bf
  99} (2019) no.~7, 074004}, \href{http://arxiv.org/abs/1902.05087}{{\tt
  arXiv:1902.05087 [hep-ph]}}.

\bibitem{Salazar:2019ncp}
F.~Salazar and B.~Schenke, ``{Diffractive dijet production in impact parameter
  dependent saturation models},''
  \href{http://dx.doi.org/10.1103/PhysRevD.100.034007}{{\em Phys. Rev. D} {\bf
  100} (2019) no.~3, 034007}, \href{http://arxiv.org/abs/1905.03763}{{\tt
  arXiv:1905.03763 [hep-ph]}}.

\bibitem{Iancu:2021rup}
E.~Iancu, A.~H. Mueller, and D.~N. Triantafyllopoulos, ``{Probing Parton
  Saturation and the Gluon Dipole via Diffractive Jet Production at the
  Electron-Ion Collider},''
  \href{http://dx.doi.org/10.1103/PhysRevLett.128.202001}{{\em Phys. Rev.
  Lett.} {\bf 128} (2022) no.~20, 202001},
  \href{http://arxiv.org/abs/2112.06353}{{\tt arXiv:2112.06353 [hep-ph]}}.

\bibitem{Hatta:2022lzj}
Y.~Hatta, B.-W. Xiao, and F.~Yuan, ``{Semi-inclusive diffractive deep inelastic
  scattering at small x},''
  \href{http://dx.doi.org/10.1103/PhysRevD.106.094015}{{\em Phys. Rev. D} {\bf
  106} (2022) no.~9, 094015}, \href{http://arxiv.org/abs/2205.08060}{{\tt
  arXiv:2205.08060 [hep-ph]}}.

\bibitem{Beuf:2022kyp}
G.~Beuf, H.~H\"anninen, T.~Lappi, Y.~Mulian, and H.~M\"antysaari,
  ``{Diffractive deep inelastic scattering at NLO in the dipole picture: The
  $q\bar{q}g$ contribution},''
  \href{http://dx.doi.org/10.1103/PhysRevD.106.094014}{{\em Phys. Rev. D} {\bf
  106} (2022) no.~9, 094014}, \href{http://arxiv.org/abs/2206.13161}{{\tt
  arXiv:2206.13161 [hep-ph]}}.

\bibitem{Iancu:2022lcw}
E.~Iancu, A.~H. Mueller, D.~N. Triantafyllopoulos, and S.~Y. Wei, ``{Gluon
  dipole factorisation for diffractive dijets},''
  \href{http://dx.doi.org/10.1007/JHEP10(2022)103}{{\em JHEP} {\bf 10} (2022)
  103}, \href{http://arxiv.org/abs/2207.06268}{{\tt arXiv:2207.06268
  [hep-ph]}}.

\bibitem{Good:1960ba}
M.~L. Good and W.~D. Walker, ``{Diffraction disssociation of beam particles},''
  \href{http://dx.doi.org/10.1103/PhysRev.120.1857}{{\em Phys. Rev.} {\bf 120}
  (1960)  1857--1860}.

\bibitem{Mantysaari:2016ykx}
H.~M\"antysaari and B.~Schenke, ``{Evidence of strong proton shape fluctuations
  from incoherent diffraction},''
  \href{http://dx.doi.org/10.1103/PhysRevLett.117.052301}{{\em Phys. Rev.
  Lett.} {\bf 117} (2016) no.~5, 052301},
  \href{http://arxiv.org/abs/1603.04349}{{\tt arXiv:1603.04349 [hep-ph]}}.

\bibitem{Mantysaari:2016jaz}
H.~M\"antysaari and B.~Schenke, ``{Revealing proton shape fluctuations with
  incoherent diffraction at high energy},''
  \href{http://dx.doi.org/10.1103/PhysRevD.94.034042}{{\em Phys. Rev. D} {\bf
  94} (2016) no.~3, 034042}, \href{http://arxiv.org/abs/1607.01711}{{\tt
  arXiv:1607.01711 [hep-ph]}}.

\bibitem{Demirci:2022wuy}
S.~Demirci, T.~Lappi, and S.~Schlichting, ``{Proton hot spots and exclusive
  vector meson production},''
  \href{http://dx.doi.org/10.1103/PhysRevD.106.074025}{{\em Phys. Rev. D} {\bf
  106} (2022) no.~7, 074025}, \href{http://arxiv.org/abs/2206.05207}{{\tt
  arXiv:2206.05207 [hep-ph]}}.

\bibitem{Marquet:2010cf}
C.~Marquet and H.~Weigert, ``{New observables to test the Color Glass
  Condensate beyond the large- $N_c$ limit},''
  \href{http://dx.doi.org/10.1016/j.nuclphysa.2010.05.056}{{\em Nucl. Phys. A}
  {\bf 843} (2010)  68--97}, \href{http://arxiv.org/abs/1003.0813}{{\tt
  arXiv:1003.0813 [hep-ph]}}.

\bibitem{Iancu:2022gpw}
E.~Iancu and Y.~Mulian, ``{Dihadron production in DIS at NLO: the real
  corrections},'' \href{http://arxiv.org/abs/2211.04837}{{\tt arXiv:2211.04837
  [hep-ph]}}.

\bibitem{McLerran:1993ka}
L.~D. McLerran and R.~Venugopalan, ``{Gluon distribution functions for very
  large nuclei at small transverse momentum},''
  \href{http://dx.doi.org/10.1103/PhysRevD.49.3352}{{\em Phys. Rev.} {\bf D49}
  (1994)  3352--3355},
\href{http://arxiv.org/abs/hep-ph/9311205}{{\tt arXiv:hep-ph/9311205}}.

\bibitem{McLerran:1993ni}
L.~D. McLerran and R.~Venugopalan, ``{Computing quark and gluon distribution
  functions for very large nuclei},''
  \href{http://dx.doi.org/10.1103/PhysRevD.49.2233}{{\em Phys. Rev.} {\bf D49}
  (1994)  2233--2241},
\href{http://arxiv.org/abs/hep-ph/9309289}{{\tt arXiv:hep-ph/9309289}}.

\bibitem{Dumitru:2011vk}
A.~Dumitru, J.~Jalilian-Marian, T.~Lappi, B.~Schenke, and R.~Venugopalan,
  ``{Renormalization group evolution of multi-gluon correlators in high energy
  QCD},'' \href{http://dx.doi.org/10.1016/j.physletb.2011.11.002}{{\em
  Phys.Lett.} {\bf B706} (2011)  219--224},
\href{http://arxiv.org/abs/1108.4764}{{\tt arXiv:1108.4764 [hep-ph]}}.

\bibitem{Iancu:2011ns}
E.~Iancu and D.~Triantafyllopoulos, ``{Higher-point correlations from the
  JIMWLK evolution},'' {\em JHEP} {\bf 1111} (2011)  105,
\href{http://arxiv.org/abs/1109.0302}{{\tt arXiv:1109.0302 [hep-ph]}}.

\bibitem{Iancu:2011nj}
E.~Iancu and D.~Triantafyllopoulos, ``{JIMWLK evolution in the Gaussian
  approximation},'' {\em JHEP} {\bf 1204} (2012)  025,
\href{http://arxiv.org/abs/1112.1104}{{\tt arXiv:1112.1104 [hep-ph]}}.

\bibitem{Alvioli:2012ba}
M.~Alvioli, G.~Soyez, and D.~Triantafyllopoulos, ``{Testing the Gaussian
  Approximation to the JIMWLK Equation},''
  \href{http://dx.doi.org/10.1103/PhysRevD.87.014016}{{\em Phys.Rev.} {\bf D87}
  (2013)  014016},
\href{http://arxiv.org/abs/1212.1656}{{\tt arXiv:1212.1656 [hep-ph]}}.

\bibitem{Iancu:2002tr}
E.~Iancu, K.~Itakura, and L.~McLerran, ``{Geometric scaling above the
  saturation scale},''
  \href{http://dx.doi.org/10.1016/S0375-9474(02)01010-2}{{\em Nucl. Phys.} {\bf
  A708} (2002)  327--352},
\href{http://arxiv.org/abs/hep-ph/0203137}{{\tt arXiv:hep-ph/0203137}}.

\bibitem{Mueller:2002zm}
A.~Mueller and D.~Triantafyllopoulos, ``{The Energy dependence of the
  saturation momentum},''
  \href{http://dx.doi.org/10.1016/S0550-3213(02)00581-3}{{\em Nucl.Phys.} {\bf
  B640} (2002)  331--350},
\href{http://arxiv.org/abs/hep-ph/0205167}{{\tt arXiv:hep-ph/0205167
  [hep-ph]}}.

\bibitem{Munier:2003vc}
S.~Munier and R.~B. Peschanski, ``{Geometric scaling as traveling waves},''
  \href{http://dx.doi.org/10.1103/PhysRevLett.91.232001}{{\em Phys. Rev. Lett.}
  {\bf 91} (2003)  232001},
\href{http://arxiv.org/abs/hep-ph/0309177}{{\tt arXiv:hep-ph/0309177}}.

\bibitem{Balitsky:2006wa}
I.~Balitsky, ``{Quark contribution to the small-x evolution of color dipole},''
  \href{http://dx.doi.org/10.1103/PhysRevD.75.014001}{{\em Phys.Rev.} {\bf D75}
  (2007)  014001},
\href{http://arxiv.org/abs/hep-ph/0609105}{{\tt arXiv:hep-ph/0609105
  [hep-ph]}}.

\bibitem{Kovchegov:2006vj}
Y.~V. Kovchegov and H.~Weigert, ``{Triumvirate of Running Couplings in Small-x
  Evolution},'' \href{http://dx.doi.org/10.1016/j.nuclphysa.2006.10.075}{{\em
  Nucl.Phys.} {\bf A784} (2007)  188--226},
\href{http://arxiv.org/abs/hep-ph/0609090}{{\tt arXiv:hep-ph/0609090
  [hep-ph]}}.

\bibitem{Balitsky:2008zza}
I.~Balitsky and G.~A. Chirilli, ``{Next-to-leading order evolution of color
  dipoles},'' \href{http://dx.doi.org/10.1103/PhysRevD.77.014019}{{\em
  Phys.Rev.} {\bf D77} (2008)  014019},
\href{http://arxiv.org/abs/0710.4330}{{\tt arXiv:0710.4330 [hep-ph]}}.

\bibitem{Beuf:2014uia}
G.~Beuf, ``{Improving the kinematics for low-x QCD evolution equations in
  coordinate space},'' \href{http://dx.doi.org/10.1103/PhysRevD.89.074039}{{\em
  Phys.Rev.} {\bf D89} (2014)  074039},
\href{http://arxiv.org/abs/1401.0313}{{\tt arXiv:1401.0313 [hep-ph]}}.

\bibitem{Iancu:2015vea}
E.~Iancu, J.~Madrigal, A.~Mueller, G.~Soyez, and D.~Triantafyllopoulos,
  ``{Resumming double logarithms in the QCD evolution of color dipoles},''
  \href{http://dx.doi.org/10.1016/j.physletb.2015.03.068}{{\em Phys.Lett.} {\bf
  B744} (2015)  293--302},
\href{http://arxiv.org/abs/1502.05642}{{\tt arXiv:1502.05642 [hep-ph]}}.

\bibitem{Ducloue:2019ezk}
B.~Duclou{\'e}, E.~Iancu, A.~H. Mueller, G.~Soyez, and D.~N.
  Triantafyllopoulos, ``{Non-linear evolution in QCD at high-energy beyond
  leading order},'' \href{http://dx.doi.org/10.1007/JHEP04(2019)081}{{\em JHEP}
  {\bf 04} (2019)  081},
\href{http://arxiv.org/abs/1902.06637}{{\tt arXiv:1902.06637 [hep-ph]}}.

\bibitem{Iancu:2020jch}
E.~Iancu, A.~H. Mueller, D.~N. Triantafyllopoulos, and S.~Y. Wei, ``{Saturation
  effects in SIDIS at very forward rapidities},''
  \href{http://dx.doi.org/10.1007/JHEP07(2021)196}{{\em JHEP} {\bf 07} (2021)
  196}, \href{http://arxiv.org/abs/2012.08562}{{\tt arXiv:2012.08562
  [hep-ph]}}.

\bibitem{Dumitru:2016jku}
A.~Dumitru and V.~Skokov, ``{$\cos(4\phi)$ azimuthal anisotropy in small-$x$
  DIS dijet production beyond the leading power TMD limit},''
  \href{http://dx.doi.org/10.1103/PhysRevD.94.014030}{{\em Phys. Rev. D} {\bf
  94} (2016) no.~1, 014030}, \href{http://arxiv.org/abs/1605.02739}{{\tt
  arXiv:1605.02739 [hep-ph]}}.

\bibitem{Altinoluk:2019fui}
T.~Altinoluk, R.~Boussarie, and P.~Kotko, ``{Interplay of the CGC and TMD
  frameworks to all orders in kinematic twist},''
  \href{http://dx.doi.org/10.1007/JHEP05(2019)156}{{\em JHEP} {\bf 05} (2019)
  156}, \href{http://arxiv.org/abs/1901.01175}{{\tt arXiv:1901.01175
  [hep-ph]}}.

\bibitem{Altinoluk:2019wyu}
T.~Altinoluk and R.~Boussarie, ``{Low $x$ physics as an infinite twist (G)TMD
  framework: unravelling the origins of saturation},''
  \href{http://dx.doi.org/10.1007/JHEP10(2019)208}{{\em JHEP} {\bf 10} (2019)
  208}, \href{http://arxiv.org/abs/1902.07930}{{\tt arXiv:1902.07930
  [hep-ph]}}.

\bibitem{Fujii:2020bkl}
H.~Fujii, C.~Marquet, and K.~Watanabe, ``{Comparison of improved TMD and CGC
  frameworks in forward quark dijet production},''
  \href{http://dx.doi.org/10.1007/JHEP12(2020)181}{{\em JHEP} {\bf 12} (2020)
  181}, \href{http://arxiv.org/abs/2006.16279}{{\tt arXiv:2006.16279
  [hep-ph]}}.

\bibitem{Boussarie:2020vzf}
R.~Boussarie and Y.~Mehtar-Tani, ``{Gauge invariance of transverse momentum
  dependent distributions at small $x$},''
  \href{http://dx.doi.org/10.1103/PhysRevD.103.094012}{{\em Phys. Rev. D} {\bf
  103} (2021) no.~9, 094012}, \href{http://arxiv.org/abs/2001.06449}{{\tt
  arXiv:2001.06449 [hep-ph]}}.

\bibitem{Hatta:2020bgy}
Y.~Hatta, B.-W. Xiao, F.~Yuan, and J.~Zhou, ``{Anisotropy in Dijet Production
  in Exclusive and Inclusive Processes},''
  \href{http://dx.doi.org/10.1103/PhysRevLett.126.142001}{{\em Phys. Rev.
  Lett.} {\bf 126} (2021) no.~14, 142001},
  \href{http://arxiv.org/abs/2010.10774}{{\tt arXiv:2010.10774 [hep-ph]}}.

\bibitem{Dominguez:2008aa}
F.~Dominguez, C.~Marquet, and B.~Wu, ``{On multiple scatterings of mesons in
  hot and cold QCD matter},''
  \href{http://dx.doi.org/10.1016/j.nuclphysa.2009.03.008}{{\em Nucl. Phys. A}
  {\bf 823} (2009)  99--119}, \href{http://arxiv.org/abs/0812.3878}{{\tt
  arXiv:0812.3878 [nucl-th]}}.

\bibitem{Mulders:2000sh}
P.~J. Mulders and J.~Rodrigues, ``{Transverse momentum dependence in gluon
  distribution and fragmentation functions},''
  \href{http://dx.doi.org/10.1103/PhysRevD.63.094021}{{\em Phys. Rev. D} {\bf
  63} (2001)  094021}, \href{http://arxiv.org/abs/hep-ph/0009343}{{\tt
  arXiv:hep-ph/0009343}}.

\end{thebibliography}

\providecommand{\href}[2]{#2}\begingroup\raggedright\endgroup

\end{document}